\documentclass[preprint2]{proto}

\usepackage{times}
\usepackage{natbib}
\usepackage{lipsum}
\usepackage{upgreek}
\usepackage{threeparttable}
\begin{document}
%

%
\title{\textbf{\LARGE Earth as an Exoplanet}}

\author {\textbf{\large Tyler D. Robinson}}
\affil{\small\em Northern Arizona University}

\author {\textbf{\large Christopher T. Reinhard}}
\affil{\small\em Georgia Institute of Technology}
%

%
\begin{abstract}
\begin{list}{ } {\rightmargin 10mm}
\baselineskip = 11pt
\parindent=1pc
{
\noindent Earth is the only planet known to harbor life and, as a result, the search for habitable and inhabited planets beyond the Solar System commonly focuses on analogs to our planet.  However, Earth's atmosphere and surface environment have evolved substantially in the last 4.5 billion years.  A combination of {\it in situ} geological and biogeochemical modeling studies of our planet have provided glimpses of environments that, while technically belonging to our Earth, are seemingly alien worlds.  For modern Earth, observations from ground-based facilities, satellites, and spacecraft have yielded a rich collection of data that can be used to effectively view our planet within the context of exoplanet characterization.  Application of planetary and exoplanetary remote sensing techniques to these datasets then enables the development of approaches for detecting signatures of habitability and life on other worlds.  In addition, an array of models have been used to simulate exoplanet-like datasets for the distant Earth, thereby providing insights that are often complementary to those from existing observations.  Understanding the myriad ways Earth has been habitable and inhabited, coupled with remote sensing approaches honed on the distant Earth, provides a key guide to recognizing potentially life-bearing environments in other planetary systems.

\vspace{3mm}
{\it Look again at that dot.  That's here.  That's home.  That's us. [\ldots] [E]very saint and sinner in the history of our species lived there --- on a mote of dust suspended in a sunbeam.} - Carl Sagan
}

\vspace{8mm}
\end{list}
\end{abstract}
%

%

\section{INTRODUCTION}
\label{sec:intro}
%

The quest for both habitable and inhabited worlds beyond Earth is key to understanding the potential distribution of life in the Universe.  This ongoing search seeks to answer profound questions: Are we alone? How unique is Earth?  Should the hunt for life beyond Earth uncover a multitude of habitable worlds and few (if any) inhabited ones, humanity would begin to understand just how lonely and fragile our situation is.  On the other hand, if our hunt yields a true diversity of inhabited worlds, then we would learn something fundamental about the tenacity of life in the cosmos.

But how will we recognize a distant habitable world, and how would we know if this environment hosts some form of life?  A key opportunity for understanding the remote characterization of habitability and life comes from studying our own planet --- Earth will always be our best example of a habitable and inhabited world.  Thus, by studying our planet within the context of exoplanet exploration and characterization, we develop ideas, approaches, and tools suitable for remotely detecting the signs of (near) global surface habitability and a vigorous planet-wide biosphere.  While habitable exoplanets are unlikely to look exactly like Earth, these worlds will probably share some important characteristics with our own, including the presence of oceans, clouds, surface inhomogeneities, and, potentially, life.  Studying globally-averaged observations of Earth within the context of remote sensing therefore provides insights into the ideal measurements to identify planetary habitability from data-limited exoplanet observations.  

Of course, Earth is not a static environment.  Life emerged on our planet into an environment completely unlike the Earth we understand today.  The subsequent evolution of our planet --- an intimate coupling between life and geochemical processes --- produced worlds seemingly alien to modern Earth.  Ranging from ice-covered ``Snowball Earth'' scenarios to the likely oxygen-free and, potentially, intermittently hazy atmosphere of the Archean (3.8--2.5 giga-annum [Ga]), each evolutionary stage of our planet offers a unique opportunity to understand habitable, life-bearing worlds distinct from the present Earth.


The chapter presented here summarizes studies of Earth within the context of exoplanet characterization.  Following a brief synopsis of the current state of exoplanet science, we review our understanding of the evolution of Earth, and its associated appearance, over the last four billion years.  Then, using this understanding of Earth through time, we review how key remotely-detectable biosignatures for our planet may have changed over geological timescales.  We then shift our emphasis to modern Earth, where existing observational datasets and modeling tools that can be used to explore ideas related to characterizing Earth-like planets from a distance.  Finally, we present an overview of what has been learned by studying Earth as an exoplanet, summarizing approaches to remote characterization of potentially habitable or inhabited worlds.  For further reading, we note that an entire book on studies of the distant Earth has been published by \citet{vazquezetal2010}.

\bigskip
\noindent
\subsection{Current State of Exoplanet Science}
\label{subsec:exoplanetsci}
\bigskip

Following the first detection of an exoplanet around a Sun-like star \citep{mayor&queloz1995} and of an exoplanet atmosphere \citep{charbonneauetal2002} over a decade ago, the field of exoplanetary science has been marked by two clear trends --- the steady discovery of increasingly smaller worlds on longer-period orbits, and the ever-increasing quality of observational data suitable for characterizing worlds around other stars.  Due to advances in exoplanet detection using a variety of techniques, we now know that, on average, every star in the Milky Way galaxy hosts at least one exoplanet \citep{cassanetal2012}.  Furthermore, due in large part to the success of the {\it Kepler} mission \citep{boruckietal2010}, we understand that occurrence rates of potentially Earth-like worlds orbiting within the Habitable Zone of Main Sequence stars are relatively large, with estimates spanning roughly 10--50\% \citep{dressing&charbonneau2013,petiguraetal2013,batalha2014,foremanmacketal2014,burkeetal2015,dressing&charbonneau2015,kopparapuetal2018}.  Excitingly, and especially for low-mass stellar hosts, surveys have revealed a number of nearby potentially Earth-like exoplanets, such as Proxima Centauri b \citep{angladaescudeetal2016} or the worlds in the TRAPPIST-1 system \citep{gillonetal2016,gillonetal2017}.

The subsequent characterization of exoplanet atmospheres has largely been accomplished using transit and/or secondary eclipse spectroscopy \citep[for a review, see][]{kreidberg2018}.  The former relies on the wavelength-dependent transmittance of an exoplanet atmosphere \citep{seager&sasselov2000,brown2001,hubbardetal2001}, which causes a transiting world to block more (for lower transmittance) or less (for higher transmittance) light when crossing the disk of its host.  By comparison, secondary eclipse spectroscopy measures the planet-to-star flux ratio by observing the combined star and exoplanet spectrum prior to the planet disappearing behind its host star (i.e., secondary eclipse).  As with any burgeoning field, some findings related to exoplanet atmospheres remain controversial or have undergone substantial revision \citep{lineetal2014,hansenetal2014,diamondloweetal2014}. Nevertheless, using these techniques astronomers have probed the atmospheres of a striking variety of exoplanets, spanning so-called hot Jupiters \citep{grillmairetal2008,swainetal2008,pontetal2008,swainetal2009,singetal2009,madhusudhan&seager2009}, as well as mini-Neptunes and super-Earths \citep{stevensonetal2010,beanetal2010,lineetal2013b,kreidbergetal2014a,knutsonetal2014a,knutsonetal2014b,ehrenreichetal2014,fraineetal2014,singetal2016,stevensonetal2016,dewitetal2018}.  

Unfortunately, transit or secondary eclipse spectroscopy is not well-suited to studying the atmospheres of potentially Earth-like planets orbiting within the Habitable Zone of their Sun-like\footnotemark hosts, due to the long orbital periods, small transit probabilities, and low signal sizes for such worlds.  Here, direct (or high-contrast) imaging will likely be the leading observational approach, and, as a result, the material below focuses primarily on directly observing Earth (in both reflected light and thermal emission).  For exoplanets, direct imaging involves blocking the light of a bright central host star in order to resolve and observe faint companions to that star \citep{traub&oppenheimer2010}.  Both internal and external occulting technologies are under active study \citep{guyonetal2006,cashetal2007,shaklanetal2010,mawetetal2012}, and ground-based telescopes equipped with coronagraphs already enable the characterization of hot gas giant exoplanets orbiting young, nearby stars \citep{maroisetal2008,skemeretal2012,macintoshetal2015}.

\footnotetext{``Sun-like'' typically refers to main sequence stars with spectral type F, G, or K. Such stars range from 30\% larger and $6\times$ more luminous than our Sun down to 30\% smaller and $13\times$ less luminous than the Sun. Sun-like stars are often contrasted to ``late type'' stars, which are red, cool main sequence M stars that can be smaller than 10\% the size of our Sun and can have luminosities as small as 1--2\% that of the Sun.}

\bigskip
\noindent
\subsection{The Future of Rocky Exoplanets}
\label{subsec:exoplanetfuture}
\bigskip

A number of planned or under-study missions will improve and expand our ability to characterize exoplanet atmospheres and surfaces. Foremost among these is NASA's {\it James Webb Space Telescope} (\emph{JWST}) \citep{gardneretal2006}, which is expected to provide high-quality transit and secondary eclipse spectra of many tens of targets over the duration its designed five year mission \citep{beichmanetal2014}.  Some of these observations will probe lower-mass, potentially rocky exoplanets \citep{demingetal2009,batalhaetal2015}.  Critically, {\it JWST} may even be capable of characterizing temperate Earth-sized planets orbiting low-mass stars \citep{kaltenegger&traub2009,cowanetal2015,barstowetal2016}, though the ability to conduct such studies depends largely on the behavior and size of systematic noise sources \citep{greeneetal2016}.

Following {\it JWST}, NASA will launch the Wide-Field InfraRed Survey Telescope \citep[{\it WFIRST};][]{spergeletal2013}.  It is anticipated that {\it WFIRST} will be equipped with a Coronagraphic Instrument (CGI) capable of visible-light imaging and, potentially, spectroscopy or spectro-photometry of exoplanets \citep{noeckeretal2016}.  Key outcomes of this mission will include a demonstration of high-precision coronagraphy in space, as well as the study of a small handful of cool, gas giant exoplanets \citep{marleyetal2014,hu2014,burrows2014,lupuetal2016,traubetal2016,nayaketal2017}.  However, the planned capabilities of {\it WFIRST}/CGI will make observations of Earth-like planets extremely unlikely \citep{robinsonetal2016}.

Exoplanet direct imaging missions that could build on the technological successes of {\it WFIRST} are already under investigation.  Included here are the {\it WFIRST} starshade rendezvous concept \citep{seageretal2015}, the Habitable Exoplanet (HabEx) imaging mission \citep{mennessonetal2016,gaudietal2018}, and the Large Ultraviolet-Visible-InfraRed (LUVOIR) explorer \citep{petersonetal2017,robergeetal2018}.  While the scope and capabilities of these mission concepts are varied \citep{starketal2016}, a central goal unites these designs --- to detect and characterize Pale Blue Dots around our nearest stellar neighbors.

%
\bigskip
\section{EARTH AS AN EVOLVING EXOPLANET}
\label{sec:early}
\bigskip

\begin{figure*}
 \epsscale{1.5}
 \plotone{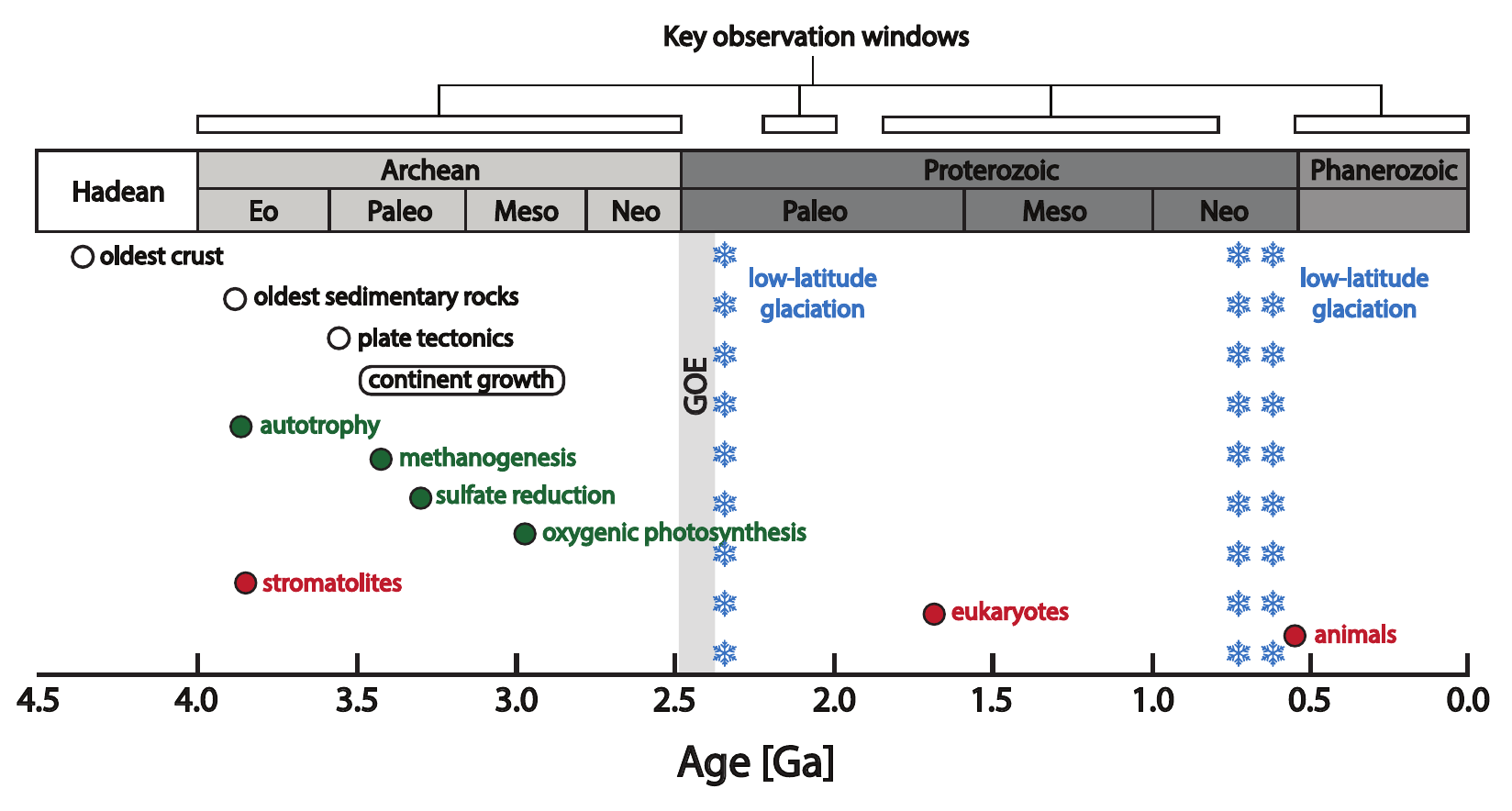}
 \caption{\small Summary timeline of geologic history for Earth, with major divisions of the geologic timescale (top) related to quantitative ages (bottom) according to billions of years before the present (Ga). Major early geologic events are shown by open white circles, including the development of the oldest crust and sedimentary rocks, the emergence of plate tectonics, and an interval of significant growth/exposure of continental crust above sea level. Filled green circles denote approximate geochemical constraints on major metabolic innovations, including carbon fixation (autotrophy), methane production, sulfate reduction, and oxygenic photosynthesis. A subset of major events from the fossil record is given by filled red circles, including the first putative stromatolites (layered sedimentary structures induced by microorganisms), the emergence of eukaryotes (large, complex cells with a membrane-bound nucleus), and the first multicellular animal fossils. Also shown are the initial oxygenation of Earth's atmosphere --- the Great Oxidation Event (GOE) --- and three occurrences of unusually intense low-latitude glaciation ('Snowball Earth' events). Above the geologic timeline we summarize our key 'observation windows' examined in Figure~\ref{fig:earthatm} and Table~\ref{tbl:earth_atm_evol}.} 
 \label{fig:geotimeline}
\end{figure*}

Earth, however, has not always been the Pale Blue Dot we see today. Indeed, the Earth system has evolved considerably over time (Figure~\ref{fig:geotimeline}).  These changes have in turn impacted both the habitability of Earth surface environments \citep[e.g.,][]{kasting&catling2003} and the remote detectability of Earth's biosphere \citep{kalteneggeretal2007,meadows2008,reinhardetal2017,rugheimer&kaltenegger2018}.  In particular, the atmospheric abundances of almost all potential biosignature gases (e.g., CH$_4$, O$_2$, O$_3$, N$_2$O, CO$_2$) have changed by many orders of magnitude throughout Earth's history.  The timing and magnitude of these changes have been controlled by often complex interactions between biological, geologic, and stellar factors.  At the same time, Earth's climate system and surface habitability have changed significantly, as influenced by both long-term trends in solar energy flux, catastrophic climate destabilization during low-latitude ``Snowball Earth'' glaciations, and major impact events. 

Despite these dramatic changes, all life on Earth appears to share a single origin that is perhaps nearly as ancient as Earth itself \citep{foxetal1980,hedges2002}.  Earth's history thus allows us to explore the long-term evolutionary factors controlling the production and maintenance of remotely detectable signatures of habitability and life against the backdrop of a continuously inhabited planet.  Fully illuminating this history requires integration of geologic observations, geochemical data, and constraints from theoretical models --- and provides a unique opportunity to develop predictive frameworks that can be leveraged in the search for living planets beyond Earth.

In this section, our focus is on observations and models aimed at constraining surface habitability and atmospheric composition through time on Earth, with a particular eye toward key habitability indicators and atmospheric biosignatures.  Critically, data and models can be used to constrain the surface habitability and atmospheric composition of Earth through time, enabling an assessment of the strength of different biosignature features and habitability markers and how these may change through planetary evolution.  In addition, Earth's geologic history provides a series of empirical tests of our understanding of Earth-like planets as integrated systems --- with different periods of Earth's evolutionary history serving as analogs for alien, yet habitable, worlds for which we have biological and geological constraints.


\bigskip
\noindent
\subsection{Geological Constraints on Evolving Climate}
\label{subsec:earlygeoclim}
\bigskip

Understanding the evolution of Earth's climate system is critical for diagnosing how potential observational discriminants of Earth's habitability may have changed with time.  Interestingly, with some notable exceptions (see below) Earth's climate appears to have been clement for the vast majority of its history.  For instance, isotopic evidence from the oldest minerals on Earth suggest the presence of liquid water at Earth's surface by 4.3 billion years ago (Ga) \citep{mojzsisetal2001,ushikuboetal2008}, and a consistent if fragmentary marine sedimentary rock record attests to a large-scale fluid-mediated rock cycle for the last ~3.8 billion years \citep[e.g.,][]{rosingetal1996}.  Other lines of evidence \citep{knauth&epstein1976,robert&chaussidon2006,gaucheretal2008} have been used to suggest that surface temperatures during much of the Archean Eon (3.8-2.5 Ga) were hot, perhaps as high as 70$^\circ$C.  More recent estimates suggest much cooler (but still quite warm) temperatures between 25--40$^\circ$C \citep{hrenetal2009,blakeetal2010}. 

Geologic evidence for glacial deposits can provide first-order information about evolving climate, particularly if the approximate location and altitude of glacial activity can be constrained.  The most well-established sedimentary evidence for early glacial activity near sea level is found in the Mozaan Group, South Africa at $\sim\!\!2.9$ Ga \citep{youngetal1998}.  Given existing paleolatitude constraints of around 45--50$^\circ$ \citep{koppetal2005}, these deposits suggest a climate similar to or colder than that of the Pleistocene Earth (a relatively cold epoch of repeated glaciations spanning 2.59--0.012~mega-annum [Ma]).  More recently, sedimentological evidence of glacial activity has been reported from the $\sim$3.5 Ga Overwacht Group, Barberton Greenstone Belt, South Africa \citep{dewit&furnes2016}, although their glacial origin is less definitive than that of the Mozaan Group sediments \citep[e.g.,][]{viljoen&viljoen1969}.  The reconstructed paleolatitudes of these deposits are between $\sim\!\!20$--40$^\circ$, so if they are indeed glaciogenic in origin they would imply a relatively cold early Archean climate.

Firm evidence for glacial activity near sea level does not reappear until after the Archean-Proterozoic boundary, with a series of glacial deposits observed in North America \citep{young2001} and South Africa \citep{rasmussenetal2013} between $\sim\!\!2.4-2.3$ Ga.  Glaciogenic deposits found on the Kaapval craton, South Africa, recently dated to 2.426$\pm$0.003~Ga \citep{gumsleyetal2017}, show evidence for being deposited at low latitudes \citep{evansetal1997}, leading to the suggestion that these deposits record a Paleoproterozoic ``Snowball Earth'' --- classically envisaged as a catastrophic destabilization of the climate system during which runaway ice-albedo feedback causes the advance of ice sheets to the tropics, and a virtual shutdown of the hydrologic cycle \citep{budyko1969,sellers1969}.  The temporal correspondence between this apparently intense ice age and the initial accumulation of O$_2$ in Earth's atmosphere (Figure~\ref{fig:geotimeline}) has led to the intriguing suggestion that the climate system was transiently destabilized by a sharp drop in atmospheric CH$_4$ attendant to rising atmospheric O$_2$ \citep{kasting2005}.

Recent evidence suggests that these intense ice ages were followed by a transient period of atmospheric oxygenation, after which the climate system appears to have been relatively stable with little firm evidence for glaciation between $\sim\!\!1.8-0.8$ billion years ago (here referred to as the 'mid-Proterozoic').  A notable exception to this comes in the form of putative glacial deposits from the Vazante Group, east-central Brazil \citep{azmyetal2008,geboyetal2013}, though their age is somewhat enigmatic \citep{geboyetal2013,rodriguesetal2012}.  These deposits indicate that at least portions of the mid-Proterozoic were not entirely ice-free, but their deposition at relatively high paleolatitude \citep{tohveretal2006} renders their broader climatic implications somewhat difficult to interpret.  

The close of the Proterozoic Eon (2.5--0.541 Ga) bore witness to perhaps the most severe climate perturbations in Earth's history, the Neoproterozoic ``Snowball Earth'' events \citep{hoffmanetal1998}.  Recent high-resolution geochronology delineates two major glacial episodes, the protracted Sturtian glaciation (lasting between 717--660 Ma) and the shorter Marinoan glaciation (terminating at 635 Ma), with a relatively brief interglacial period lasting less than 25 million years \citep{rooneyetal2015}.  While understanding the intensity, dynamics, and biogeochemical impacts of these glaciations remain areas of active research [extensively reviewed in \citet{hoffman&schrag2002}, \citet{pierrehumberetal2011}, and \citet{hoffmanetal2017}], it is clear that this period marks a dramatic perturbation to planetary climate and would have represented a significant and protracted shift in the remotely detectable indicators of Earth's habitability.

The Phanerozoic Eon (e.g., the last 541 million years) has been marked by at least three large-scale ice ages \citep{delabroye&vecoli2010,veevers&powell1987,zachosetal2001}. These events have been linked with faunal turnover and mass extinction \citep{raymond&metz2004}, and in some cases reflect major milestones in the evolution of Earth's biosphere such as the earliest colonization of the land surface by simple plant life at $\sim\!\!470$  Ma \citep{lentonetal2012} and the extensive production and burial of organic matter by burgeoning terrestrial ecosystems around 300 Ma \citep{feulner2017}.  At the same time, the most recent half-billion years of Earth's history shows evidence for significant transient perturbations to Earth's carbon cycle and climate system on a wide range of timescales \citep{zachosetal2008,honischetal2012}, often associated with dramatic changes to the diversity and abundance of macroscopic life \citep{irwin1994,payneetal2004}.  Nevertheless, despite large changes to carbon fluxes into and out of the ocean-atmosphere system, Phanerozoic Earth's climate has consistently avoided the sort of catastrophic climate destabilization witnessed during the late Proterozoic.  

In sum, Earth's geologic record suggests that the establishment of a robust hydrosphere, with liquid water oceans and low-temperature aqueous weathering of exposed crust, occurred very shortly after Earth's formation.  In addition, surface temperatures have generally been stable and relatively warm for the vast majority of Earth's history, despite long-term changes in solar insolation and dramatic changes to atmospheric composition (see below).  However, this history also highlights the importance of internal feedbacks within the climate system in structuring planetary habitability on Earth (and thus the likelihood of remote detection) over time.  For example, the ``Snowball Earth'' glaciations suggest that an Earth-like planet that spends its lifetime safely within the Habitable Zone of its host star can still undergo catastrophic climate destabilization, and that both the timing and duration of these events can be unpredictable \citep[e.g.,][]{rooneyetal2015}.  This is in marked contrast to the regular, periodic mode of climate instability predicted for Earth-like planets near the outer edge of the Habitable Zone \citep{haqqmisraetal2016}.  In addition, the contrasting timescales of the two Neoproterozoic glaciations imply a wide range of potential effects on the long-term maintenance of remotely detectable biosignatures, placing significant impetus on better understanding the large-scale biogeochemistry of ``Snowball Earth'' conditions (see below).

\bigskip
\noindent
\subsection{Geological Constraints on Evolving Atmospheric Chemistry}
\label{subsec:earlygeochem}
\bigskip

Geologic and geochemical data also provide a window into the dramatic evolution of Earth's atmospheric chemistry, with implications for both the habitability of surface environments and the remote detectability of atmospheric biosignatures.  For the last $\sim\!\!2$ million years, atmospheric composition on Earth can be tracked directly by analyzing the composition of volatiles trapped in ice \citep{wolff&spahni2007}.  Prior to this, biogeochemists and planetary scientists seeking to reconstruct the composition of Earth's atmosphere must rely on some form of 'proxy' --- an indirect indicator of atmospheric composition.  For example, in many species of plant the spatial density of cells on the leaf that are used to exchange gases with the environment --- 'stomata' --- scales coherently with the abundance of CO$_2$ in the growth environment \citep[e.g.,][]{royer2001}.  Paleobotanists can thus use the stomatal density of fossil plant leaves as a proxy for atmospheric CO$_2$ abundance in Earth's past.  Our focus in this chapter is on major changes to atmospheric gas species that are important for regulating global climate (e.g., CO$_2$, CH$_4$, N$_2$, and possibly H$_2$) and species that are potentially promising biosignature gases (e.g., O$_2$, O$_3$, CH$_4$, N$_2$O).  Some species, most notably methane and organic hazes, serve dual roles as both arbiters of climate and potential biosignatures.  

\begin{figure*}
 \epsscale{1.5}
 \plotone{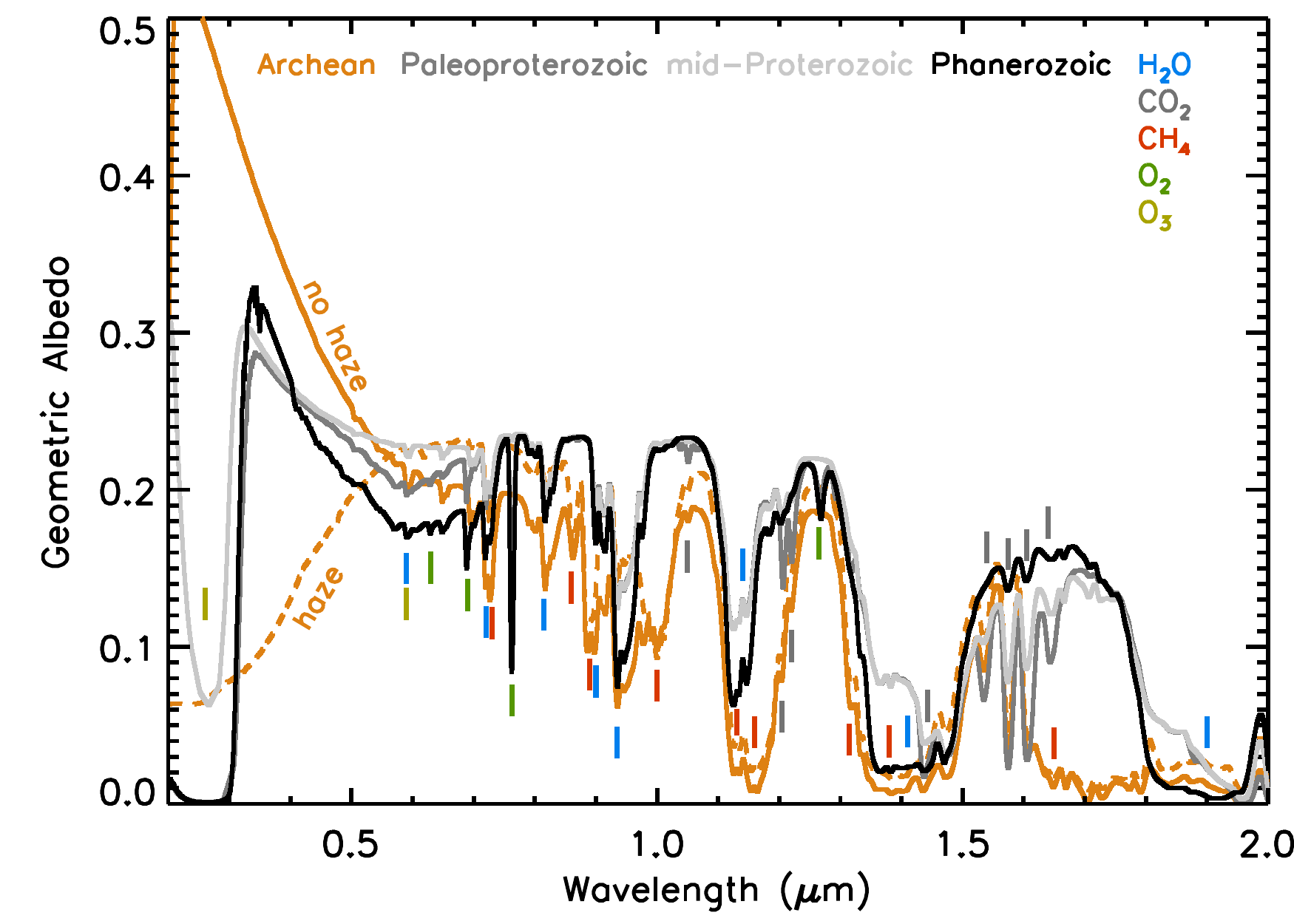}
 \caption{\small Simulated spectra of Earth at key evolutionary stages.  Colors indicate time period: Archean (orange), Paleoproterozoic (dark gray), mid-Proterozoic (light gray), and Phanerozoic (black).  Both hazy and haze-free Archean models are shown, and all models include fractional water cloud coverage.  Key absorption features are indicated.  Original sources for spectra are \citet[][Archean]{arneyetal2016}, \citet[][Phanerozoic]{robinsonetal2011}, and E.~Schwieterman (personal communication, Paleoproterozoic and mid-Proterozoic).}
 \label{fig:earthevol}
\end{figure*}

There are four broad temporal intervals of Earth's evolutionary history that are relevant to our purposes: the Archean (4.0 -- 2.5 Ga), the early Paleoproterozoic (more specifically the interval between 2.2 -- 2.0 Ga), the mid-Proterozoic (specifically the interval between 1.8 -- 0.8 Ga), and the Phanerozoic (roughly 0.5 Ga to the present) (see Figure~\ref{fig:geotimeline}).  It is important to bear in mind that these all represent extremely long periods of time, and that there is likely to be higher-order variability within each interval. Nevertheless, model-derived spectra of Earth during these key evolutionary stages demonstrate how varying atmospheric compositions have led to dramatically different spectral appearances for our planet over time (Figure~\ref{fig:earthevol}). 

\begin{figure*}
 \epsscale{1.5}
 \plotone{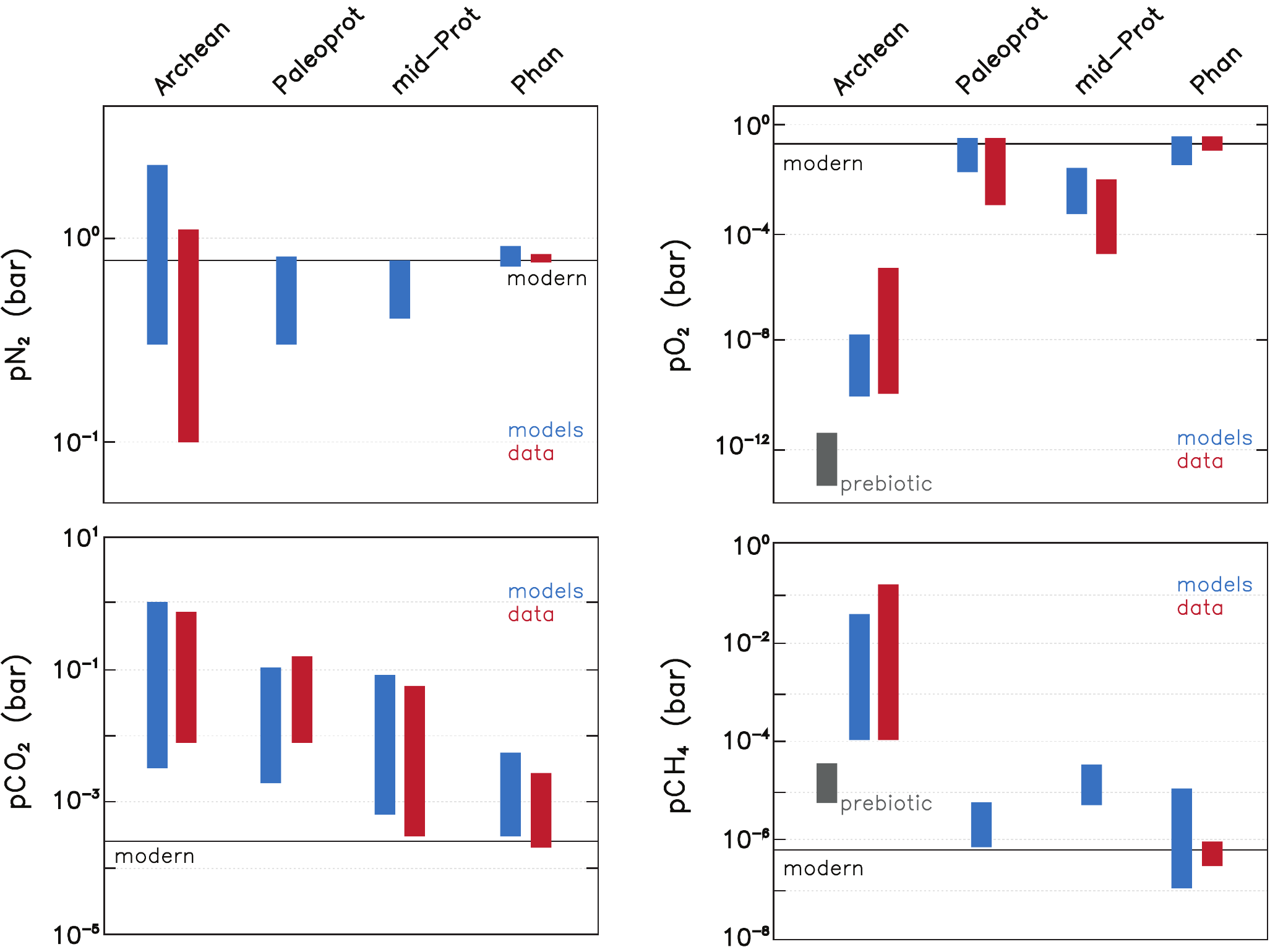}
 \caption{\small Summary of theoretical and empirical constraints on the abundances of N$_2$, O$_2$, CO$_2$, and CH$_4$ in Earth's atmosphere for the four major time periods discussed in the text (Archean, Paleoproterozoic, mid-Proterozoic, and Phanerozoic).   Blue bars show reconstructions from models, while red bars show inferences based on empirical data.  Also shown for the Archean are model-based estimates of prebiotic O$_2$ and CH$_4$ levels (grey bars).  The ranges are meant to be inclusive, and some of the variability in a given time period should be considered to arise from time-dependent variability rather than uncertainty [e.g., \citet{olsonetal2018b}].  Constraints are as described in Table~\ref{tbl:earth_atm_evol}.} 
 \label{fig:earthatm}
\end{figure*}

\bigskip
\noindent
\subsubsection{Atmospheric Oxygen (O$_2$) and Methane (CH$_4$) on Earth Through Time}
\label{subsec:earthgeoo2ch4}
\bigskip

The atmospheric abundances of O$_2$ CH$_4$ on Earth have changed considerably over time (Figure~\ref{fig:earthatm}).  In particular, the abundant O$_2$ in Earth's modern atmosphere is a relatively recent phenomenon --- atmospheric O$_2$ abundance has increased over Earth's history by many orders of magnitude, from the very low levels of the Hadean/Archean, through a period of intermediate values during the Proterozoic, eventually rising to the high levels we observe on Earth today.  Because both O$_2$ and CH$_4$ are largely generated and recycled through biological processes, and are spectrally active, they both represent potentially useful atmospheric biosignatures.  Levels of O$_2$ and CH$_4$ are also linked mechanistically through the redox state of Earth's atmosphere, such that periods of Earth's history characterized by low atmospheric O$_2$ tend to feature elevated atmospheric CH$_4$, and vice-versa (Figure~\ref{fig:earthatm}).  In this section, we discuss the evolution of atmospheric O$_2$ and CH$_4$ on Earth as constrained by a range of geologic and geochemical indicators, with an eye toward better understanding how the detectability of Earth's biosphere has changed over time. 

Sulfur (S) isotope distributions in marine sedimentary rocks of Archean age suggest extremely low atmospheric O$_2$ levels, constrained to an upper limit of $\sim\!\!10^{-6}$ bar \citep{farquhar2000,pavlov&kasting2002} but most likely below $\sim\!\!10^{-8}$ bar \citep{claireetal2006,zahnleetal2006}.  These anomalies also suggest high abundance of some reducing gas for effective production of S$_8$ in the atmosphere, with CH$_4$ as the most likely candidate \citep{zahnleetal2006}.  More recently, coherent time-dependent changes in S isotope systematics have been tied to the transient production and breakdown of atmospheric organic hazes during the late Archean \citep{zerkleetal2012,izonetal2017}.  Given the surface CH$_4$ fluxes required to maintain persistent organic hazes, their presence may serve as an effective biosignature in reducing planetary atmospheres such as that of the Archean Earth \citep{arneyetal2018}.   This isotopic evidence for a broadly reducing, low-O$_2$ ocean-atmosphere system is consistent with a wide range of other geochemical and sedimentological observations \citep{rye&holland1998,rasmussen&buick1999,planavskyetal2010,croweetal2014}.  

Sedimentological and isotopic evidence records the initial accumulation of O$_2$ in Earth's atmosphere during the ``Great Oxidation Event'' (GOE) at $\sim$2.3 billion years ago \citep[e.g.,][]{holland1984,holland2002,luoetal2016} (Figure~\ref{fig:geotimeline}).  More recently, geologic and geochemical evidence has led to the hypothesis of a protracted, but ultimately transient, period of ocean-atmosphere oxygenation following the GOE \citep[reviewed in][]{lyonsetal2014}.  In particular, marine sedimentary carbonate rocks record an extended interval of $^{13}$C enrichment --- the so-called 'Lomagundi Event' \citep{karhu&holland1996,bekkeretal2001,melezhiketal2007} --- which implies a massive release of O$_2$ to the ocean-atmosphere system according to conventional models of Earth's carbon cycle \citep{kump&arthur1999}.  This period also records the earliest extensive marine deposits of sulfate-bearing evaporite minerals \citep{schroderetal2008}, increasingly large sulfur (S) isotope fractionations \citep{canfield2005}, a dramatic increase in the ratio of oxidized iron in marine shales \citep{bekker&holland2012}, significant enrichments of redox-sensitive metals in anoxic marine sediments \citep{canfieldetal2013,partinetal2013,reinhardetal2013}, and the first economic phosphorite deposits \citep{leplandetal2014}. 

A wide range of geochemical proxies point to a subsequent return to relatively low ocean-atmosphere oxygen levels during the mid-Proterozoic, between $\sim\!\!1.8-0.8$ billion years ago, perhaps to levels that would have rendered O$_2$ difficult to detect remotely (see Section~\ref{sec:remotebiosigs}).  In particular, the disappearance of sulfate evaporites and large phosphorite deposits from the rock record \citep{schroderetal2008,reinhardetal2017b}, a drop in ferric to total iron ratios in marine shales \citep{bekker&holland2012}, the S isotope systematics of marine sedimentary rocks \citep{planavskyetal2012,scottetal2014}, and the incomplete retention of Fe and Mn in ancient soil horizons \citep{zbindenetal1988} all point to a decrease in ocean-atmosphere oxygen levels following the Lomagundi Event.  More recent geochemical observations have buttressed this view \citep{planavskyetal2014prot,coleetal2016,hardistyetal2017,tangdetal2016b,partinetal2013,reinhardetal2013,sheenetal2018,bellefroid2018}.  The majority of geochemical observations are consistent with a background {\it p}O$_2$ value at or well below $\sim\!\!10^{-2}$ bar \citep{lyonsetal2014} .  However, sulfur isotopic anomalies in sedimentary rocks characteristic of the Archean do not return during this interval, suggesting that atmospheric {\it p}O$_2$ remained above $\sim\!\!10^{-6}$ bar, atmospheric {\it p}CH$_4$ remained well below $\sim\!\!10^{-2}-10^{-3}$ bar, or both \citep[e.g.,][]{zahnleetal2006}. Precisely quantifying atmospheric {\it p}O$_2$ during this period remains a significant outstanding challenge, and Archean oxygen levels, perhaps paradoxically, are perhaps better constrained than those of the Proterozoic

The late Proterozoic bore witness to significant changes to ocean-atmosphere redox, before, during, and after the ``Snowball Earth'' glaciation events \citep[reviewed in][]{lyonsetal2014}.  Indeed, there is some evidence for a shift in ocean-atmosphere redox immediately preceding the first low-latitude glaciation \citep{planavskyetal2014prot,thomsonetal2015}, implicating time-dependent changes to Earth's oxygen cycle as a potentially important component of climate destabilization in both the Paleoproterozoic and Neoproterozoic (Figure~\ref{fig:geotimeline}).  The ultimate result of these upheavals appears to have been an oxygenation of the ocean-atmosphere system to a degree approaching that of the modern Earth.  For most of Phanerozoic time (541 million years ago to the present), atmospheric {\it p}O$_2$ appears to have remained within the fire window\footnotemark of between $\sim\!\!0.15-0.35$ bar \citep{belcher&mcelwain2008,glasspool&scott2010}, although atmospheric {\it p}O$_2$ during the Paleozoic prior to the rise of land plants is somewhat poorly constrained and may have been below 0.1~bar \citep[e.g.,][]{bergmanetal2004,lentonetal2018}.  However, essentially all available geologic, geochemical, and biological observations are consistent with a well-oxygenated ocean-atmosphere system for the last 500--600 million years \citep{lyonsetal2014}.

\footnotetext{The ``fire window'' refers to a range of atmospheric {\it p}O$_2$ levels that are constrained by widespread charcoal in the Phanerozoic geologic record as well as the persistence of plant life in this same record.  
At the lower bound of the window, burning of plant material would be rare and charcoal would not be widely produced.  At the upper bound of the window, plant burning would be global and could not be extinguished 
\citep{scott&glasspool2006}.}

\bigskip
\noindent
\subsubsection{Atmospheric Carbon Dioxide (CO$_2$) on Earth Through Time}
\label{subsec:earthgeoco2}
\bigskip

The abundance of CO$_2$ in Earth's atmosphere has also changed by orders of magnitude throughout Earth's history (Figure~\ref{fig:earthatm}).  These changes are deeply intertwined with Earth's overall habitability through the carbonate-silicate cycle, or "Walker feedback" \citep{walker1981}, which is hypothesized to regulate atmospheric CO$_2$ levels through temperature-dependent weathering of the silicate crust.  The operation of such a feedback is central to our understanding of planetary habitability, and indeed forms the basis for the "Habitable Zone" concept \citep{kasting1993}.  Earth's history provides some evidence in favor of this framework, with atmospheric CO$_2$ abundance generally decreasing over time in the face of a long-term increase in solar luminosity (Figure~\ref{fig:earthatm}). However, more precisely quantifying the strength and transient dynamics of this feedback, and the boundary conditions under which it may break down, are critical ongoing tasks for researchers with an interest in planetary habitability.  In this section, we discuss geologic and geochemical constraints on atmospheric CO$_2$ abundance on Earth through time, while Section~\ref{subsec:earlymodelchem} below discusses models of the long-term carbon cycle and the impact of evolving atmospheric CO$_2$ on global climate. 

Archean atmospheric {\it p}CO$_2$ is not very well-constrained, though all existing data are consistent with values that were elevated above those of the modern Earth, perhaps by 2--3 orders of magnitude. Mineral assemblages observed within riverine sediments provide a lower limit on atmospheric {\it p}CO$_2$ of roughly 10$^{-3}$ bar at 3.2~Ga \citep{hessleretal2004}, but these observations are also consistent with CO$_2$ levels an order of magnitude or more higher than this.  Similarly, secondary mineral assemblages in ancient soil horizons (paleosols) that formed between $\sim\!\!2.7-2.5$ Ga have been interpreted to indicate {\it p}CO$_2$ values between $\sim\!\!10^{-3}-10^{-2}$ bar during the late Archean \citep{ryeetal1995,sheldon2006}.  A more recent model suggests much higher Archean {\it p}CO$_2$, up to or exceeding $\sim\!\!10^{-1}$ bar \citep{kanzaki&murakami2015}, though estimates according to this method can vary over many orders of magnitude at any given time due to uncertainties in assumed soil formation timescales.  

Broadly, geologic observations tend to suggest a drop in atmospheric CO$_2$ levels between the Paleoproterozoic and the mid-Proterozoic. Paleosols formed between 2.5--1.8~Ga yields a somewhat more consistent picture, with estimates from both of the most recent {\it p}CO$_2$ reconstruction techniques yielding values on the order of $\sim\!\!10^{-3}-10^{-2}$ bar \citep{sheldon2006,kanzaki&murakami2015}, although the estimates of \citet{kanzaki&murakami2015} for similarly aged paleosols are in general a factor of 2--5 higher than those of \citet{sheldon2006}.  There is only one well-studied paleosol near the Archean-Proterozoic boundary, making it difficult to establish with confidence whether significant changes in atmospheric {\it p}CO$_2$ occurred transiting the Archean-Proterozoic boundary.  Moving into the mid-Proterozoic, reconstructions based on paleosols \citep{sheldon2013}, carbon isotopes \citep{kaufman&xiao2003}, and fossils of cyanobacteria \citep{kah&riding2007} are all broadly consistent with atmospheric {\it p}CO$_2$ values on the order of $\sim\!\!3\times10^{-3}$ bar, though individual estimates vary between values roughly equivalent to the modern Earth to high values of $\sim\!\!10^{-2}$ bar.  

A much wider range of potential {\it p}CO$_2$ proxies exists for the Phanerozoic, including a higher-resolution paleosol record, the carbon isotope compositions of plant and phytoplankton fossils, the density of stomata on fossilized leaves, and geochemical proxies of ocean pH \citep[reviewed in][]{royer2014}.  Although these approaches are all undergoing continual refinement, they generally point to a range for atmospheric {\it p}CO$_2$ during the Phanerozoic between roughly $\sim\!\!10^{-4}$--10$^{-3}$ bar, with significant time-dependent shifts associated with major biospheric innovations and changes in climate (see above).  An important caveat to this record is that constraints become very patchy during early Paleozoic time (e.g., prior to round 400 million years ago), a period during which Earth system models indicate atmospheric CO$_2$ levels were higher than at any other time during the Phanerozoic (Berner \& Kothavala, 2001).  However, recent model inversions suggest that atmospheric {\it p}CO$_2$ has never been significantly above $\sim\!\!5$x10$^{-3}$ bar during the last 500 million years \citep{royeretal2014,lentonetal2018} (see Section~\ref{subsec:earlymodelclim}) .

\bigskip
\noindent
\subsubsection{Atmospheric Nitrogen (N$_2$) on Earth Through Time}
\label{subsec:earthgeon2}
\bigskip

Earth's modern atmosphere is $\sim\!\!80$\% molecular nitrogen (N$_2$) by volume.  A range of geologic processes could potentially lead to large changes in atmospheric N$_2$ on Earth, but the empirical trajectory of N$_2$ levels is poorly known at present.  This is important, because on Earth the abundance of N$_2$ in the atmosphere largely controls overall atmospheric pressure, which can significantly impact overall atmospheric opacity to infrared radiation through pressure-induced broadening of absorption lines for greenhouse gases, linking atmospheric pressure and global surface temperature \citep[e.g.,][]{goldblattetal2009}. In addition, one of the few known mechanisms for producing an abiotic O$_2$-rich planetary atmosphere, and thus a potential biosignature "false positive" for O$_2$ \citep{meadows2018}, relies on low atmospheric pressure \citep{wordsworth&pierrehumbert2014}.  As a result, there is strong impetus to better understand the evolutionary history of N$_2$ on Earth as a means to better understand the factors controlling planetary N$_2$ cycling on other Earth-like planets.

A current frontier in reconstructing the evolution of Earth's atmosphere is developing constraints on atmospheric pressure, as linked most directly with changes in atmospheric N$_2$ abundance.  As discussed in Section~\ref{subsec:earlymodelclim}, mass balance calculations suggest that N$_2$ levels may have varied significantly from the present level of $\sim$0.8~bar, with potentially non-trivial impacts on climate \citep{goldblattetal2009}.  Recent approaches toward reconstructing overall atmospheric pressure have included estimating air density based on the diameter of fossilized raindrop imprints in a 2.7-billion-year-old tuff from the Ventersdorp Supergroup, South Africa \citep{sometal2012}, and estimating total barometric pressure via the size distribution of vesicles in a lava flow from roughly the same age preserved in the Pilbara Craton, Australia \citep{sometal2016}.  The technique based on raindrop imprints implies that atmospheric pressure at 2.7~Ga was between $\sim$0.5--2.0~bar, though placing an upper limit with this method is difficult \citep{kavanagh&goldblatt2015}.  The vesicle approach provides a much more stringent upper limit of around 0.5~bar.  \citet{martyetal2013} attempted to estimate {\it p}N$_2$ directly by analyzing the isotopic composition of nitrogen and argon in 3.5 and 3.0 billion-year-old fluids trapped in hydrothermal quartz from the Pilbara craton, Australia, deriving mixing arrays between end-member hydrothermal fluids of variable composition with a single end-member for air-saturated Archean seawater.  Their analysis indicates that {\it p}N$_2$ was not significantly above $\sim$0.5--1.0~bar during the early Archean.  \citet{nishizawaetal2007} provide a {\it p}N$_2$ estimate of $\sim$3 bar from fluid inclusions in the same unit, but the N$_2$/$^{36}$Ar values from their samples indicate that their data do not capture the low-N$_2$ end-member analyzed by \citet{martyetal2013} and thus likely overestimate ambient {\it p}N$_2$.  In any case, uncertainties in all current approaches and the fragmentary nature of the archives required for their application are such that current data are consistent with atmospheric N$_2$ abundance a factor of two or more above/below modern, rendering the potential climate impacts of N$_2$ somewhat enigmatic at present but important to consider (see below).

\begin{deluxetable}{llllllll}
\tabletypesize{\small}
\tablecaption{Empirical and Theoretical Constraints on Earth's Evolving Atmospheric Composition \label{tbl:earth_atm_evol}}
\tablewidth{0pt}
\tablehead{Period & Case & {\it p}N$_2$ [bar] & {\it p}O$_2$ [bar] & {\it p}CO$_2$ [bar] & {\it p}CH$_4$ [bar] & Source(s) \\ }
\startdata
{\it prebiotic}          &            &                        &                                      &                        &                                                                      &                       \\
                         & model high & \multicolumn{1}{c}{--} & \multicolumn{1}{c}{3.0 $\times$ 10$^{-12\dagger}$} & \multicolumn{1}{c}{--} & \multicolumn{1}{c}{3.0 $\times$ 10$^{-5}$} & \multicolumn{1}{l}{\it 1,2} \\
                         & model low  & \multicolumn{1}{c}{--} & \multicolumn{1}{c}{3.0 $\times$ 10$^{-14\ddagger}$} & \multicolumn{1}{c}{--} & \multicolumn{1}{c}{5.0 $\times$ 10$^{-6}$} & \multicolumn{1}{l}{\it 1,3} \\
                         & data high  & \multicolumn{1}{c}{--} & \multicolumn{1}{c}{--} & \multicolumn{1}{c}{--} & \multicolumn{1}{c}{--} & \multicolumn{1}{l}{--} \\
                         & data high  & \multicolumn{1}{c}{--} & \multicolumn{1}{c}{--} & \multicolumn{1}{c}{--} & \multicolumn{1}{c}{--} & \multicolumn{1}{l}{--} \\
                         &            &             &          &                                      &                        &                                             &                       \\
{\it Archean}            & model high & \multicolumn{1}{c}{2.3} & \multicolumn{1}{c}{2.0 $\times$ 10$^{-8}$} & \multicolumn{1}{c}{1.0 $\times$ 10$^{0}$} & \multicolumn{1}{c}{3.5 $\times$ 10$^{-2}$} & \multicolumn{1}{l}{\it 4-12} \\
                         & model low  & \multicolumn{1}{c}{0.3} & \multicolumn{1}{c}{1.0 $\times$ 10$^{-10}$} & \multicolumn{1}{c}{3.0 $\times$ 10$^{-3}$} & \multicolumn{1}{c}{1.0 $\times$ 10$^{-4}$} & \multicolumn{1}{l}{\it 5-13} \\
                         & data high  & \multicolumn{1}{c}{1.1} & \multicolumn{1}{c}{2.0 $\times$ 10$^{-6}$} & \multicolumn{1}{c}{7.0 $\times$ 10$^{-1}$} & \multicolumn{1}{c}{1.4 $\times$ 10$^{-1\parallel}$} & \multicolumn{1}{l}{\it 14-16} \\
                         & data low   & \multicolumn{1}{c}{0.1$^{\ell}$} & \multicolumn{1}{c}{--} & \multicolumn{1}{c}{7.0 $\times$ 10$^{-3}$} & \multicolumn{1}{c}{1.0 $\times$ 10$^{-4}$} & \multicolumn{1}{l}{\it 7,17-19} \\                
                         &            &             &          &                                      &                        &                                             &                       \\
{\it Paleoproterozoic}   & model high & \multicolumn{1}{c}{0.8} & \multicolumn{1}{c}{3.0 $\times$ 10$^{-1}$} & \multicolumn{1}{c}{1.0 $\times$ 10$^{-1}$} & \multicolumn{1}{c}{5.0 $\times$ 10$^{-6}$} & \multicolumn{1}{l}{\it 8-10,13,20,21} \\
                         & model low  & \multicolumn{1}{c}{0.3} & \multicolumn{1}{c}{2.0 $\times$ 10$^{-2}$} & \multicolumn{1}{c}{2.0 $\times$ 10$^{-3}$} & \multicolumn{1}{c}{7.0 $\times$ 10$^{-7}$} & \multicolumn{1}{l}{\it 8-10,13,21} \\
                         & data high  & \multicolumn{1}{c}{--} & \multicolumn{1}{c}{3.0 $\times$ 10$^{-1}$} & \multicolumn{1}{c}{1.5 $\times$ 10$^{-1}$} & \multicolumn{1}{c}{--} & \multicolumn{1}{l}{\it 16,22} \\
                         & data low   & \multicolumn{1}{c}{--} & \multicolumn{1}{c}{1.0 $\times$ 10$^{-3}$} & \multicolumn{1}{c}{7.0 $\times$ 10$^{-3}$} & \multicolumn{1}{c}{--} & \multicolumn{1}{l}{\it 18,22,23} \\                
                         &            &             &          &                                      &                        &                                             &                       \\
{\it mid-Proterozoic}    & model high & \multicolumn{1}{c}{0.8} & \multicolumn{1}{c}{2.0 $\times$ 10$^{-2}$} & \multicolumn{1}{c}{8.0 $\times$ 10$^{-2}$} & \multicolumn{1}{c}{3.0 $\times$ 10$^{-5}$} & \multicolumn{1}{l}{\it 8-10,13,24-26} \\
                         & model low  & \multicolumn{1}{c}{0.4} & \multicolumn{1}{c}{6.0 $\times$ 10$^{-4}$} & \multicolumn{1}{c}{6.0 $\times$ 10$^{-4}$} & \multicolumn{1}{c}{5.0 $\times$ 10$^{-6\zeta}$} & \multicolumn{1}{l}{\it 8-10,13,25,27} \\
                         & data high  & \multicolumn{1}{c}{--} & \multicolumn{1}{c}{8.0 $\times$ 10$^{-3}$} & \multicolumn{1}{c}{5.5 $\times$ 10$^{-2}$} & \multicolumn{1}{c}{--} & \multicolumn{1}{l}{\it 23,28,29} \\
                         & data low   & \multicolumn{1}{c}{--} & \multicolumn{1}{c}{2.0 $\times$ 10$^{-5}$} & \multicolumn{1}{c}{3.0 $\times$ 10$^{-4}$} & \multicolumn{1}{c}{--} & \multicolumn{1}{l}{\it 30,31} \\      
                         &            &             &          &                                      &                        &                                             &                       \\
{\it Phanerozoic}        & model high & \multicolumn{1}{c}{0.9} & \multicolumn{1}{c}{3.0 $\times$ 10$^{-1}$} & \multicolumn{1}{c}{5.5 $\times$ 10$^{-3}$} & \multicolumn{1}{c}{1.0 $\times$ 10$^{-5}$} & \multicolumn{1}{l}{\it 32-36} \\
                         & model low  & \multicolumn{1}{c}{0.7} & \multicolumn{1}{c}{4.0 $\times$ 10$^{-2}$} & \multicolumn{1}{c}{2.8 $\times$ 10$^{-4}$} & \multicolumn{1}{c}{1.0 $\times$ 10$^{-7}$} & \multicolumn{1}{l}{\it 32-36} \\
                         & data high  & \multicolumn{1}{c}{0.8} & \multicolumn{1}{c}{1.5 $\times$ 10$^{-1}$} & \multicolumn{1}{c}{2.8 $\times$ 10$^{-3}$} & \multicolumn{1}{c}{8.0 $\times$ 10$^{-7}$} & \multicolumn{1}{l}{\it 37-39} \\
                         & data low   & \multicolumn{1}{c}{0.8} & \multicolumn{1}{c}{3.0 $\times$ 10$^{-1}$} & \multicolumn{1}{c}{1.9 $\times$ 10$^{-4}$} & \multicolumn{1}{c}{3.5 $\times$ 10$^{-7}$} & \multicolumn{1}{l}{\it 39-41} \\      
\enddata
\tablerefs{$^{1}$\citet{haqqmisraetal2011},$^{2}$\citet{emmanuel&ague2007},$^{3}$\citet{tianetal2011},$^{4}$\citet{goldblattetal2009},$^{5}$\citet{claireetal2006},$^{6}$\citet{zahnleetal2006},$^{7}$\citet{kurzweiletal2013},$^{8}$\citet{kasting1987},$^{9}$\citet{halevy&bachan2017},$^{10}$\citet{krissansentottonetal2018},$^{11}$\citet{kharechaetal2005},$^{12}$\citet{ozakietal2018},$^{13}$\citet{stuekenetal2016},$^{14}$\citet{martyetal2013},$^{15}$\citet{pavlov&kasting2002},$^{16}$\citet{kanzaki&murakami2015},$^{17}$\citet{sometal2016},$^{18}$\citet{sheldon2006},$^{19}$\citet{drieseetal2011},$^{20}$\citet{bachan&kump2015},$^{21}$\citet{haradaetal2015},$^{22}$\citet{bekker&holland2012},$^{23}$\citet{rye&holland1998},$^{24}$\citet{laakso&schrag2017},$^{25}$\citet{catlingetal2007},$^{26}$\citet{zhaoetal2018},$^{27}$\citet{olsonetal2016},$^{28}$\citet{kaufman&xiao2003},$^{29}$\citet{kah&riding2007},$^{30}$\citet{planavskyetal2014prot},$^{31}$\citet{sheldon2013},$^{32}$\citet{berner2006},$^{33}$\citet{royer2014},$^{34}$\citet{lentonetal2018},$^{35}$\citet{bartdorffetal2008},$^{36}$\citet{beerlingetal2009},$^{37}$\citet{belcher&mcelwain2008},$^{38}$\citet{royer2014},$^{39}$\citet{wolff&spahni2007},$^{40}$\citet{glasspool&scott2010},$^{41}$\citet{galbraith&eggleston2017}}
\tablenotetext{\dagger}{Assumes {\it p}CO$_2$ = 2 bar.}
\tablenotetext{\ddagger}{Assumes {\it p}CO$_2$ = 0.02 bar.}
\tablenotetext{\parallel}{Assuming data high {\it p}CO$_2$ and CH$_4$/CO$_2$ = 0.2.}
\tablenotetext{\ell}{Assuming predominantly N$_2$ atmosphere.}
\tablenotetext{\zeta}{Assuming {\it p}O$_2$ = 10$^{-3}$ bar, [SO$_4^{2-}$] = 500 µmol/kg.}
\tablecomments{All values are approximate.  See primary references for assumptions and caveats not noted here.}
\end{deluxetable}

\bigskip
\noindent
\subsection{Model Constraints on Evolving Climate}
\label{subsec:earlymodelclim}
\bigskip

Standard stellar evolution models \citep{gough1981} predict that the Sun was 20--30\% less luminous than it is today during the Hadean and Archean.  With a planetary greenhouse effect equivalent to that of the modern Earth, this would lead to below-freezing global average surface temperatures prior to $\sim\!\!2$ Ga, in stark contrast to observations from Earth's rock record --- an inconsistency often referred to as the "Faint Young Sun Paradox" \citep{sagan&mullen1972,feulner2012}.  The observation of a generally clement or even warm climate during the Hadean and Archean (see above) thus implies that the composition of Earth's atmosphere was very different from that of the modern (barring major changes in orbital parameters or non-chemical albedo effects).  Indeed, as discussed above, there is persuasive geological and geochemical evidence that the composition of Earth's atmosphere was very different during the Hadean, Archean, and Proterozoic.  

The most prominent solutions to this problem invoke a larger inventory of greenhouse gases in Earth's early atmosphere. \citet{sagan&mullen1972} explored a reducing NH$_3$--CH$_4$--H$_2$S--H$_2$O--CO$_2$ greenhouse, with a dominant role for NH$_3$.  However, the rapid photolysis of NH$_3$ in the upper atmosphere would have required a very large source at Earth's surface, and would have in turn resulted in the production of rather extreme amounts of N$_2$ on geologically rapid timescales \citep{kuhn&atreya1979}.  More recently, it has been suggested that the photolysis of NH$_3$ in the upper atmosphere may have been mitigated somewhat by absorption of UV photons by a fractal organic haze \citep{wolf&toon2010}, an idea that warrants additional scrutiny of the relative altitudes of peak NH$_3$ photolysis and haze absorption in future work \citep{wolf&toon2010}.  In any case, subsequent work has tended to focus on CH$_4$-CO$_2$-H$_2$O greenhouses and, more recently, the possible radiative effects of high H$_2$ \citep{kasting2005,haqqmisraetal08}.

The factors regulating planetary climate on the prebiotic Earth are not particularly well-understood.  Direct constraints are few, but models predict that the abundance of CH$_4$ in Earth's prebiotic atmosphere would have been low \citep{kasting2005,emmanuel&ague2007}, while a recent inversion using a geologic carbon cycle model yields a median {\it p}CO$_2$ estimate of 0.3 bar with a 95\% confidence interval of 0.03--1.0 bar \citep{krissansentottonetal2018}.  Greenhouses dominated by H$_2$O and CO$_2$ with {\it p}CO$_2$ values on the lower end of this range would be unlikely to exhibit clement surface temperatures under Hadean or early Archean solar luminosity, but both 1-D radiative-convective and 3-D global climate models predict that values at the upper end of this range would result in surface temperatures well above freezing under early and late Archean solar luminosity \citep{kasting&ackerman1986,kasting1987,charnayetal2013,wolf&toon2013,wolf&toon2014}.  Significant additional warming may have been provided by collision-induced absorption by H$_2$-N$_2$ under plausible prebiotic conditions \citep{wordsworth&pierrehumbert2013}, though the strength of this would have depended strongly on atmospheric H$_2$ and N$_2$ abundance, both of which are poorly constrained for the prebiotic atmosphere.

The emergence of a biosphere on Earth would have had a significant impact on atmospheric chemistry and climate.  In particular, primitive microbial metabolisms such as methanogenesis, acetogenesis, and anoxygenic photosynthesis would have dramatically increased fluxes of CH$_4$ to Earth's atmosphere.  The implications of this for climate are twofold.  First, CH$_4$ is an important greenhouse gas in its own right, providing another means toward offsetting decreased solar luminosity that would be particularly effective in a reducing atmosphere \citep{pavlovetal2000,haqqmisraetal08}.  Second, the potential for large biogenic CH$_4$ fluxes introduces an additional feedback on climate via formation of an organic haze in the atmosphere \citep{zahnle1986,pavlovetal2001,haqqmisraetal08}.  Photochemical models \citep{haqqmisraetal08,zerkleetal2012,arneyetal2016} and laboratory experiments \citep{dewittetal2009,traineretal2004,traineretal2006} predict that once the CH$_4$/CO$_2$ ratio of the atmosphere increases beyond $\sim\!\!0.1$ hydrocarbon aerosols will begin to form, with optical thicknesses at visible and UV wavelengths that increase rapidly at CH$_4$/CO$_2$ values between $\sim\!\!0.1$--1 \citep{domagalgoldmanetal2008,haqqmisraetal08}.  The formation of these hazes can lead to significant cooling, even at relatively low CH$_4$/CO$_2$ values of $\sim\!\!0.1-0.2$ \citep{arneyetal2016}.  These combined effects would have been important in regulating climate and surface temperature during both the Archean and Proterozoic.  

Taken together, models suggest that clement or even warm surface temperatures could have been maintained on the prebiotic Earth by a CO$_2$--H$_2$O greenhouse, potentially supplemented by collision- and pressure-induced warming at greater H$_2$ and N$_2$ abundance \citep{kasting1987,goldblattetal2009,wordsworth&pierrehumbert2013,krissansentottonetal2018}.  Following the emergence of a surface biosphere, 1-D radiative-convective and 3-D global climate models, coupled ecosystem-biogeochemistry models, and the geologic record are all consistent in suggesting that surface temperatures well above freezing could have been maintained throughout the Archean with a CH$_4$--H$_2$O--CO$_2$ greenhouse and optically thin haze \citep{haqqmisraetal08,zerkleetal2012,charnayetal2013,wolf&toon2013,izonetal2017,ozakietal2018,krissansentottonetal2018}.  However, additional factors beyond changes to the atmospheric greenhouse may also have been important, and together could lead to significant warming.  For example, changes to the average size of cloud droplets attendant to fewer cloud condensation nuclei (CCN) in the early atmosphere would have resulted in more effective rainout and fewer low clouds at low-mid latitudes \citep{charnayetal2013,wolf&toon2014} and possibly thinner clouds overall \citep{goldblatt&zahnle2011,charnayetal2013}, both of which would have resulted in significant warming.  Maintaining a habitable climate during Earth's earliest history is thus not particularly challenging, as all of these factors were potentially in play.  However, achieving the highest estimates of Archean temperature with plausible atmospheric CO$_2$ and CH$_4$ levels remains a challenge.

Existing models of the long-term carbon cycle and surface temperature are broadly consistent with geochemical evidence for elevated {\it p}CO$_2$ during the Paleoproterozoic \citep[e.g.,][]{halevy&bachan2017,krissansentottonetal2018}.  However, the initial rise of atmospheric O$_2$ at $\sim\!\!2.3$ Ga would have destabilized the Archean atmospheric greenhouse \citep{claireetal2006,zahnleetal2006,haqqmisraetal08}, potentially leading to dramatic effects on climate during the earliest Paleoproterozoic (see above).  Understanding climate dynamics and coherently modeling climate and biogeochemistry during and after the GOE and through the Lomagundi Event remain outstanding challenges, but existing models suggest large changes to Earth surface temperature, the global carbon cycle, and atmospheric composition \citep{claireetal2006,haradaetal2015}.

There is limited geologic evidence for glacial conditions during the mid-Proterozoic (see above), despite a solar luminosity $\sim$10\% lower than that of the modern Earth \citep{gough1981}.  Recent coupled 3-D climate modeling indicates that global glaciation should occur under these conditions if atmospheric {\it p}CO$_2$ drops to around 10$^{-3}$ bar \citep{fiorella&sheldon2017}, and extensive glaciation at high and middle latitudes should occur even above {\it p}CO$_2$ values an order of magnitude higher than the modern Earth unless surface temperatures are buffered by some other greenhouse gas.  Nitrous oxide (N$_2$O) would be unlikely to provide the requisite radiative forcing on its own, given the relatively low atmospheric {\it p}O$_2$ during the mid-Proterozoic and the large biological nitrogen fluxes required \citep{robersonetal2011}, but may have a marginal impact on surface temperatures \citep{buick2007,stantonetal2018}.  Methane (CH$_4$) is another candidate, but 3-D models of ocean biogeochemistry \citep{olsonetal2016} interpreted in light of 3-D climate modeling \citep{fiorella&sheldon2017} indicate that an ocean-only CH$_4$ cycle would have been unable to maintain an ice-free climate during the mid-Proterozoic.  One possible solution to this would be a significant microbial CH$_4$ flux from terrestrial ecosystems \citep{zhaoetal2018}.  Alternatively, atmospheric {\it p}CO$_2$ may have been somewhat higher than geochemical proxies suggest \citep{krissansentottonetal2018}, or other changes to factors like cloud droplet radius or surface albedo may have contributed to stabilizing relatively warm temperatures \citep{fiorella&sheldon2017}.  Lastly, at least periods of the mid-Proterozoic may not have been entirely ice-free (see above).  In any case, climate models, geochemical proxies, and models of marine/terrestrial biogeochemistry yield a picture of a relatively weak H$_2$O--CO$_2$ greenhouse buffered by CH$_4$ levels on the order of $\sim\!\!10^{-5}$ bar or slightly less, depending on the importance of terrestrial CH$_4$ cycling.  

The dynamics of Earth's climate during the intense ice ages of the late Proterozoic are much more well-studied than their Paleoproterozoic counterparts, and the reader is here referred to two recent comprehensive reviews on the subject \citep{pierrehumberetal2011,hoffmanetal2017}.  However, the biogeochemical dynamics associated with these perturbations are more poorly understood, particularly with regard to Earth's O$_2$ and CH$_4$ cycles, and this represents an important topic of future work.  For example, although impacts to the Earth's "oxidized" carbon cycle have been explored in a range of models \citep{lehiretal2008geo,lehiretal2008bgs,millsetal2011}, it remains unclear what role Earth's CH$_4$ cycle may have played in the inception or recovery from low-latitude glaciation, if any \citep{schragetal2002,pierrehumberetal2011,olsonetal2016}.  In any case, the Sturtian glacial episode in particular is estimated to have lasted for roughly 50 million years \citep{rooneyetal2014}, with the attendant impacts on atmospheric biosignatures and the remote detectability of any surviving biosphere almost completely unknown.

The Phanerozoic climate system, though perhaps relatively stable in the scope of Earth's entire history, has been extremely dynamic, with at least three major ice ages and intervening periods of relatively warm, largely ice-free conditions (see above).  Though they differ in their tectonic boundary conditions, scope/intensity, and overall biological impact these climate shifts are generally considered to have been driven primarily by variations in atmospheric CO$_2$, together with internal climate system feedbacks and modulated by changes in Earth's orbital parameters \citep{zachosetal2001,royeretal2004,herrmannetal2004,montanez&poulsen2013}.  The Phanerozoic climate system is thus thought to have been controlled largely by an H$_2$O--CO$_2$ greenhouse, buffered by volcanic outgassing, organic carbon weathering under a high-O$_2$ atmosphere, and solar luminosity roughly equivalent to that of the modern Earth.  A notable exception to this may have occurred during the Carboniferous 'coal swamp' era roughly 300 million years ago, during which biogeochemical and climate models predict significant radiative forcing from atmospheric CH$_4$  \citep{bartdorffetal2008,beerlingetal2009}.  Taken together, low-order and 3-D climate models are consistent with geologic and geochemical records in indicating that Earth's surface has been consistently habitable for the last $\sim$500 million years, with long-term average surface temperatures largely within the range of $\sim\!\!15-25^{\circ}$C \citep{royeretal2004,lentonetal2018}.

\bigskip
\noindent
\subsection{Model Constraints on Evolving Atmospheric Chemistry}
\label{subsec:earlymodelchem}
\bigskip

Theoretical models have also provided a great deal of insight into the evolving redox state and major background gas composition of Earth's atmosphere.  Earth's prebiotic atmosphere is generally thought to have been mildly reducing, composed predominantly of N$_2$--CO$_2$--H$_2$O, with variable H$_2$ and CH$_4$ and only trace amounts of species like O$_2$, O$_3$, and N$_2$O.  Photochemical models assuming that hydrogen escaped the Archean atmosphere at the diffusion limit \citep[e.g.,][]{hunten1973} predict that ground-level atmospheric O$_2$ would have been on the order of $\sim\!\!10^{-14}$-$10^{-12}$~bar, with H$_2$ on the order of $\sim\!\!10^{-4}$-$10^{-3}$ bar depending on the assumed volcanic outgassing rate \citep{kasting&walker1981,haqqmisraetal2011}.  However, it has been suggested that lower exobase temperatures in an O$_2$-poor, CO$_2$-rich atmosphere would have significantly decreased the efficiency of thermal (Jeans) escape at the top of the atmosphere, with the result that rates of hydrogen escape would have been controlled instead by extreme ultraviolet (EUV) energy fluxes from the young Sun \citep{tianetal2005}.  Balancing these ``energy-limited'' escape rates with plausible volcanic H$_2$ outgassing rates results in H$_2$ mixing ratios as high as $\sim$0.3~bar, with important ramifications for prebiotic chemistry \citep{tianetal2005} and early climate \citep{wordsworth&pierrehumbert2013}.  There remains some debate as to whether heating of the exobase by gases other than O$_2$ and/or other nonthermal escape processes could promote more efficient escape \citep{catling2006,tianetal2006}.  Nevertheless, it remains plausible that Earth's prebiotic atmosphere was relatively H$_2$-rich. 

The evolution of the earliest biosphere would have dramatically transformed atmospheric chemistry.  In particular, the emergence of microbial methanogenesis and the evolution of the most primitive forms of anoxygenic photosynthesis, both of which are thought to be very ancient based on geochemical \citep{tice&lowe2004,uenoetal2006} and phylogenetic \citep{xiongetal2000,wolfe&fournier2018} evidence, would have consumed H$_2$ through reactions such as:

\begin{center}
CO$_2$ + 4H$_2$ $\rightarrow$ CH$_4$ + 2H$_2$O

2H$_2$ + CO$_2$ $\rightarrow$ CH$_2$O + H$_2$O
\end{center}

Both processes would have the net effect of decreasing the atmospheric H$_2$/CH$_4$ ratio, to an extent that would be limited by the availability of energy and nutrients \citep{kharechaetal2005,ozakietal2018}.  In particular, globally integrated rates of anoxygenic photosynthesis would most likely have been limited by the availability of electron donors (H$_2$, Fe$^{2+}$, H$_2$S), in contrast to oxygenic photosynthesis which can use water as an electron donor (see below). Models of the Earth's primitive biosphere can produce extremely high atmospheric CH$_4$ levels, on the order of $\sim\!\!10^{-2}$ bar (Figure~\ref{fig:earthatm}), depending on electron donor flux, levels of available nutrients, and photosynthetic community assemblage \citep{kharechaetal2005,ozakietal2018}.  However, these models neglect microbial anaerobic oxidation of methane (AOM), under the presumption that Archean seawater sulfate levels were extremely low \citep[e.g.,][]{croweetal2014}, and to some extent plausible upper limits on atmospheric CH$_4$ are constrained by the climate effects of hydrocarbon haze formation at elevated CH$_4$/CO$_2$ ratios such that atmospheric CH$_4$ should not be treated in isolation \citep{ozakietal2018}.  Nevertheless, existing models consistently predict that the emergence of Earth's earliest biosphere would have dramatically shifted the atmospheric CH$_4$/H$_2$ ratio, readily supporting atmospheric CH$_4$ levels that would have the potential for remote observation \citep{reinhardetal2017,arneyetal2018}.

Mass balance calculations \citep{goldblattetal2009} and time-dependent biogeochemical models \citep{stuekenetal2016} are consistent with elevated atmospheric N$_2$ throughout the Archean, but can also accommodate long-term decrease in {\it p}N$_2$ through the Hadean and Archean depending on the assumed history of CO$_2$ outgassing and the mechanics coupling organic C and N burial \citep{stuekenetal2016}.  Similarly, existing models of the carbon cycle that are coupled to long-term stellar evolution are broadly consistent with geochemical data in indicating high atmospheric {\it p}CO$_2$ and a secular decline through the Hadean and Archean \citep{kasting1987,sleep&zahnle2001,halevy&bachan2017,charnayetal2017,krissansentottonetal2018}.  The particular trajectories depend on assumptions regarding secular changes in heat flow, the mechanisms regulating seafloor weathering, and the timing and magnitude of major changes to the Earth surface CH$_4$ cycle.  Fully understanding the impact of atmospheric N$_2$ and CO$_2$ on the detectability of habitability markers and atmospheric biosignatures on the Hadean/early Archean Earth will require both more precise geochemical/paleobarometric constraints and further development of approaches that effectively couple models of long-term biogeochemistry and climate to dynamic models of ocean-atmosphere redox.

The evolution of oxygenic photosynthesis resulted in an autotrophic biosphere that could use water as an electron donor, freeing global autotrophy from electron donor limitation and dramatically increasing the potential energy flux through the biosphere:

\begin{center}
CO$_2$ + H$_2$O $\rightarrow$ CH$_2$O + O$_2$
\end{center}

The timing of this event is still debated, but is likely to have occurred at some point prior to the late Archean \citep{kurzweiletal2013,planavskyetal2014arch,magnaboscoetal2018}.  Following the emergence of biological oxygen production, atmospheric chemistry would have been controlled largely by the balance of fluxes between O$_2$ and CH$_4$ produced by the surface biosphere and the consumption of O$_2$ by reducing volcanic/metamorphic gases and weathering reactions with reduced phases in Earth's upper crust \citep{catling&claire2005}.  Low-order biogeochemical models and 1-D models of atmospheric chemistry are consistent with constraints from the geologic record (discussed above) in suggesting low atmospheric {\it p}O$_2$, on the order of $\sim\!\!10^{-10}-10^{-8}$ bar, and relatively high atmospheric {\it p}CH$_4$, around $10^{-4}-10^{-3}$ bar, on the Archean Earth after the emergence of an oxygenic biosphere \citep{goldblattetal2006,claireetal2006,zahnleetal2006,daines&lenton2016}.

A number of potential drivers have been suggested for the GOE, including secular tectonic processes \citep{kump&barley2007,holland2009,gaillardetal2011}, changes in global biological fluxes \citep{koppetal2005,konhauser2009}, and time-integrated hydrogen escape from the atmosphere \citep{catlingetal2001,claireetal2006}.  It is likely that to some extent all of these factors were important.  Atmospheric {\it p}O$_2$ may have changed by many orders of magnitude moving across the GOE, with most models predicting a geologically instantaneous rise from $\sim\!\!10^{-8}$ bar to as high as $\sim\!\!10^{-2}$ bar regardless of the underlying mechanism \citep{claireetal2006,goldblattetal2006}.  Both the timing and magnitude of this event are consistent with existing isotopic records \citep[e.g.,][]{luoetal2016}.  However, biogeochemical models suggest that the imbalance in the global redox budget required to transit the GOE need not necessarily have been large \citep{claireetal2006,goldblattetal2006}.  

The GOE effectively represented a shift in the trace redox gas in Earth's atmosphere from O$_2$ to CH$_4$.  Although models do indeed predict an initial drop in atmospheric {\it p}CH$_4$ during the GOE, the ultimate establishment of a substantial stratospheric O$_3$ layer attendant to rising ground-level {\it p}O$_2$ is predicted to shield CH$_4$ from destruction in the troposphere.  This allows atmospheric CH$_4$ levels to rebound in photochemical models to $\sim\!\!10^{-4}$ bar in the Proterozoic \citep{claireetal2006,goldblattetal2006}.  However, models that include ocean biogeochemistry and microbial consumption of CH$_4$ with O$_2$ and SO$_4^{2-}$ generally result in lower steady-state atmospheric {\it p}CH$_4$ following the GOE, typically on the order of $\sim\!\!10^{-5}$ bar or lower \citep{catlingetal2007,daines&lenton2016,olsonetal2016}.  These results depend strongly on assumed atmospheric {\it p}O$_2$ and the ocean reservoir of SO$_4^{2-}$ \citep{olsonetal2016}, and are not currently equipped to deal with the potentially important impacts of a terrestrial biosphere \citep{zhaoetal2018}.  A full exploration of this problem will require coupled, open-system models of photochemistry and ocean/terrestrial microbial metabolism.  Nevertheless, the abundance of CH$_4$ in Earth's atmosphere following the GOE was likely significantly lower than that of the Archean Earth, and in particular would have been orders of magnitude below that of the earliest Archean and Hadean Earth prior to the evolution of oxygenic photosynthesis \citep{kharechaetal2005,ozakietal2018}.  Unfortunately, a quantitative geologic or geochemical indicator of atmospheric {\it p}CH$_4$ at levels below those of the Archean has not yet been developed, and atmospheric {\it p}CH$_4$ is very difficult to track empirically throughout the remainder of Earth's history.

Although the differences in mean state before and after the GOE are relatively well understood, the dynamics of climate and atmospheric chemistry in the immediate aftermath of the GOE are not.  Some models predict that this change to Earth's surface redox balance would have had significant climate impacts \citep{claireetal2006,haqqmisraetal08}, one result of which may have been an ultimately transient but quantitatively dramatic elevation in atmospheric {\it p}O$_2$ \citep{haradaetal2015}.  This scenario would be consistent with emerging geochemical evidence for elevated atmospheric O$_2$ (and thus O$_3$), possibly for 100-million-year timescales, during the Paleoproterozoic (see above).  The protracted, but ultimately transient, rise in atmospheric {\it p}O$_2$ implies a significant drop in atmospheric CH$_4$ levels \citep{haradaetal2015}, and potentially a substantial drop in atmospheric {\it p}CO$_2$ unless buffered by a sedimentary rock cycle very different from that of the modern Earth \citep[e.g.,][]{bachan&kump2015}.  In any case, long-term (e.g, mean state) carbon cycle and climate models that are entirely uncoupled or only implicitly coupled to the O$_2$ cycle are broadly consistent with the current geochemical constraints for atmospheric {\it p}CO$_2$ during the Paleoproterozoic discussed above \citep{sleep&zahnle2001,halevy&bachan2017,krissansentottonetal2018}.   

The initial accumulation of O$_2$ in Earth's atmosphere appears to have been followed by a subsequent return to relatively low ocean-atmosphere oxygen levels during the mid-Proterozoic (between $\sim\!\!1.8-0.8$ billion years ago).  However, the absence of non-mass-dependent S isotope fractionations in marine sediments and the apparent absence of reduced detrital minerals in fluvial settings indicate atmospheric {\it p}O$_2$ remained above $\sim\!\!10^{-6}$ bar.  On long timescales, the modern atmospheric O$_2$ level is maintained dynamically by the balance between net O$_2$ sources (principally the burial of organic carbon and reduced sulfur into the Earth's upper crust) and net O$_2$ sinks (largely the subsequent exhumation and oxidative weathering of organic carbon and reduced sulfur, along with reactions between O$_2$ and reduced metamorphic and volcanic gases).  However, there are strong nonlinearities in the scaling relationships between these fluxes and the amount of O$_2$ in the atmosphere.  In addition, the major sink fluxes on the modern Earth decrease in magnitude as atmospheric {\it p}O$_2$ drops, while the major source fluxes increase.  As a result, not all atmospheric {\it p}O$_2$ values are equally stable, and understanding the internal processes and feedbacks capable of maintaining atmospheric O$_2$ levels above those characteristic of the Archean but well below those of the modern Earth remains an outstanding question \citep{lyonsetal2014,dainesetal2017}.  

As discussed above, models of mid-Proterozoic climate and biogeochemistry suggest a relatively weak H$_2$O--CO$_2$ greenhouse buffered by modest CH$_4$ levels (Figure~\ref{fig:earthatm}, Table~\ref{tbl:earth_atm_evol}).  In particular, long-term carbon cycle models suggest atmospheric {\it p}CO$_2$ of around $\sim\!\!10^{-3}$--10$^{-2}$ bar during the mid-Proterozoic, while models of marine/terrestrial biogeochemistry suggest atmospheric {\it p}CH$_4$ values on the order of $\sim\!\!10^{-6}$--10$^{-5}$ bar, together consistent with most geochemical constraints and a largely ice-free climate state \citep{fiorella&sheldon2017}.  That said, some paleosol reconstructions approach roughly modern {\it p}CO$_2$ values \citep[e.g.,][]{sheldon2013}, which is difficult to reconcile with evidence for a largely ice-free Earth surface for most of the mid-Proterozoic unless Earth's greenhouse was impacted strongly by fluxes of CH$_4$ from a terrestrial microbial biosphere \citep{zhaoetal2018}.  A full picture of Earth's mid-Proterozoic atmosphere awaits a comprehensive model that couples open-system carbon cycling with a balanced redox budget and dynamic O$_2$-CH$_4$ cycle, but existing data and models are consistent with this period of Earth's history representing a potential "false negative" for conventional biosignature techniques \citep[e.g.,][]{reinhardetal2017} --- a period through which the spectral features of most canonical biosignature gases would have been relatively weak, perhaps for geologic timescales.

Though there is accumulating geologic and geochemical evidence that the extreme low-latitude glaciations of the late Proterozoic were associated with significant changes to ocean-atmosphere redox and atmospheric chemistry \citep{hoffmanetal1998,canfieldetal2007,sahooetal2012,coxetal2013,planavskyetal2014prot,thomsonetal2015,hoffmanetal2017}, the relative timing and mechanistic links remain somewhat obscure.  Simple biogeochemical models indicate that low-latitude glacial episodes can readily drive a secular transition from low- to high-oxygen steady states at sufficiently high {\it p}CO$_2$ thresholds for deglaciation \citep{laakso&schrag2017}.  For example, a deglaciation threshold of {\it p}CO$_2$ $\sim\!\!~0.1$ bar is sufficient to drive a permanent transition in atmospheric {\it p}O$_2$ from 10$^{-3}$ to 10$^{-1}$ bar \citep{laakso&schrag2017} during deglaciation.  Glacial CO$_2$ levels of this order are readily achievable even in models that allow for efficient ocean-atmosphere gas exchange and seafloor weathering \citep[e.g.,][]{lehiretal2008geo}.  Efforts to better understand the temporal polarity and mechanistic details linking climate destabilization and nonlinear changes to atmospheric chemistry during both the Paleoproterozoic and Neoproterozoic represent an important avenue of future work.  Nevertheless, significant changes in the redox state of Earth's ocean-atmosphere system are strongly implicated as having been both cause and consequence of sporadic perturbations to Earth's habitability.

Biogeochemical models are generally consistent in suggesting a high-O$_2$, low-CH$_4$, and moderate CO$_2$ atmosphere throughout the Phanerozoic (541 Ma to the present).  Both atmospheric O$_2$ and atmospheric CO$_2$ have been controlled by the combined effects of roughly modern solar luminosity, time-dependent variability in volcanic degassing, rock uplift, changes to the major ion chemistry of seawater, and the emergence and expansion of terrestrial ecosystems \citep{berner1991,berner2006,royeretal2014,lentonetal2018}.  Long-term atmospheric CH$_4$ levels have been controlled largely by the evolutionary and climate dynamics controlling biogenic CH$_4$ fluxes from terrestrial ecosystems \citep{bartdorffetal2008}.  Despite some discrepancies between different models in estimates of atmospheric {\it p}O$_2$ during the earliest part of the Phanerozoic, most models indicate ranges for atmospheric {\it p}O$_2$, {\it p}CO$_2$, and {\it p}CH$_4$ between $\sim\!\!0.1-0.3$ bar, $\sim\!\!10^{-4}-10^{-3}$ bar, and $\sim\!\!10^{-7}-10^{-5}$ bar, respectively, all of which are consistent with existing geologic and geochemical constraints (Figure~\ref{fig:earthatm}, Table~\ref{tbl:earth_atm_evol}).  Similar models for atmospheric {\it p}N$_2$ through time suggest values close to that of the modern Earth for most of the Phanerozoic, though direct geologic constraints on this are lacking for all but the most recent periods of Earth's history.

Observations from the geologic record and results from quantitative models are united in suggesting extensive changes to the Earth system over time, including the chemistry of the ocean-atmosphere system, the dynamics of long-term climate, and the size and scope of Earth's biosphere.  The contours of this evolution provide important information for exoplanet characterization efforts.  In particular, simulating and predicting observations across the spectrum of habitable worlds represented in Earth's evolutionary history provides a series of test cases for evaluating putative discriminants of habitability and life on Earth-like planets beyond our solar system, and can potentially provide important insight into the challenges associated with deciphering exo-Earth observations.

%
\section{REMOTE DETECTABILITY OF EARTH'S BIOSPHERE THROUGH TIME}
\label{sec:remotebiosigs}
%

How would Earth's evolving climate and atmospheric chemistry have appeared to a remote observer? We focus here on a subset of the most prominent biosignatures that may be remotely detectable --- namely, atmospheric oxygen (O$_2$), its photochemical byproduct ozone (O$_3$), methane (CH$_4$), hydrocarbon haze, and nitrous oxide (N$_2$O) \citep[e.g.,][]{schwietermanetal2018}.  We also include two major habitability indicators --- water vapor (H$_2$O) and carbon dioxide (CO$_2$) --- the latter of which is required for climate system stabilization via the carbonate-silicate geochemical cycle \citep{walker1981}. The true detectability of any particular biosignature or habitability indicator will depend on the magnitude of the signal produced (related to, e.g., a species' atmospheric abundance and the atmospheric opacity produced by a specific feature), the parameters controlling observational precision (e.g., stellar host, distance to the target, instrument and astrophysical noise sources), and the wavelength range accessible to the instrument being used. Our discussion is thus not meant to be exhaustive or definitive, but is instead meant to provide some context for motivating remote observations of the modern Earth and simulated observations of different periods of Earth's evolutionary history.  As a guide to this discussion, Table~\ref{tbl:features} contains a detailed listing of the wavelengths, widths, and strengths of spectral features for key biosignature and habitability indicator gases.

The most prominent features of molecular oxygen (O$_2$) are at 0.76 and 0.69~$\upmu$m, the Fraunhofer~A and B bands, respectively. The Fraunhofer~A band is the stronger of the two bands, but is likely to be relatively weak unless atmospheric {\it p}O$_2$ is above the few percent level \citep{desmaraisetal2002,reinhardetal2017,schwietermanetal2018}. When considered in the context of Earth's evolution, it is clear that O$_2$ spectral features were likely non-existent during the Archean, and may have been weak during much of the Proterozoic. However, absorption by O$_2$ at both the Fraunhofer~A and B bands would have been relatively strong through essentially all of Phanerozoic time and possibly during a protracted interval in the Paleoproterozoic.

\begin{deluxetable}{lccccc}
\tabletypesize{\small}
\tablecaption{Spectral Feature Details for Key Biosignature and Habitability Indicator Gases \label{tbl:features}}
\tablewidth{0pt}
\tablehead{Species & $\lambda$\tablenotemark{a} & $\Delta\lambda$\tablenotemark{b} & $\lambda/\Delta\lambda$ & Opacity\tablenotemark{c} & Optical Depth\tablenotemark{d} \\
                   &      ($\upmu$m)            &           ($\upmu$m)             &                         & (cm$^2$ molecule$^{-1}$) &                              }
\startdata
     O$_2$         &         0.629              &         $3.0\times10^{-3}$       &          210            &   $1.5\times10^{-25}$    &    $6.7\times10^{-1}$          \\
     O$_2$         &         0.689              &         $4.1\times10^{-3}$       &          170            &   $3.9\times10^{-24}$    &    $1.7\times10^{+1}$          \\
     O$_2$         &         0.762              &         $5.1\times10^{-3}$       &          150            &   $5.8\times10^{-23}$    &    $2.6\times10^{+2}$          \\
     O$_2$         &         0.865              &         $6.6\times10^{-3}$       &          130            &   $1.6\times10^{-27}$    &    $7.3\times10^{-3}$          \\
     O$_2$         &         1.07               &         $4.3\times10^{-3}$       &          250            &   $2.7\times10^{-27}$    &    $1.2\times10^{-2}$          \\
     O$_2$         &         1.27               &         $5.8\times10^{-3}$       &          220            &   $7.8\times10^{-25}$    &    $3.5\times10^{+0}$          \\
     O$_2$         &         6.30               &         $3.2\times10^{-1}$       &          19             &   $3.2\times10^{-27}$    &    $1.4\times10^{-2}$          \\
                   &                            &                                  &          ---            &                          &                                \\
     O$_3$         &         0.256              &         $3.9\times10^{-2}$       &          6.5            &   $1.2\times10^{-17}$    &    $1.1\times10^{+2}$          \\
     O$_3$         &         0.600              &         $1.2\times10^{-1}$       &          5.0            &   $4.8\times10^{-21}$    &    $4.4\times10^{-2}$          \\
     O$_3$         &         2.48               &         $2.1\times10^{-3}$       &          1100           &   $1.5\times10^{-21}$    &    $1.3\times10^{-2}$          \\
     O$_3$         &         3.27               &         $6.9\times10^{-3}$       &          470            &   $1.2\times10^{-20}$    &    $1.1\times10^{-1}$          \\
     O$_3$         &         3.59               &         $4.7\times10^{-2}$       &          77             &   $1.8\times10^{-21}$    &    $1.6\times10^{-2}$          \\
     O$_3$         &         4.74               &         $8.7\times10^{-2}$       &          54             &   $7.6\times10^{-20}$    &    $6.9\times10^{-1}$          \\
     O$_3$         &         5.80               &         $1.6\times10^{-1}$       &          37             &   $3.8\times10^{-21}$    &    $3.4\times10^{-2}$          \\
     O$_3$         &         9.58               &         $3.5\times10^{-1}$       &          27             &   $7.3\times10^{-19}$    &    $6.7\times10^{+0}$          \\
     O$_3$         &         14.3               &         $1.0\times10^{+0}$       &          14             &   $2.6\times10^{-20}$    &    $2.3\times10^{-1}$          \\
                   &                            &                                  &          ---            &                          &                                \\
    CH$_4$         &         0.510              &         $5.0\times10^{-3}$       &          100            &   $4.3\times10^{-27}$    &     $1.5\times10^{-7}$         \\
    CH$_4$         &         0.542              &         $5.0\times10^{-3}$       &          110            &   $4.6\times10^{-26}$    &     $1.5\times10^{-7}$         \\
    CH$_4$         &         0.576              &         $8.0\times10^{-3}$       &          72             &   $1.3\times10^{-26}$    &     $1.5\times10^{-7}$         \\
    CH$_4$         &         0.598              &         $8.0\times10^{-3}$       &          75             &   $9.0\times10^{-27}$    &     $1.5\times10^{-7}$         \\
    CH$_4$         &         0.619              &         $8.0\times10^{-3}$       &          77             &   $2.2\times10^{-25}$    &     $1.5\times10^{-7}$         \\
    CH$_4$         &         0.667              &         $1.5\times10^{-2}$       &          44             &   $5.6\times10^{-26}$    &     $1.5\times10^{-7}$         \\
    CH$_4$         &         0.703              &         $1.1\times10^{-2}$       &          64             &   $1.1\times10^{-25}$    &     $1.5\times10^{-7}$         \\
    CH$_4$         &         0.726              &         $1.0\times10^{-2}$       &          73             &   $1.4\times10^{-24}$    &     $1.5\times10^{-7}$         \\
    CH$_4$         &         0.798              &         $2.5\times10^{-2}$       &          32             &   $4.4\times10^{-25}$    &     $1.5\times10^{-7}$         \\
    CH$_4$         &         0.840              &         $1.0\times10^{-2}$       &          84             &   $3.3\times10^{-25}$    &     $1.5\times10^{-7}$         \\
    CH$_4$         &         0.861              &         $1.1\times10^{-2}$       &          78             &   $1.9\times10^{-24}$    &     $1.5\times10^{-7}$         \\
    CH$_4$         &         0.887              &         $1.8\times10^{-2}$       &          49             &   $1.1\times10^{-23}$    &     $1.5\times10^{-7}$         \\
    CH$_4$         &         1.00               &         $3.7\times10^{-2}$       &          27             &   $6.4\times10^{-24}$    &     $1.5\times10^{-7}$         \\
    CH$_4$         &         1.13               &         $1.3\times10^{-2}$       &          87             &   $3.2\times10^{-22}$    &     $1.1\times10^{-2}$         \\
    CH$_4$         &         1.16               &         $7.4\times10^{-3}$       &          160            &   $7.7\times10^{-22}$    &     $2.6\times10^{-2}$         \\
    CH$_4$         &         1.33               &         $1.4\times10^{-4}$       &         9400            &   $1.9\times10^{-21}$    &     $2.6\times10^{-2}$         \\
    CH$_4$         &         1.65               &         $1.3\times10^{-2}$       &         120             &   $1.8\times10^{-20}$    &     $6.2\times10^{-1}$         \\
    CH$_4$         &         1.67               &         $1.4\times10^{-3}$       &        1200             &   $1.4\times10^{-20}$    &     $4.9\times10^{-1}$         \\
    CH$_4$         &         1.68               &         $1.3\times10^{-2}$       &         130             &   $8.1\times10^{-21}$    &     $2.8\times10^{-1}$         \\
    CH$_4$         &         2.20               &         $1.2\times10^{-3}$       &         1800            &   $8.1\times10^{-21}$    &     $2.8\times10^{-1}$         \\
    CH$_4$         &         2.31               &         $6.6\times10^{-2}$       &          35             &   $2.9\times10^{-20}$    &     $1.0\times10^{+0}$         \\
    CH$_4$         &         2.37               &         $2.3\times10^{-2}$       &         100             &   $3.7\times10^{-20}$    &     $1.3\times10^{+0}$         \\
    CH$_4$         &         2.59               &         $1.9\times10^{-2}$       &         130             &   $2.5\times10^{-21}$    &     $8.7\times10^{-2}$         \\
    CH$_4$         &         3.32               &         $9.4\times10^{-2}$       &          35             &   $2.1\times10^{-18}$    &     $7.1\times10^{+1}$         \\
    CH$_4$         &         7.66               &         $2.3\times10^{-1}$       &          34             &   $7.7\times10^{-19}$    &     $2.6\times10^{+1}$         \\
                   &                            &                                  &          ---            &                          &                                \\
    N$_2$O         &        1.52                &         $1.1\times10^{-2}$       &          140            &   $1.2\times10^{-22}$    &     $7.9\times10^{-4}$         \\
    N$_2$O         &        1.67                &         $1.3\times10^{-2}$       &          130            &   $8.0\times10^{-23}$    &     $5.1\times10^{-4}$         \\
    N$_2$O         &        1.70                &         $1.4\times10^{-2}$       &          120            &   $2.9\times10^{-23}$    &     $1.8\times10^{-4}$         \\
    N$_2$O         &        1.77                &         $1.5\times10^{-2}$       &          120            &   $7.5\times10^{-23}$    &     $4.8\times10^{-4}$         \\
    N$_2$O         &        1.96                &         $1.8\times10^{-2}$       &          110            &   $2.2\times10^{-22}$    &     $1.4\times10^{-3}$         \\
    N$_2$O         &        1.99                &         $2.0\times10^{-2}$       &          100            &   $2.3\times10^{-22}$    &     $1.5\times10^{-3}$         \\
    N$_2$O         &        2.04                &         $2.1\times10^{-2}$       &          100            &   $4.0\times10^{-23}$    &     $2.6\times10^{-4}$         \\
    N$_2$O         &        2.11                &         $2.1\times10^{-2}$       &          100            &   $3.3\times10^{-21}$    &     $2.1\times10^{-2}$         \\
    N$_2$O         &        2.16                &         $2.4\times10^{-2}$       &          91             &   $4.6\times10^{-22}$    &     $2.9\times10^{-3}$         \\
    N$_2$O         &        2.26                &         $2.5\times10^{-2}$       &          90             &   $5.2\times10^{-21}$    &     $3.3\times10^{-2}$         \\
    N$_2$O         &        2.46                &         $3.1\times10^{-2}$       &          78             &   $2.2\times10^{-22}$    &     $1.4\times10^{-3}$         \\
    N$_2$O         &        2.61                &         $3.3\times10^{-2}$       &          80             &   $6.6\times10^{-21}$    &     $4.2\times10^{-2}$         \\
    N$_2$O         &        2.67                &         $3.5\times10^{-2}$       &          75             &   $3.1\times10^{-21}$    &     $2.0\times10^{-2}$         \\
    N$_2$O         &        2.87                &         $4.0\times10^{-2}$       &          71             &   $1.5\times10^{-19}$    &     $9.5\times10^{-1}$         \\
    N$_2$O         &        2.97                &         $4.6\times10^{-2}$       &          65             &   $7.3\times10^{-21}$    &     $4.7\times10^{-2}$         \\
    N$_2$O         &        3.58                &         $5.6\times10^{-2}$       &          63             &   $2.5\times10^{-20}$    &     $1.6\times10^{-1}$         \\
    N$_2$O         &        3.90                &         $7.6\times10^{-2}$       &          52             &   $1.1\times10^{-19}$    &     $6.8\times10^{-1}$         \\
    N$_2$O         &        4.06                &         $8.3\times10^{-2}$       &          49             &   $2.5\times10^{-20}$    &     $1.6\times10^{-1}$         \\
    N$_2$O         &        4.31                &         $9.9\times10^{-2}$       &          43             &   $2.2\times10^{-21}$    &     $1.4\times10^{-2}$         \\
    N$_2$O         &        4.50                &         $9.9\times10^{-2}$       &          45             &   $4.4\times10^{-18}$    &     $2.8\times10^{+1}$         \\
    N$_2$O         &        5.32                &         $8.7\times10^{-2}$       &          61             &   $1.6\times10^{-20}$    &     $1.0\times10^{-1}$         \\
    N$_2$O         &        5.72                &         $1.4\times10^{-1}$       &          42             &   $6.4\times10^{-22}$    &     $4.1\times10^{-3}$         \\
    N$_2$O         &        6.12                &         $2.0\times10^{-1}$       &          31             &   $4.1\times10^{-22}$    &     $2.6\times10^{-3}$         \\
    N$_2$O         &        7.78                &         $3.1\times10^{-1}$       &          25             &   $7.4\times10^{-19}$    &     $4.7\times10^{+0}$         \\
    N$_2$O         &        8.56                &         $3.7\times10^{-1}$       &          25             &   $2.6\times10^{-20}$    &     $1.6\times10^{-1}$         \\
    N$_2$O         &        9.47                &         $4.7\times10^{-1}$       &          20             &   $2.7\times10^{-23}$    &     $1.7\times10^{-4}$         \\
    N$_2$O         &        10.7                &         $5.5\times10^{-1}$       &          19             &   $1.4\times10^{-22}$    &     $8.8\times10^{-4}$         \\
    N$_2$O         &        14.4                &         $9.6\times10^{-1}$       &          15             &   $7.4\times10^{-22}$    &     $4.7\times10^{-3}$         \\
    N$_2$O         &        17.0                &         $1.1\times10^{+0}$       &          15             &   $6.4\times10^{-19}$    &     $4.1\times10^{+0}$         \\
                   &                            &                                  &          ---            &                          &                                \\
    H$_2$O         &        0.653               &         $1.0\times10^{-2}$       &          65             &   $1.2\times10^{-23}$    &     $1.1\times10^{+0}$         \\
    H$_2$O         &        0.722               &         $1.1\times10^{-2}$       &          64             &   $1.3\times10^{-22}$    &     $1.2\times10^{+1}$         \\
    H$_2$O         &        0.823               &         $1.5\times10^{-2}$       &          55             &   $1.7\times10^{-22}$    &     $1.7\times10^{+1}$         \\
    H$_2$O         &        0.940               &         $2.1\times10^{-2}$       &          45             &   $2.2\times10^{-21}$    &     $2.1\times10^{+2}$         \\
    H$_2$O         &         1.14               &         $2.9\times10^{-2}$       &          39             &   $4.9\times10^{-21}$    &     $4.8\times10^{+2}$         \\
    H$_2$O         &         1.38               &         $4.3\times10^{-2}$       &          32             &   $6.6\times10^{-20}$    &     $6.5\times10^{+3}$         \\
    H$_2$O         &         1.89               &         $8.9\times10^{-2}$       &          21             &   $1.0\times10^{-19}$    &     $9.8\times10^{+3}$         \\
    H$_2$O         &         2.65               &         $1.8\times10^{-1}$       &          15             &   $8.5\times10^{-19}$    &     $8.3\times10^{+4}$         \\
    H$_2$O         &         3.17               &         $3.0\times10^{-1}$       &          11             &   $6.4\times10^{-21}$    &     $6.3\times10^{+2}$         \\
    H$_2$O         &         3.68               &         $2.2\times10^{-1}$       &          17             &   $5.2\times10^{-23}$    &     $5.0\times10^{+0}$         \\
    H$_2$O         &         6.27               &         $1.1\times10^{+0}$       &          5.7            &   $1.2\times10^{-18}$    &     $1.2\times10^{+5}$         \\
                   &                            &                                  &          ---            &                          &                                \\
    CO$_2$         &         1.43               &         $9.8\times10^{-3}$       &         150             &   $2.6\times10^{-22}$    &     $1.8\times10^{+0}$         \\
    CO$_2$         &         1.58               &         $1.1\times10^{-2}$       &         140             &   $7.6\times10^{-23}$    &     $5.3\times10^{-1}$         \\
    CO$_2$         &         1.61               &         $1.2\times10^{-2}$       &         130             &   $7.6\times10^{-23}$    &     $5.3\times10^{-1}$         \\
    CO$_2$         &         1.96               &         $1.7\times10^{-2}$       &         110             &   $1.9\times10^{-21}$    &     $1.3\times10^{+1}$         \\
    CO$_2$         &         2.01               &         $1.9\times10^{-2}$       &         110             &   $5.6\times10^{-21}$    &     $3.9\times10^{+1}$         \\
    CO$_2$         &         2.06               &         $1.7\times10^{-2}$       &          120            &   $1.2\times10^{-21}$    &     $8.3\times10^{+0}$         \\
    CO$_2$         &         2.69               &         $3.4\times10^{-2}$       &          79             &   $2.6\times10^{-19}$    &     $1.8\times10^{+3}$         \\
    CO$_2$         &         2.77               &         $3.6\times10^{-2}$       &          76             &   $1.7\times10^{-19}$    &     $1.2\times10^{+3}$         \\
    CO$_2$         &         4.26               &         $8.6\times10^{-2}$       &          50             &   $1.6\times10^{-17}$    &     $1.1\times10^{+5}$         \\
    CO$_2$         &         7.31               &         $2.5\times10^{-1}$       &          29             &   $2.9\times10^{-24}$    &     $2.1\times10^{-2}$         \\
    CO$_2$         &         7.94               &         $3.0\times10^{-1}$       &          27             &   $2.6\times10^{-24}$    &     $1.8\times10^{-2}$         \\
    CO$_2$         &         9.40               &         $4.2\times10^{-1}$       &          22             &   $1.3\times10^{-22}$    &     $9.4\times10^{-1}$         \\
    CO$_2$         &         10.4               &         $5.1\times10^{-1}$       &          20             &   $8.5\times10^{-23}$    &     $0.6\times10^{-1}$         \\
    CO$_2$         &         15.0               &         $9.6\times10^{-1}$       &          16             &   $4.4\times10^{-18}$    &     $3.1\times10^{+4}$         \\
\enddata
\tablenotetext{a}{Feature central wavelength.}
\tablenotetext{b}{Feature full-width at half-max (assuming $p=1$~bar and $T=288$~K).}
\tablenotetext{c}{Peak opacity at assumed pressure and temperature.}
\tablenotetext{d}{Assuming modern Earth (vertical) column densities and adopting peak opacity values.}
\end{deluxetable}

Ozone (O$_3$), which is produced photochemically by O$_2$ and thus scales with atmospheric O$_2$ abundance, is an extremely useful biosignature in the context of Earth evolution. Ozone offers strong spectral features across a range of wavelengths, including diagnostic features in the mid-infrared (at 9.6~$\upmu$m), the visible/near-infrared (the Chappuis band between 0.55 and 0.65~$\upmu$m), and in the ultraviolet (the Hartley-Huggins bands centered near 0.26~$\upmu$m). The latter of these is of particular note, as it is sensitive to extremely low peak O$_3$ abundances of $\sim\!\!1$ ppm or less. Although all of these features would be extremely weak at the vanishingly low atmospheric O$_2$/O$_3$ levels characteristic of the Archean, the Hartley-Huggins band would have produced relatively strong, though not saturated, absorption at atmospheric {\it p}O$_2$ values approaching the lowest inferred for the mid-Proterozoic, while absorption at both the Hartley-Huggins and mid-infrared bands would have been relatively strong at the upper end of mid-Proterozoic {\it p}O$_2$ estimates \citep{seguraetal2003,reinhardetal2017,rugheimer&kaltenegger2018,olsonetal2018a}. All of these features would have been relatively strong for the last $\sim\!\!500$ million years of Earth's evolutionary history, and possibly during the early Paleoproterozoic.

Methane (CH$_4$) has a number of spectral features, including many relatively weak features spanning the visible wavelength range, stronger near-infrared features at 1.65, 2.3, and 2.4~$\upmu$m, and a strong mid-infrared band at 7.7~$\upmu$m. The visible wavelength features between 0.6 and 1.0~$\upmu$m only have appreciable depth for atmospheric CH$_4$ abundances above $\sim\!\!10^{-3}$ bar, suggesting that they may have been relatively strong during the Archean and may have been particularly promising biosignatures prior to the evolution of oxygenic photosynthesis \citep{kharechaetal2005,ozakietal2018}. The stronger near-infrared methane features would likely have been prominent for the vast majority of the Archean  \citep{reinhardetal2017}. For most of Earth's history subsequent to the Archean absorption by CH$_4$ in the mid-infrared at 7.7~$\upmu$m would have been relatively strong. Indeed, this feature is apparent even at the very low atmospheric CH$_4$ abundance of the modern Earth \citep{desmaraisetal2002}. However, overlap with a significant water vapor band may render this feature challenging to detect in some cases.

Hydrocarbon hazes --- which can be produced in reducing atmospheres with CH$_4$/CO$_2$ ratios above $\sim\!\!0.1$ --- also produce strong features and could represent a biosignature 'proxy' for biotic CH$_4$ production \citep{arneyetal2016,arneyetal2018}. Indeed, there is isotopic evidence for at least sporadic haze production on the Archean Earth (see Section~\ref{subsec:earthgeoo2ch4}), and photochemical models suggest that the surface CH$_4$ fluxes required to maintain both haze production and the clement climate state implied by Earth's rock record are most consistent with biospheric CH$_4$ production. The most prominent features for haze include a broad ultraviolet/visible absorption feature and a band near 6.5~$\upmu$m in the mid-infrared.  Both may have been relatively strong during the Archean, and the shortwave feature could have caused Earth to present as a ``Pale Orange Dot'' early in its history \citep{arneyetal2016}. However, it is unlikely that Earth's CH$_4$/CO$_2$ ratio has been high enough to produce haze after $\sim\!\!2.5$~Ga.

Nitrous oxide (N$_2$O) is a potential atmospheric biosignature, in addition to being a powerful greenhouse gas and an important component of stratospheric ozone chemistry \citep{prather&hsu2010}.  On the modern Earth, natural (non-anthropogenic) sources of N$_2$O are dominated by microbial activity in terrestrial soils and productive regions of the surface ocean \citep{matson&vitousek1990,hirschetal2006}.  Under low oxygen concentrations, N$_2$O can be produced biologically during the metabolic oxidation of ammonium (NH$_4^+$) and nitrite (NO$_2^-$) and during the reduction of nitrate (NO$_3^-$) during incomplete denitrification \citep{bianchietal2012,freingetal2012}.  The inorganic reaction of nitric oxide (NO) with dissolved ferrous iron (Fe$^{2+}$) can also produce N$_2$O in a process referred to as "chemodenitrification" \citep{wullstein&gilmour1966}.  The source NO for this reaction can be derived from either biological nitrogen fixation or abiotically through the breakdown of atmospheric N$_2$ by lightning.  Still, N$_2$O is considered a putative biosignature because production of N$_2$O from the latter is likely to be relatively small \citep{schumann&huntrieser2007}.  Importantly, the photochemical stability of N$_2$O is a relatively strong function of atmospheric O$_2$ \citep{levineetal1979,kasting&donahue1980} --- for example, sustaining a modern atmospheric N$_2$O abundance at a plausible Proterozoic atmospheric {\it p}O$_2$ of $\sim\!\!1$\% of the present atmospheric level requires a surface N$_2$O flux of roughly 30 times that on the modern Earth for a Sun-like star \citep{robersonetal2011,stantonetal2018}.  This suggests that through Earth's history atmospheric N$_2$O abundance would broadly have tracked the abundance of atmospheric O$_2$ and O$_3$.  Because even modern Earth-like N$_2$O abundances produce relatively weak spectral features \citep[][and Table~\ref{tbl:features}]{schwietermanetal2018}, N$_2$O is often not a primary focus of biosignature detectability studies.

Finally, water vapor (H$_2$O) and carbon dioxide (CO$_2$), while essential to many biological functions, are also critical signposts of a habitable world.  The former will be present in the lower atmosphere of any world with stable surface liquid water, and the latter is key to maintaining surface habitability through its properties as a greenhouse gas and its role in the carbonate-silicate cycle.  The spectrum of modern Earth is strongly sculpted by water vapor absorption features throughout the red-visible, near-infrared, and mid-infrared, and these same features would have produced strong spectral features on Earth during all non-snowball periods of its evolution.  However, even in extremely cold Snowball Earth scenarios, the water vapor bands spanning the near-infrared at at 6.3~$\upmu$m would remain apparent.  Carbon dioxide has several strong features at longer near-infrared wavelengths (most notably at 4.3~$\upmu$m) and in the mid-infrared (at 15~$\upmu$m).  Earlier in Earth's history, where higher levels of CO$_2$ would have been required to maintain habitable surface conditions, carbon dioxide features at shorter near-infrared wavelengths would have appeared much stronger \citep[e.g., bands near 1.6 and 2~$\upmu$m;][]{meadows2008,arneyetal2016,rugheimer&kaltenegger2018}.

%
\bigskip
\section{OBSERVING EARTH FROM AFAR}
\label{sec:earthobs}
\bigskip
%

Observational data for the distant Earth provide a critical opportunity to study the spectral appearance of an ocean-bearing, inhabited world.  As is discussed below, such data enable investigations into the remote characterization of the physical and chemical state of our planet through applications of Solar System planetary exploration techniques.  Zooming even further out, though, observations where Earth is treated as a single pixel --- a Pale Blue Dot --- provide insights into how we will, one day, study photometric and spectroscopic observations of potentially Earth-like exoplanets for signs of habitability and life.

\bigskip
\noindent
\subsection{Earth as a Planet: The {\it Galileo} Experiment}
\label{subsec:earthplanet}
\bigskip

The {\it Pioneer 10}/\emph{11} \citep[launched 1972;][]{bakeretal1975,gehrels1976,ingersolletal1976,klioreetal1976} and {\it Voyager 1}/\emph{2} missions \citep[launched 1977;][]{kohlase&penzo1977,haneletal1977} enabled the initial exploration of key approaches to analyzing spacecraft flyby data for many Solar System worlds.  Instruments and techniques for either acquiring or interpreting spatially-resolved observations of planets and moons using photometry or spectroscopy at wavelengths spanning the ultraviolet through the infrared were among the important developments.  From these observations planetary scientists were able to infer details about atmospheric chemistry and composition, cloud and aerosol formation and distribution, atmospheric thermal structure and circulation, surface chemical and thermal properties (for worlds with thin atmospheres), as well as planetary energy balance.

Flybys of Earth by the {\it Galileo} spacecraft \citep[launched 1989;][]{johnsonetal1992} in December of 1990 and 1992 afforded planetary scientists the first opportunity to analyze our planet using the same tools and techniques that had been (and would be) applied throughout the Solar System.  During the flybys, data were acquired using the Near-Infrared Mapping Spectrometer \citep[NIMS;][]{carlsonetal1992}, the Solid-State Imaging system \citep[SSI;][]{beltonetal1992}, the Ultraviolet Spectrometer \citep[UVS;][]{hordetal1992}, and the Plasma Wave Subsystem \citep[PWS;][]{gurnettetal1992}.  Critically, spatially-resolved imagery from the SSI was provided across eight filters \citep[][their Table II]{beltonetal1992}, and spatially-resolved spectra were acquired by NIMS over 0.7--5.2~$\upmu$m (at a wavelength resolution, $\Delta\lambda$, of 0.025~$\upmu$m longward of 1~$\upmu$m, and 0.0125~$\upmu$m shortward of 1~$\upmu$m).

\begin{figure*}
 \epsscale{2.05}
 \plotone{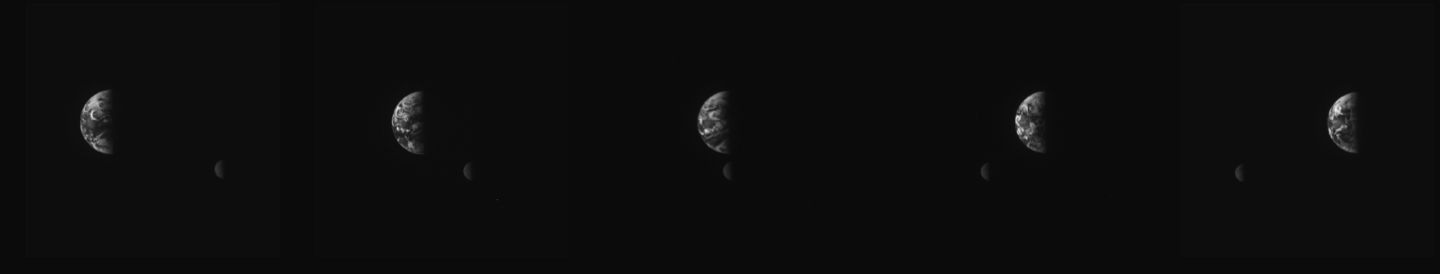}
 \plotone{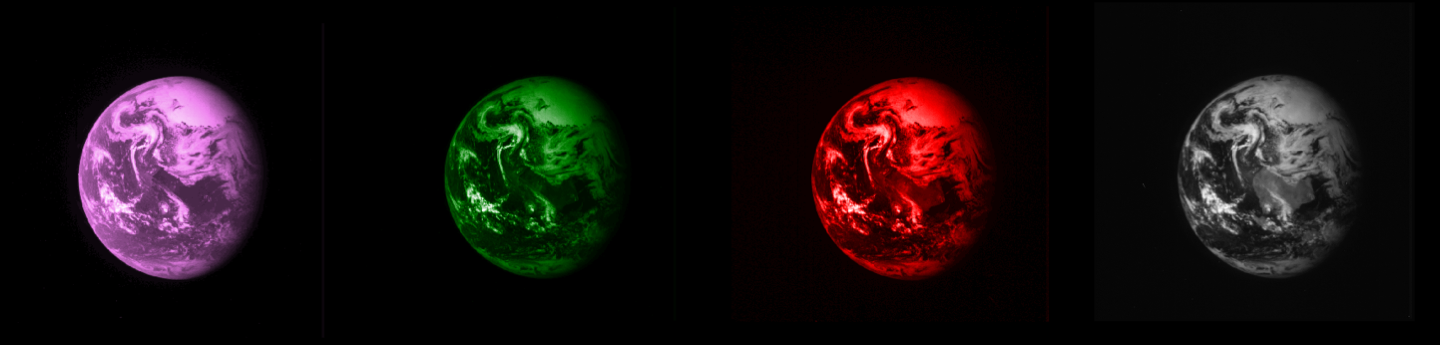}
 \caption{\small Top: Time sequence of Earth and the Moon acquired with the SSI ``infrared'' filter (961--1011~nm) after the second {\it Galileo} flyby (December, 1992).  Bottom: Images of Earth acquired with the SSI from the first {\it Galileo} flyby (December, 1990).  Filters from left to right are violet (382--427~nm), green (527--592~nm), red (641--701~nm), and ``infrared''. The sub-observer point in each image is approximately identical, and Australia is the landmass near the center of the images.  For all images south is oriented ``up,'' as it was in the true flyby geometry.}
 \label{fig:galileo_earth_ssi}
\end{figure*}

\begin{figure*}
 \epsscale{1.0}
 \plotone{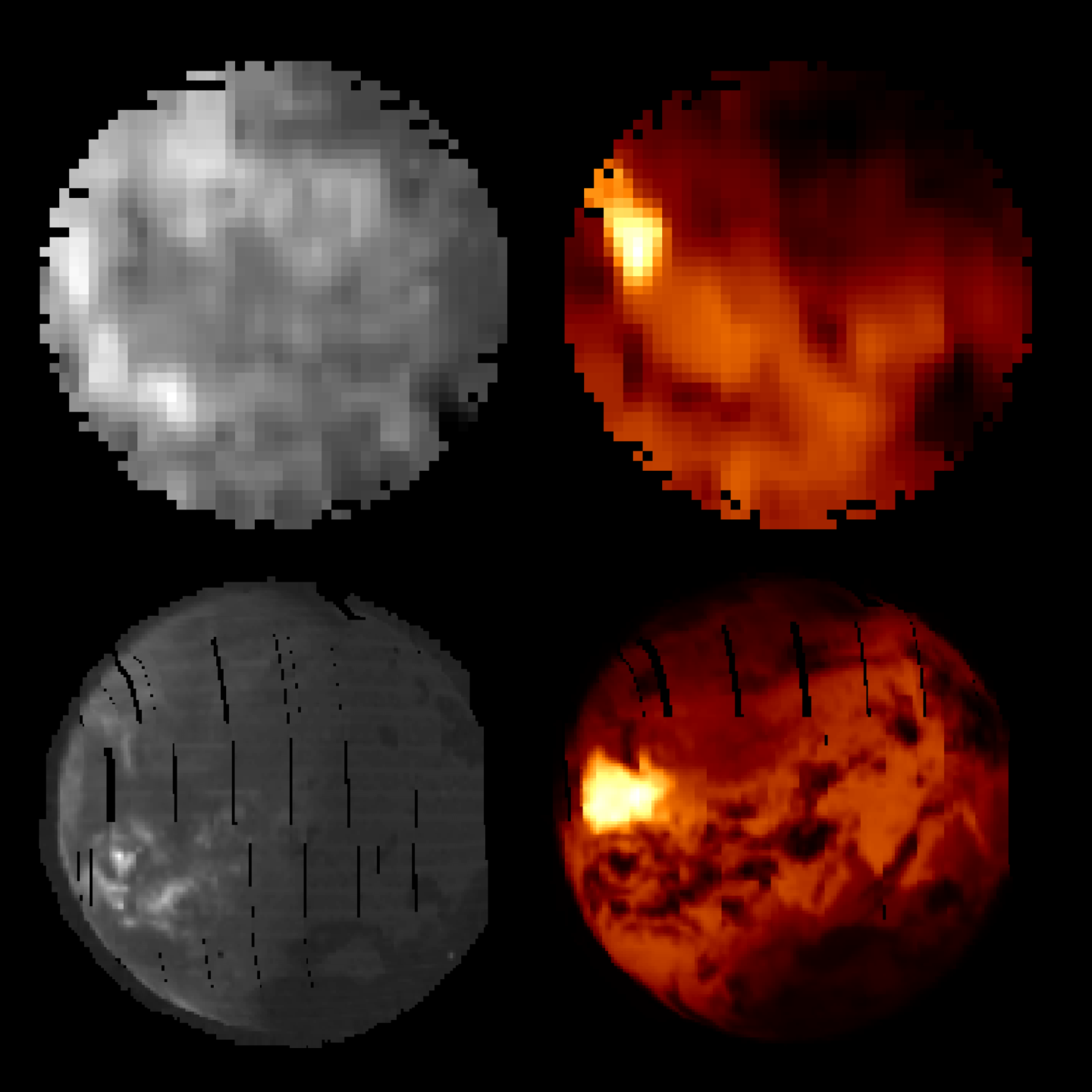}
 \caption{\small Reconstructed images of Earth using NIMS observations from the first (top) and second (bottom) {\it Galileo} Earth flybys.  Images in the left column are at 2.75~$\upmu$m while images in the right column are at 4.0~$\upmu$m.  The bright (warm) source seen in both 4.0~$\upmu$m images is likely Australia.  The spacecraft was nearer to Earth in the second flyby dataset, thereby providing better resolution across the disk.}
 \label{fig:galileo_earth_nims}
\end{figure*}

Figure~\ref{fig:galileo_earth_ssi} shows a sampling of SSI images of Earth from the first Earth flyby and a time sequence of SSI images of Earth and the Moon that were acquired after the second Earth flyby.  Many images from the second Earth flyby suffered from saturation defects.  Reconstructed NIMS images from both Earth flybys are shown in Figure~\ref{fig:galileo_earth_nims}, where certain instrument and mapping defects can be seen.  Here, the 4.0~$\upmu$m images highlight a wavelength range with relatively little atmospheric opacity and where thermal emission dominates.  By contrast, the 2.75~$\upmu$m image (located within a H$_2$O absorption band) contains both reflected and thermal contributions, which, for example, results in reflective clouds only being seen in the sunlit portions of the images.

In a landmark study, \citet{saganetal1993} used the {\it Galileo} flyby data to ascertain key details about the surface and atmospheric state of our planet.  Spectra from the NIMS instrument contain information about surface and atmospheric chemistry, and can also indicate surface thermal conditions at longer wavelengths.  Thus, \citet{saganetal1993} argued that reflective polar caps seen in SSI images were water ice, and that the surface of the planet spanned the freezing point of water (covering at least 240--290~K).  Additionally, the darkest regions in the SSI images showed signs of specular reflection, indicating that these regions were liquid oceans.  Finally, clearsky soundings of H$_2$O absorption features indicated a surface with large relative humidity (i.e., near the condensation point).  Taken altogether, these lines of evidence clearly indicate that the world under investigation is habitable, or capable of maintaining liquid water on its surface.

\begin{figure*}
 \epsscale{1.5}
 \plotone{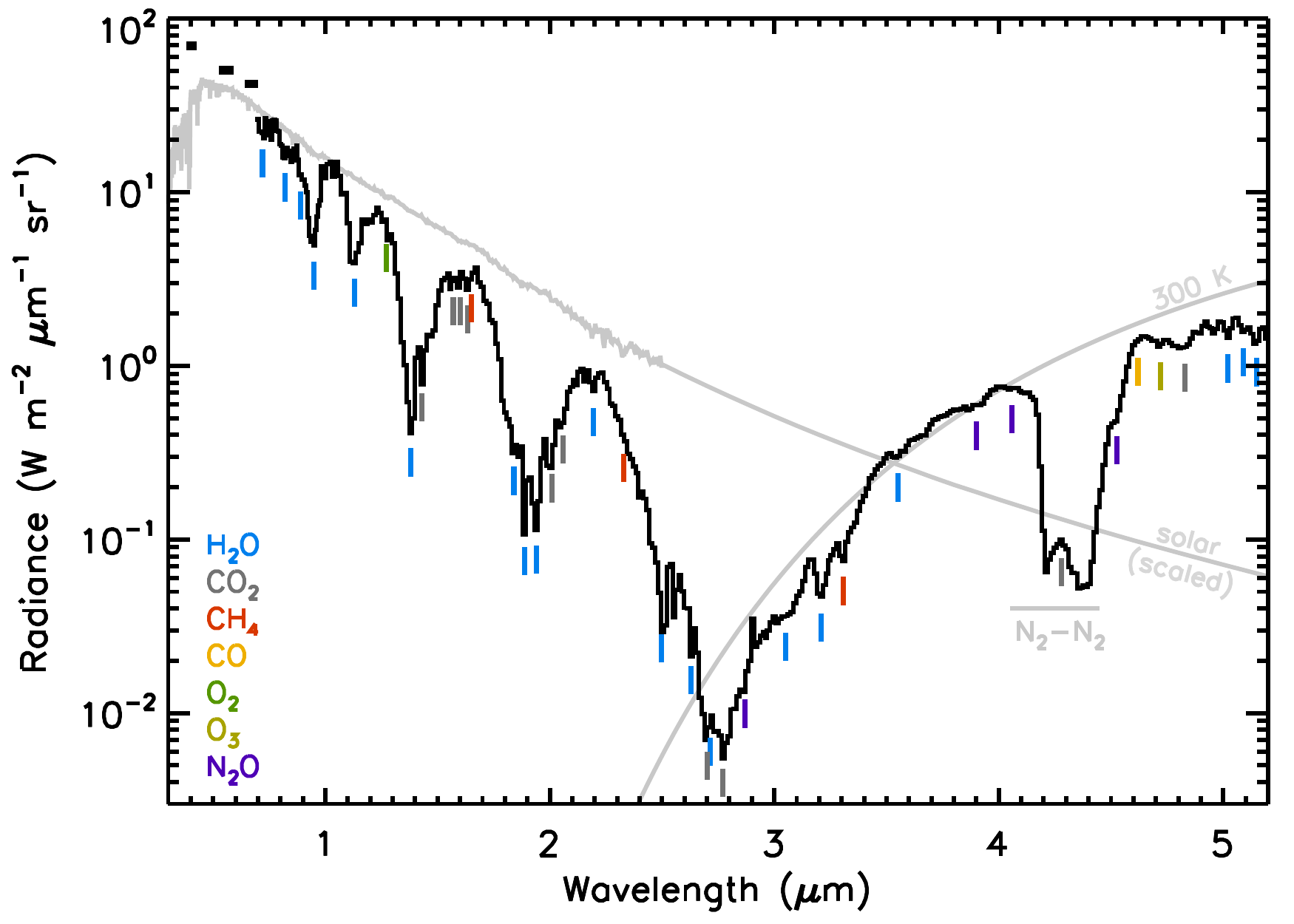}
 \caption{\small Spatially-resolved spectrum of Earth from the second {\it Galileo} flyby, acquired for a small region of the planet in the Indian Ocean and near the terminator at the time of observation.  Visible SSI photometry (in the violet, green, and red filters) and NIMS spectroscopy are both shown, and key absorption bands are indicated. Reflected light contributes significantly below about 3~$\upmu$m, while thermal emission dominates at longer wavelengths.  Light gray curves indicate a scaled solar spectrum and a 300~K blackbody.}
 \label{fig:galileo_resolved_spec}
\end{figure*}

\citet{drossartetal1993} retrieved abundances of key atmospheric constituents (CO$_2$, H$_2$O, CO, O$_3$, CH$_4$, and N$_2$O) via simple parameterized fits to resolved NIMS observations (see Figure~\ref{fig:galileo_resolved_spec}).  Here, model spectra were generated by adopting scaled Earth-like profiles for trace atmospheric species.  Thermal structure profiles were derived using the 4.3~$\upmu$m CO$_2$ band, which rely on this gas having a well-mixed vertical profile (i.e., a near-constant mixing ratio with altitude).  For a well-mixed gas, variation in an infrared absorption band can be attributed to thermal structure rather than variation in abundance with altitude.

Observations from the NIMS instrument also indicated large column densities of O$_2$ in Earth's atmosphere \citep{saganetal1993}.  It was estimated that diffusion-limited escape of hydrogen (produced from H$_2$O photolysis) would require many billions of years to build up atmospheric oxygen to the observed levels, implying an alternative source.  Using the abundances derived in \citet{drossartetal1993}, it was shown that atmospheric CH$_4$ was in a state of extreme disequilibrium with no known geological source that could supply CH$_4$ at the rate required to maintain the observed concentrations.  Certain surface regions of the planet imaged in multiple SSI filters demonstrated a sharp increase in reflectivity at wavelengths beyond 700~nm, known from ground-truth investigations to be the vegetation ``red edge'' (a rapid increase in reflectivity near 0.7~$\upmu$m that is related to pigments).  Thus, multiple lines of evidence point towards the potential for biological activity to be shaping the surface and atmospheric properties of Earth\footnotemark.

\footnotetext{The {\it Galileo} Earth dataset has also been investigated for signs of shadowing caused by trees \citep{doughty&wolf2016}, whose vertical structures create distinct bi-directional reflectance distribution functions that have been proposed as a biosignature for Earth-like exoplanets \citep{doughty&wolf2010}.}

Of course, the strongest evidence for Earth being inhabited came from the PWS dataset \citep{saganetal1993}.  Here, radio emissions from our planet were monitored as a function of time and frequency throughout the {\it Galileo} encounter.  Narrow-band emissions between 4--5~MHz (i.e., in the high frequency portion of the radio spectrum where a variety of radio communications occur), isolated in both frequency and time, were interpreted as radio transmissions from an intelligent species on our planet.  In other words, an observational approach championed in the search for extraterrestrial intelligence (SETI) --- listening at radio frequencies --- yielded the least ambiguous evidence for the inhabitance of Earth from the {\it Galileo} flyby dataset.

In the words of \citet{saganetal1993}, the {\it Galileo} Earth datasets offered a ``unique control experiment on the ability of flyby spacecraft to detect life at various stages of evolutionary development.''  Combining lines of evidence that spanned the ultraviolet, visible, infrared, and radio spectral regimes, the {\it Galileo} observations indicated a habitable planet with a diversity of surface environments and whose atmosphere (and, thus, spectrum) is strongly influenced by life.  In the context of exoplanets, however, the key question becomes: Which habitability and life signatures are lost when Earth is studied not as a resolved source but as a distant, unresolved target?

\bigskip
\noindent
\subsection{Observing the Pale Blue Dot}
\label{subsec:earthdot}
\bigskip

\begin{figure*}
 \epsscale{1.5}
 \plotone{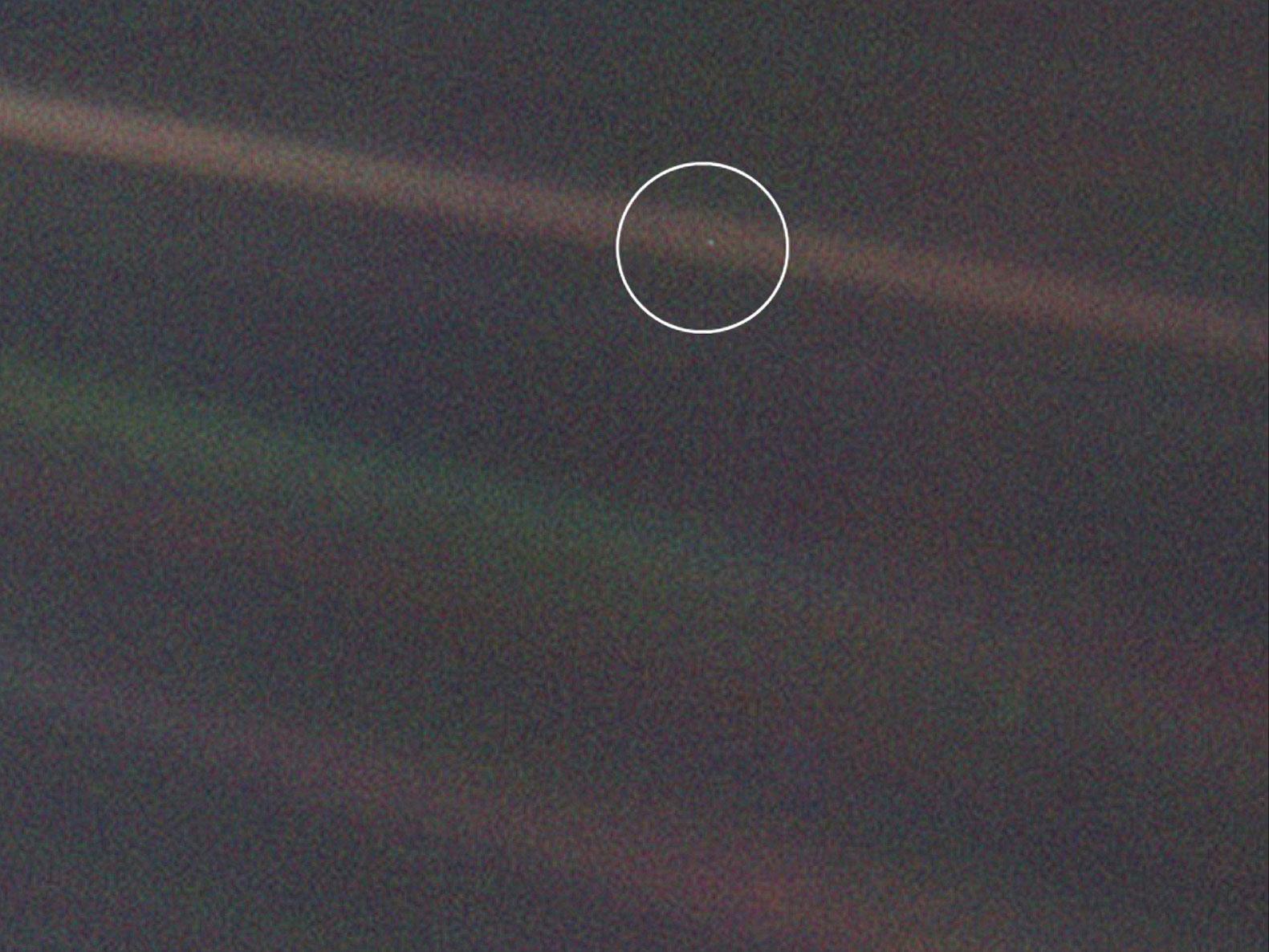}
 \caption{\small The famous ``Pale Blue Dot'' image of Earth, acquired by the {\it Voyager~1} spacecraft from a distance of 40~au.  Image credit: NASA/JPL-Caltech.}
 \label{fig:palebluedot}
\end{figure*}

A more accurate model for future observations of Earth-like exoplanets is not the {\it Galileo} flyby observations, but instead the famous ``Pale Blue Dot'' photograph of Earth taken by the {\it Voyager~1} spacecraft (see Figure~\ref{fig:palebluedot}).  Here --- even for future large telescopes --- observations will not be able to spatially resolve features on the disk of an exoplanet \citep[although time-domain data can be used to obtain some spatial resolution; e.g.,][]{cowanetal09,majeauetal2012}.  The resolution of a telescope is limited by the physics of light diffraction to an angular size of roughly $\lambda/D$, where $\lambda$ is wavelength and $D$ is telescope diameter.  For a 10-meter class telescope observing at visible wavelengths (i.e., near 500~nm), the angular resolution is at best $5\times10^{-8}$ radians (or about 10 milli-arcseconds).  Even for our nearest stellar neighbors (e.g., $\alpha$ Centauri at 1.3~parsecs) this corresponds to a spatial resolution of $2\times10^6$~km, or about three times the radius of our Sun.


Unresolved observations of planets are sometimes referred to as being ``disk-integrated.''  Here, the entire three-dimensional complexity of a planet is effectively collapsed into a single pixel.  For worlds like Earth, this means that cloud-free regions of the planet are blended with cloudy regions, warm equatorial zones are observationally mixed with cold polar caps, and continents become unresolved from oceans.  Additionally, viewing geometry plays an important role, as the portions of the planet near the limb will contribute less overall flux to an observation (since these areas are of smaller solid angular size), and, in reflected light, regions near the day/night terminator will also contribute relatively little flux owing to their lower insolation.

If we wish to extend the analysis techniques applied to the {\it Galileo} Earth flyby observations to an {\it unresolved} Pale Blue Dot, we must look towards either observational datasets for an unresolved Earth, or towards datasets or products that can mimic an unresolved Earth.  There are, in general, three approaches to obtaining (or constructing) such disk-integrated observations of our planet.  First, one can observe light reflected from Earth from the portion of the Moon that is not illuminated by the Sun \citep[i.e., so-called ``Earthshine'' observations;][]{danjon1928,dubois1947,woolfetal02,palleetal03,turnbulletal2006}. Second, one can use higher spatial resolution observations from satellites in Earth orbit to piece together a more integrated view of Earth \citep{heartyetal09,macdonald&cowan2019}.  Finally, spacecraft observations of the distant Earth --- like those acquired by {\it Galileo} --- can be integrated over the planetary disk to yield exoplanet-like datasets.  We discuss each of these approaches below, highlighting the advantages and disadvantages of each technique.

\bigskip
\noindent
\subsubsection{Earthshine}
\label{subsec:earthshine}
\bigskip

Using Earthshine from the dark portion of the Moon --- which is illuminated by Earth but not the Sun --- has a long history of revealing key details about our planet.   In the {\it Dialogue Concerning the Two Chief World Systems}, Galileo used Earthshine to deduce that ``seas would appear darker, and [\ldots] land brighter'' when observed from a distance \citep{galileo1632}.  In the first multi-year Earthshine monitoring experiment, described in \citet{danjon1928} and continued by \citet{dubois1947}, the broadband visual reflectivity of Earth was shown to vary by several tens of percent at a given phase angle (i.e., the planet-star-observer angle), and cloud variability was identified as the likely driver of these variations.

\begin{figure*}
 \epsscale{1.5}
 \plotone{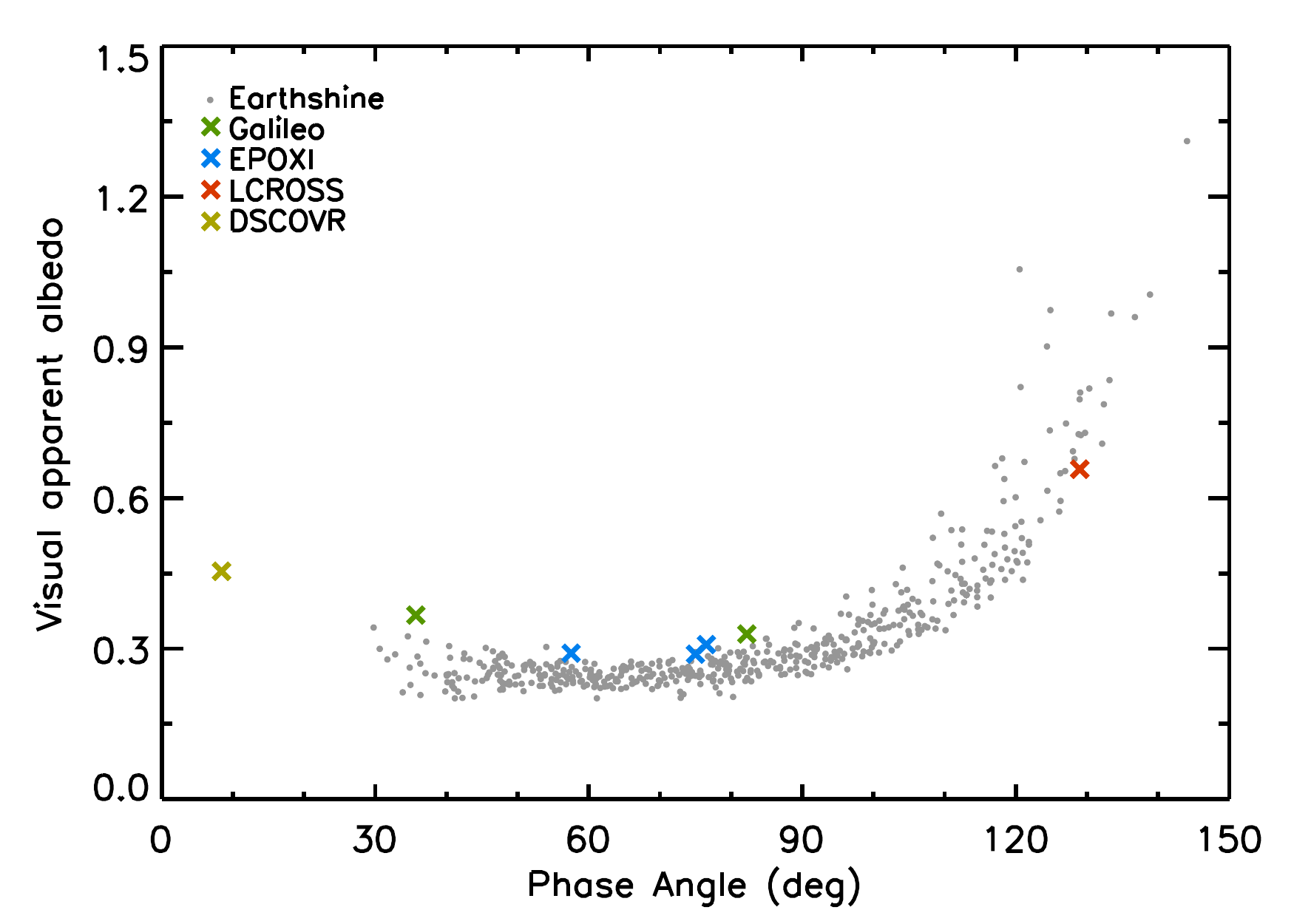}
 \caption{\small Measurements of the phase-dependent visual (400--700~nm) apparent albedo of Earth from Earthshine data spanning several years \citep[from][]{palleetal03}, and from several spacecraft missions.  Apparent albedo values larger than unity indicate stronger directional scattering than can be produced by a sphere whose surface reflects light isotropically (i.e., a Lambert sphere).  The {\it DSCOVR} datapoint is derived from the available four narrowband channels that span the visible range since an integrated 400--700~nm observation cannot be produced from {\it DSCOVR} data.}
 \label{fig:apparentalb}
\end{figure*}

Modern Earthshine observations \citep{goodeetal01,woolfetal02,palleetal2004a} have reached an impressive level of precision.  Achieving this precision requires corrections for airmass effects, the lunar phase function (i.e., how the Moon scatters light into varied directions), and variations in reflectivity across the lunar surface.  Nevertheless, it is now common for Earthshine measurements to achieve 1\% precision on a given night \citep{qiuetal2003}.  Such high-quality photometric observations have revealed variability in the visible reflectivity of Earth at daily, monthly, seasonal, and decadal timescales \citep{goodeetal01,palleetal03,palleetal2004b,palleetal2009a,palleetal2016}.  Figure~\ref{fig:apparentalb} shows a collection of phase-dependent visual (400--700~nm) apparent albedo measurements from Earthshine measurements.  Apparent albedo ($A_{\rm app}$) is defined by normalizing an observed planetary flux to that from a perfectly reflecting Lambert sphere (i.e., a sphere whose surface reflects light equally well into all directions) observed at the same phase angle, or,
\begin{equation}
    A_{\rm app} = \frac{3}{2} \frac{F_{\rm p}}{F_{\rm s}} \frac{\pi}{\sin \alpha + \left( \pi - \alpha \right) \cos \alpha} \ ,
\end{equation}
where $\alpha$ is the star-planet-observer (i.e., phase) angle, $F_{\rm p}$ is the emergent planetary flux at the top of the atmosphere, $F_{\rm s}$ is the solar/stellar flux at normal incidence on the top of the planetary atmosphere, and the flux quantities can either be wavelength-dependent (resulting in a wavelength-dependent apparent albedo) or integrated.  (Note that the factor of $3/2$ comes from the conversion between geometric and spherical albedo.) Thus, a Lambert sphere would have a constant apparent albedo as a function of phase.  Critically, then, the non-constant apparent albedo of Earth demonstrated in Figure~\ref{fig:apparentalb} reveals a weak back scattering peak at small phase angles, a region of Lambert-like scattering at intermediate phase angles, and a strong forward scattering peak at large phase angles.

\begin{figure*}
 \epsscale{1.5}
 \plotone{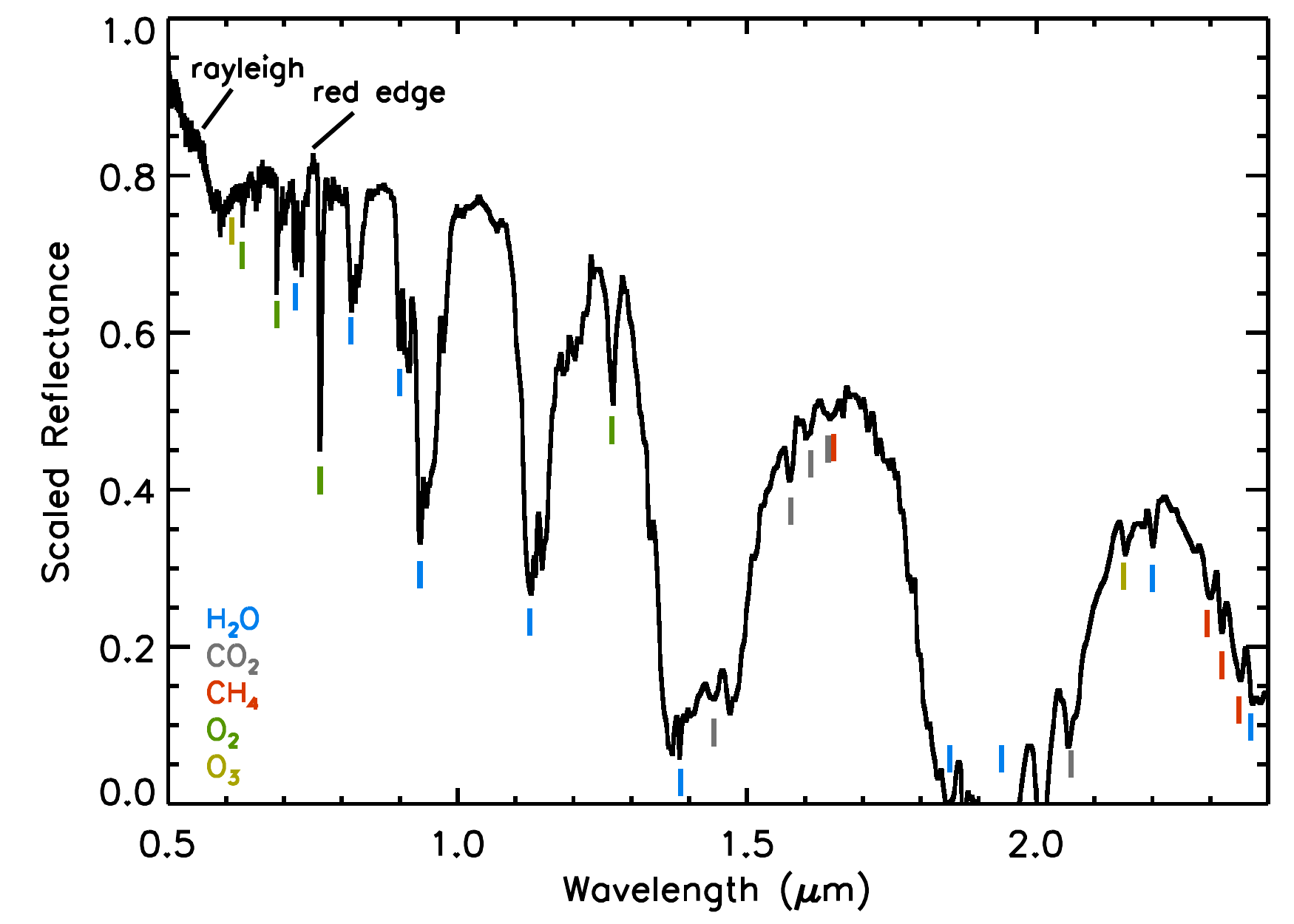}
 \caption{\small Scaled reflectance spectrum of Earth at visible and near-infrared wavelengths measured from Earthshine.  Key absorption and reflection features are indicated.  Data courtesy M.~Turnbull from \citet{turnbulletal2006}.}
 \label{fig:earthshinespec}
\end{figure*}

Spectroscopic studies of Earthshine \citep{woolfetal02} offer additional insights into Earth as an exoplanet (Figure~\ref{fig:earthshinespec}), beyond those obtained through photometric Earthshine investigations.  Using spectroscopic Earthshine data collected over several weeks or months, \citet{arnoldetal02} and \citet{seageretal2005} showed that the aforementioned vegetation red edge signature is variable in the reflectance spectrum of Earth, and can lead to sharp reflectivity increases at the 10\% level in the 600--800~nm range.  A red edge-focused Earthshine study by \citet{montanesrodriguezetal2006} found no strong signature in spectroscopic data from a single night, which highlighted the importance of cloud cover both in setting the overall brightness of Earth and in masking surface reflectance features.  In Earthshine observations that spanned 0.7--2.4~$\upmu$m, \citet{turnbulletal2006} noted a plethora of absorption features that were indicative of life, habitability, and geological activity.   Also, after accounting for how the lunar surface de-polarizes radiation, polarization-sensitive spectroscopic Earthshine observations have explored the degree to which a spectrum of Earth can be polarized (0--20\%, depending on wavelength) as well as the impact of cloud cover on this signature  \citep{sterziketal2012,milespaezetal2014}.  Finally, by investigating the Earthshine spectrum at extremely high spectral resolution ($\lambda/\Delta\lambda$; also referred to as the spectral resolving power) \citet{gonzalezmerinoetal2013} uncovered narrow spectral features due to atomic sodium in Earth's atmosphere that are either of terrestrial or meteoritic origin.

Of course, Earth-like planets around other stars may not be solely investigated using reflected-light techniques \citep[e.g.,][]{snellenetal2013}, especially in the case of potentially habitable worlds orbiting M dwarf hosts where transit or secondary eclipse observations would be the preferred approach.  Impressively, observational techniques developed for Earthshine data collection have been re-purposed to enable observations of the transmission spectrum of Earth's atmosphere.  By observing the Moon during a lunar eclipse, \citet{palleetal2009b} were able to measure light that had been transmitted through our atmosphere and reflected by the lunar surface.  These observations revealed signatures of key atmospheric and biosignature gases, and even included narrow features due to ionized calcium as well as broad pressure-induced features from O$_2$ and N$_2$ (the latter of which is typically difficult to detect due to its general lack of rotational-vibrational features).  A follow-up analysis of these data by \citet{garciamunozetal2012} showed that refractive effects in transit spectra of Earth twins would limit the atmospheric depths probed (during mid-transit) to be above about 10~km, thus providing limited information from the surface and tropospheric environments.  Also, additional high-resolution transmission spectra acquired using Earthshine-related techniques have revealed variability in the depths of H$_2$O absorption features \citep{yanetal2015}, likely tied to the condensible nature of this gas in our atmosphere.

Finally, while Earthshine techniques have been proven to be both powerful and versatile, this approach does have its shortcomings.  First, due to the ground-based nature of the observations, full diurnal cycles in the reflectivity of Earth cannot be observed except during polar night \citep{briotetal2013}.  Second, it is often difficult to calibrate Earthshine observations in a fashion that reveals the absolute brightness of our planet.  Thus, some Earthshine datasets are only reported as a scaled reflectance value, and these products are of lower utility when it comes to exoplanet detectability and characterization studies.  Finally, Earthshine cannot be used to observe thermal emission from Earth since the Moon is also self-luminous at infrared wavelengths.

\bigskip
\noindent
\subsubsection{Orbit}
\label{subsec:orbit}
\bigskip

\begin{figure*}
 \epsscale{1.5}
 \plotone{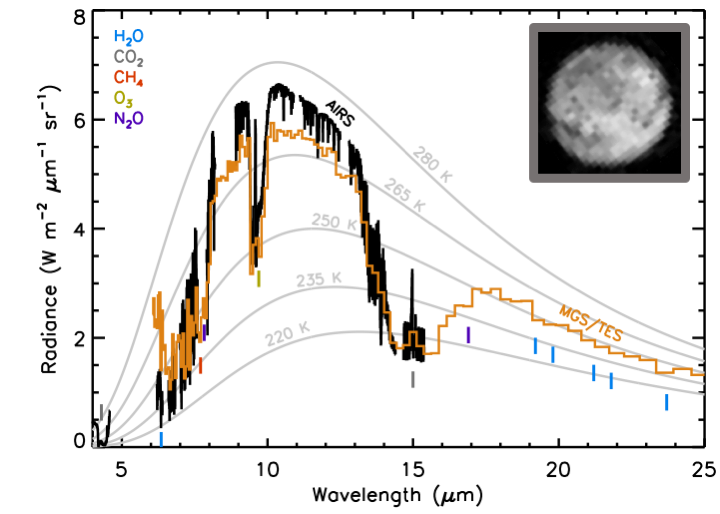}
 \caption{\small Disk-integrated thermal infrared spectra of Earth from the AIRS instrument \citep{heartyetal09} and from the {\it Mars Global Surveyor} Thermal Emission Spectrometer  \citep[{\it MGS}/TES;][]{christensenandpearl97}, where differences are due to the combined effects of seasons, climate, clouds, and viewing geometry. Key features are labeled and blackbody spectra at different emitting temperatures are shown. Inset is a broadband thermal infrared (6--10~$\upmu$m) image of Earth from the {\it LCROSS} mission \citep{robinsonetal2014b}.}
 \label{fig:earthIR}
\end{figure*}

A large suite of satellites are continuously monitoring the Earth system from space.  While most of these Earth-observing satellites only resolve a small patch of our planet in any individual observation, the collective dataset from these satellites benefits from extensive temporal, spatial, and spectral coverage.  Thus ``stitching'' together spatially-resolved radiance measurements from one (or several) observing platform(s) can enable a view of the entire disk of Earth.  This approach was pioneered by \citet{heartyetal09}, who used spatially-resolved thermal radiance observations from the Atmospheric Infrared Sounder (AIRS) instrument \citep[aboard NASA's {\it Aqua} satellite;][]{aumann03} to create disk-integrated infrared spectra of Earth (Figure~\ref{fig:earthIR}).

The practice of stitching together resolved radiance measurements from an Earth-observing satellite is, unfortunately, not straightforward.  Temporal gaps sometimes exist in these datasets where a given latitude/longitude patch of Earth has not been observed in a given 24~hr period.  Thus, if the goal is to produce a snapshot of Earth at a given time, an interpolation of existing radiance observations across time must be performed.  As the Earth climate system (as well as top-of-atmosphere radiances) is non-linear, this interpolation introduces some uncertainties.

The greater challenge to deriving whole-Earth views from resolved satellite observations, though, stems from viewing geometry constraints.  Most Earth-observing satellites are designed to acquire observations in the nadir (i.e., direct downward) direction.  For satellites observing in reflected light, the range of solar incidence angles can also be limited, especially over any given several-day period.  Thus, when stitching together a whole-disk observation, data for certain viewing geometries (e.g., patches located near the limb, where the observing geometry is quite distinct from nadir-looking) may not exist.  In this case, assumptions must be adopted for how radiance will vary with the emission angle and/or the solar incidence angle.  For example, \citet{heartyetal09} adopted a limb darkening law to transform radiances acquired at nadir to radiances appropriate for other emission angles.

An alternative approach to using directly-observed radiances from satellites is, instead, to adopt a collection of satellite-derived ``scene'' models.  These scene models describe the viewing geometry-dependent brightness of different surface categories for Earth (e.g., ocean or desert) under different cloud coverage scenarios.  In other words, these models specify the bi-directional reflectance distribution functions for a large variety of surface type and cloud coverage combinations.  Scene models can be derived from satellite observations \citep{suttlesetal1988} or can be designed to fit satellite observations \citep{manalosmithetal98}.  Combining data that describe the time-dependent distribution of clouds, snow, and ice on Earth with a set of scene models then enables the recreation of whole-disk views of our planet \citep{fordseager&turner01,palleetal03,oakley&cash09}.  Integrating these three-dimensional models over the planetary disk then yields the brightness (or reflectivity) of the Pale Blue Dot.  One key shortcoming of the scene model approach, however, is that such models are rarely spectrally resolved, and instead specify a broadband reflectivity or brightness.  Thus, such models cannot produce spectra of the disk-integrated Earth, and, instead, focus on computing broadband lightcurves for the Pale Blue Dot.

Solar occultation observations acquired from Earth orbit provide a direct measurement of the transmittance along a slant path through the atmosphere.  Initially such datasets \citep{abramsetal1999,bovensmannetal1999,bernathetal2005} provided an excellent model validation for tools designed to simulate transit spectra of Earth-like exoplanets \citep{kaltenegger&traub2009,misraetal2014b}.  However, as was recognized by \citet{robinsonetal2014a} and \citet{dalbaetal2015}, occultation observations from orbit can be directly translated into transit spectra.  Using data from the Canadian Atmospheric Chemistry Experiment - Fourier Transform Spectrometer \citep[ACE-FTS;][]{bernathetal2005}, \citet{schreieretal2018} created transit spectra of Earth spanning 2.2--13.3~$\upmu$m and demonstrated that signatures of chlorofluorocarbons appeared in the occultation-derived transit observations, in addition to more-standard features of H$_2$O, CO$_2$, CH$_4$, N$_2$O, N$_2$, NO$_2$, and O$_2$ \citep[see also,][]{macdonald&cowan2019}.

\bigskip
\noindent
\subsubsection{Spacecraft}
\label{subsec:spacecraft}
\bigskip

The ideal approach for mimicking direct observations of Earth-like exoplanets is, of course, to acquire photometry and/or spectroscopy for a truly distant Earth.  Such observations must be taken from distances beyond low-Earth or geostationary orbit, as the entire disk of the planet is not entirely visible from these vantages (e.g., only about 85\% of the disk is observable from geostationary orbit).  Thus, views of Earth from spacecraft at lunar distances or from Earth-Sun Lagrange points, or observations from interplanetary spacecraft, are all excellent sources.  Until the recent launch of the Deep Space Climate Observatory \citep[DSCOVR;][]{bieseckeretal2015} mission to the Earth-Sun L1 point, no dedicated mission existed for observing Earth from a great distance.  Thus, the majority of the spacecraft observations relevant to Earth as an exoplanet came from missions sent to other Solar System worlds.

While spacecraft observations of the distant Earth are ideal for exoplanet-themed investigations, this approach is not without its shortcomings.  First, it is difficult to find time during the main phase of a mission to dedicate towards observations of non-primary targets such as Earth.  This means that the temporal coverage of spacecraft datasets for the distant Earth is poor, with many of these datesets acquired during the cruise phase of a mission.  Second, and most unfortunately, spacecraft datasets for the distant Earth often remain unpublished.  In these circumstances, the data may have been acquired only for press or outreach purposes, or it might be that analysis and publication of these data are seen as a distraction from the main goals of a mission.  Unpublished datasets are known to exist (both from private communications and press releases) for a number of other missions including: {\it Cassini}, {\it Clementine}, {\it Lunar Reconnaissance Orbiter}, {\it Mars Express}, {\it Mars Reconnaissance Orbiter}, {\it MESSENGER}, {\it OSIRIS-REx}, {\it SELENE/Kaguya}, and {\it Venus Express}.

\begin{figure*}
 \epsscale{1.00}
 \plotone{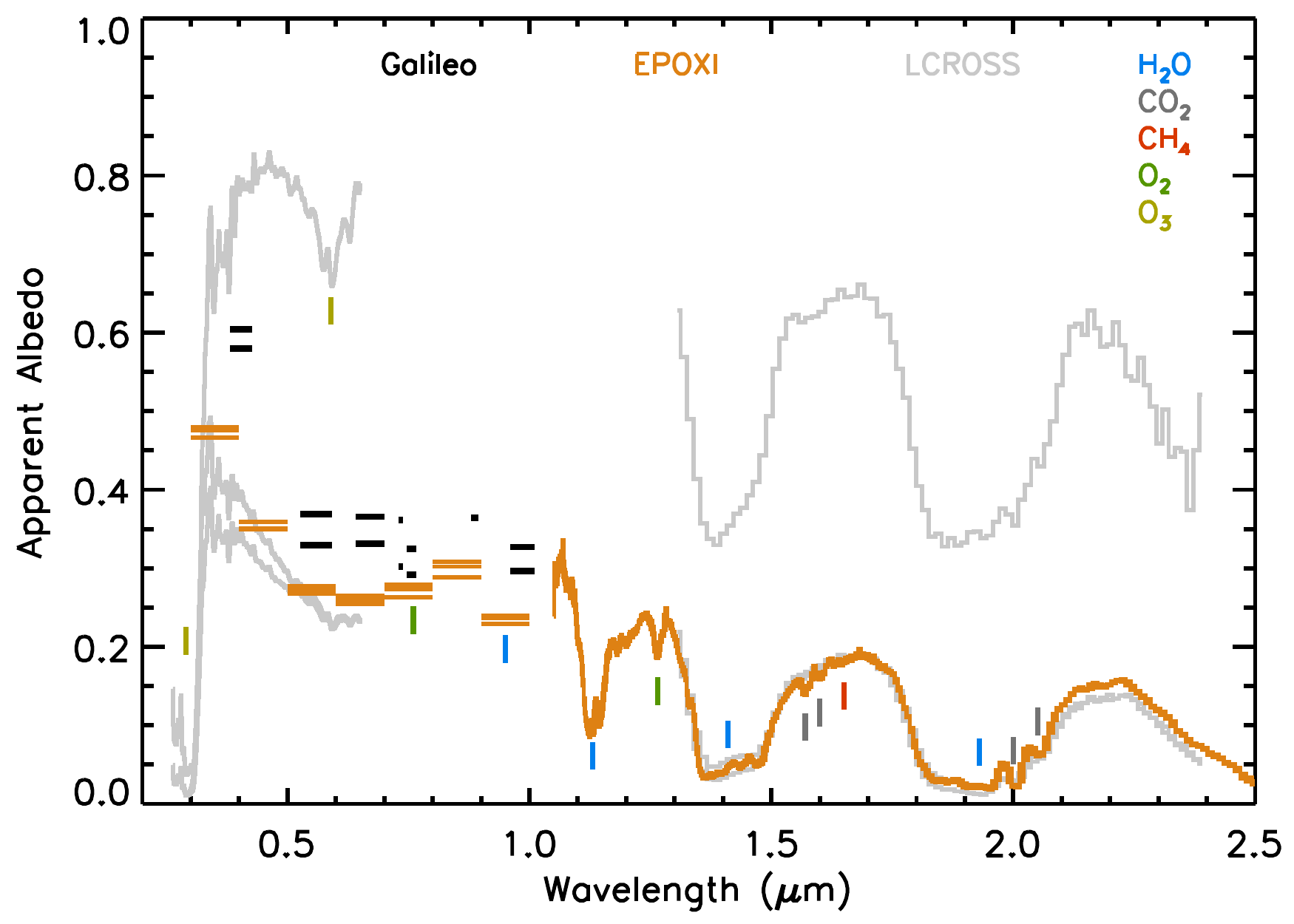}
 \plotone{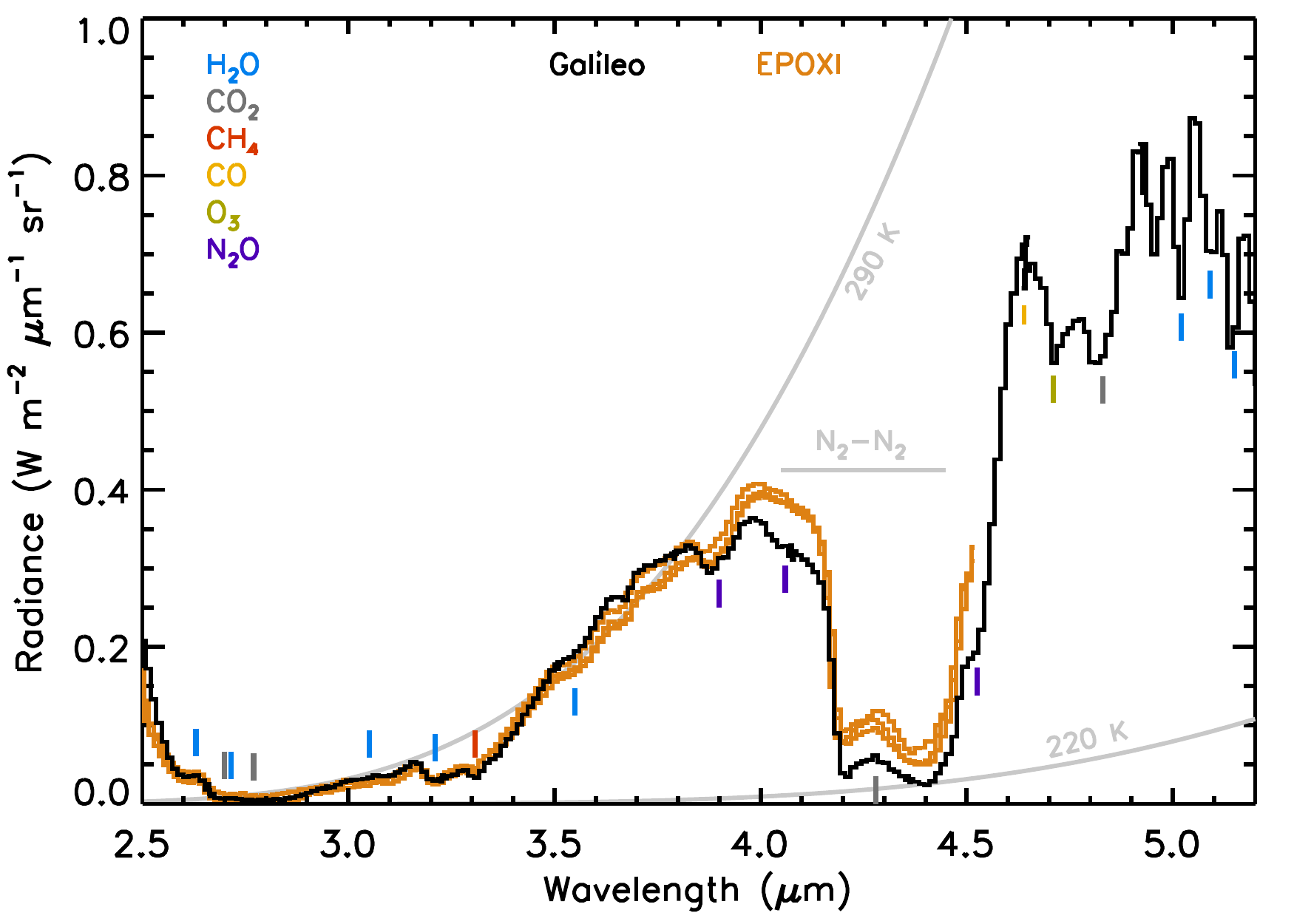}
 \caption{\small Summary of published observations of the distant Earth at ultraviolet, visible, and near-infrared wavelengths.  The first figure presents spectra of Earth's apparent albedo from {\it Galileo} \citep{saganetal1993}, {\it EPOXI} \citep{livengoodetal2011}, and {\it LCROSS} \citep{robinsonetal2014b}.  A crescent-phase observation from {\it LCROSS} is marked by large apparent albedo, which was  driven primarily by forward scattering from a glint spot.  The second figure presents near-infrared emission observations from {\it Galileo} and {\it EPOXI}, and blackbody spectra are provided.  Key absorption features are indicated in both figures.}
 \label{fig:obs_digest}
\end{figure*}

A detailing of published spacecraft-acquired datasets that are relevant to Earth as an exoplanet is shown in Table~\ref{tbl:earthobs}, emphasizing photometric and spectroscopic observations that span the ultraviolet, visible, and infrared wavelengths.  Beyond the previously-discussed {\it Galileo} Earth flyby observations, key datasets also come from a snapshot thermal infrared spectrum acquired by the {\it Mars Global Surveyor} Thermal Emission Spectrometer ({\it MGS}/TES), visible photometry and near-infrared spectroscopy spanning 24~hr on five separate dates from the {\it EPOXI} mission (which re-purposed the {\it Deep Impact} flyby spacecraft), visible spectroscopy and infrared photometry and spectroscopy taken over brief intervals on three separate dates by the {\it Lunar CRater Observation and Sensing Satellite} ({\it LCROSS}), and the aforementioned {\it DSCOVR} data (which include images taken in 10 narrowband channels spanning ultraviolet and visible wavelengths, and bolometric measurements in several channels spanning 0.2--100~$\upmu$m).  The EPOXI dataset has been used to analyze key spectral features for Earth in the near-infrared range and to quantify the vegetation red edge signature in disk-integrated observations \citep{livengoodetal2011,robinsonetal2011}, and to investigate mapping techniques for unresolved objects \citep{cowanetal09,fujiietal11,cowanetal11}.  In \citet{robinsonetal2014b}, the {\it LCROSS} Earth observations were used to quantify the impact of ocean glint and ozone absorption on phase-dependent disk-integrated visible spectroscopic data for the Pale Blue Dot.  A digest of ultraviolet, visible, and near-infrared observations is shown in Figure~\ref{fig:obs_digest}.

Finally, while existing datasets have made many valuable contributions to our understanding of the appearance of the Pale Blue Dot, major gaps still exist in our observational coverage.  Regarding Table~\ref{tbl:earthobs}, it is obvious that spacecraft observations of Earth in reflected light at crescent phases (i.e., phase angles from roughly 145--180$^{\circ}$) are lacking --- only a single dataset, from the {\it LCROSS} mission \citep{robinsonetal2014b}, exists for all phase angles beyond quadrature (which occurs at a phase angle of 90$^{\circ}$, where the planet is half illuminated).  Thermal infrared observations at moderate to high spectral resolution are also not represented.  Additionally, no datasets span a continuous timeframe of longer than roughly 24~hr, which hinders studies of rotational variability.  Lastly, excepting the few {\it LCROSS} pointings \citep{robinsonetal2014b}, existing visible-wavelength datasets only contain photometry, so spectroscopy below about 1~$\upmu$m is not well represented.

\begin{deluxetable}{lcccll}
\tabletypesize{\small}
\tablecaption{Published Spacecraft Datasets for Earth as an Exoplanet \label{tbl:earthobs}}
\tablewidth{0pt}
\tablehead{Spacecraft & Date\tablenotemark{a} & Phase Angle(s) &     Wavelength          & Resolution\tablenotemark{b} & Source(s) \\
                      &                       &                &     ($\upmu$m)          &                             &    }
\startdata
              & 1990-12-10 & 35$^\circ$    &              & $\Delta\lambda=0.01$--0.44~$\upmu$m (vis) & \\
{\it Galileo} & 1992-12-09 & 82$^\circ$    &  0.38--5.2   & & \citet{saganetal1993,drossartetal1993} \\
              & 1992-12-16 & 89$^\circ$    &              & $\Delta\lambda=0.025$~$\upmu$m (NIR)      & \\
              &            &               &              &--- & \\
{\it MGS}/TES & 1996-11-23 & n/a           &    6--50     & $\lambda/\Delta\lambda=15$--170           & \citet{christensenandpearl97} \\
              &            &               &              &--- & \\
              & 2008-03-18 & 58$^\circ$    &              & & \\
              & 2008-05-28 & 75$^\circ$    &              & $\Delta\lambda\approx 0.1$~$\upmu$m (vis) & \citet{livengoodetal2011}; \\
{\it EPOXI}   & 2008-06-04 & 77$^\circ$    & 0.37--4.54   & & \citet{cowanetal11,fujiietal11}; \\
              & 2009-03-27 & 87$^\circ$    &              & $\lambda/\Delta\lambda=215$--730 (NIR)    & \citet{robinsonetal2011} \\
              & 2009-10-04 & 86$^\circ$    &              & & \\
              &            &               &              &--- & \\
              & 2009-08-01 & 23$^\circ$    &              & $\lambda/\Delta\lambda=300$--800 (vis)    & \\
{\it LCROSS}  & 2009-08-17 & 129$^\circ$   & 0.26--13.5   & $\Delta\lambda=0.3$, 0.8~$\upmu$m (NIR)   & \citet{robinsonetal2014b} \\
              & 2009-09-18 & 75$^\circ$    &              & $\Delta\lambda=4$, 7.5~$\upmu$m (thermal) & \\
              &            &               &              &--- & \\
{\it DSCOVR}  & ongoing    & 4--12$^\circ$ & 0.318--0.780 & $\Delta\lambda=1$--3~nm                   & \citet{bieseckeretal2015,yangetal2018} \\
\enddata
\tablenotetext{a}{Based on UT at start of observations.}
\tablenotetext{b}{Abbreviating visible range ($\sim\!0.4$--1~$\upmu$m) as ``vis'' and near-infrared range ($\sim\!1$--5~$\upmu$m) as ``NIR.''}
\end{deluxetable}

%
\bigskip
\section{MODELING THE PALE BLUE DOT}
\label{sec:earthmodels}
\bigskip
%

Techniques for simulating observations of the distant Earth provide a complementary approach to spacecraft, orbital, and Earthshine observations.  Especially once validated against observational datasets, models of the disk-integrated Earth enable the exploration of the Pale Blue Dot across a wide range of wavelengths and spectral resolutions, and can also fill in the various gaps that exist between different observational approaches.  Currently, a hierarchy of Earth models exists, spanning simple reflectance tools to complex three-dimensional models whose outputs cover the ultraviolet through the far-infrared.

\bigskip
\noindent
\subsection{One-Dimensional Approaches}
\label{subsec:1dmodels}
\bigskip

One-dimensional models of the Pale Blue Dot capture the vertical structure of Earth's atmosphere, but omit any latitudinal or longitudinal structure in the atmosphere and surface.  Such simplifications enable these one-dimensional approaches to be computationally efficient, and often allow for higher spectral resolution in model outputs.  Nevertheless, key details about the fractional distribution of clouds and various surface types on Earth must be accounted for, either through data-informed weighting factors or through tuning parameters.

\citet{traub&jucks2002} presented one of the earliest models of the Pale Blue Dot.  This one-dimensional tool spanned the ultraviolet through thermal infrared, and included absorption and emission from key atmospheric species.  Radiation multiple scattering was neglected, and modeled observations in reflected light were generated by linearly combining spectral components (including Rayleigh, clear sky, high cloud, and others).  At visible wavelengths, disk-integrated observations were simulated using a single solar zenith angle (i.e., the Sun was placed at a zenith angle of 60$^\circ$ over a plane-parallel atmosphere).  Both \citet{woolfetal02} and \citet{turnbulletal2006} used the \citet{traub&jucks2002} model to analyze Earthshine spectra.  By fitting the reflected-light spectral components in the \citet{traub&jucks2002} model to the Earthshine data, these authors determined that the most important aspects of their reflected-light observations were a clearsky component, a gray high cloud continuum, and Rayleigh scattering.  More recently, the \citet{traub&jucks2002} model has been used to study the spectral evolution of Earth through time \citep{kalteneggeretal2007,rugheimer&kaltenegger2018}, including a comparison to the previously mentioned {\it EPOXI} dataset \citep{rugheimeretal2013}.

A multiple-scattering one-dimensional model, developed by \citet{martintorres2003}, was adopted by \citet{montanesrodriguezetal2006} to help understand the signature of the vegetation red edge in Earthshine spectra.  In this work, the good match between the Earthshine data and the simulations was attributed to the scattering treatment within the model.  Also, the \citet{montanesrodriguezetal2006} study developed a sophisticated approach to capturing the latitudinal and longitudinal distribution of clouds and surface types on Earth.  Specifically, disk-averaged cloud and surface coverage maps were derived from Earth science data products, including appropriate weighting factors for the solar and lunar geometry.

\bigskip
\noindent
\subsection{Three-Dimensional Models}
\label{subsec:3dmodels}
\bigskip

\begin{figure*}
 \epsscale{1.25}
 \plotone{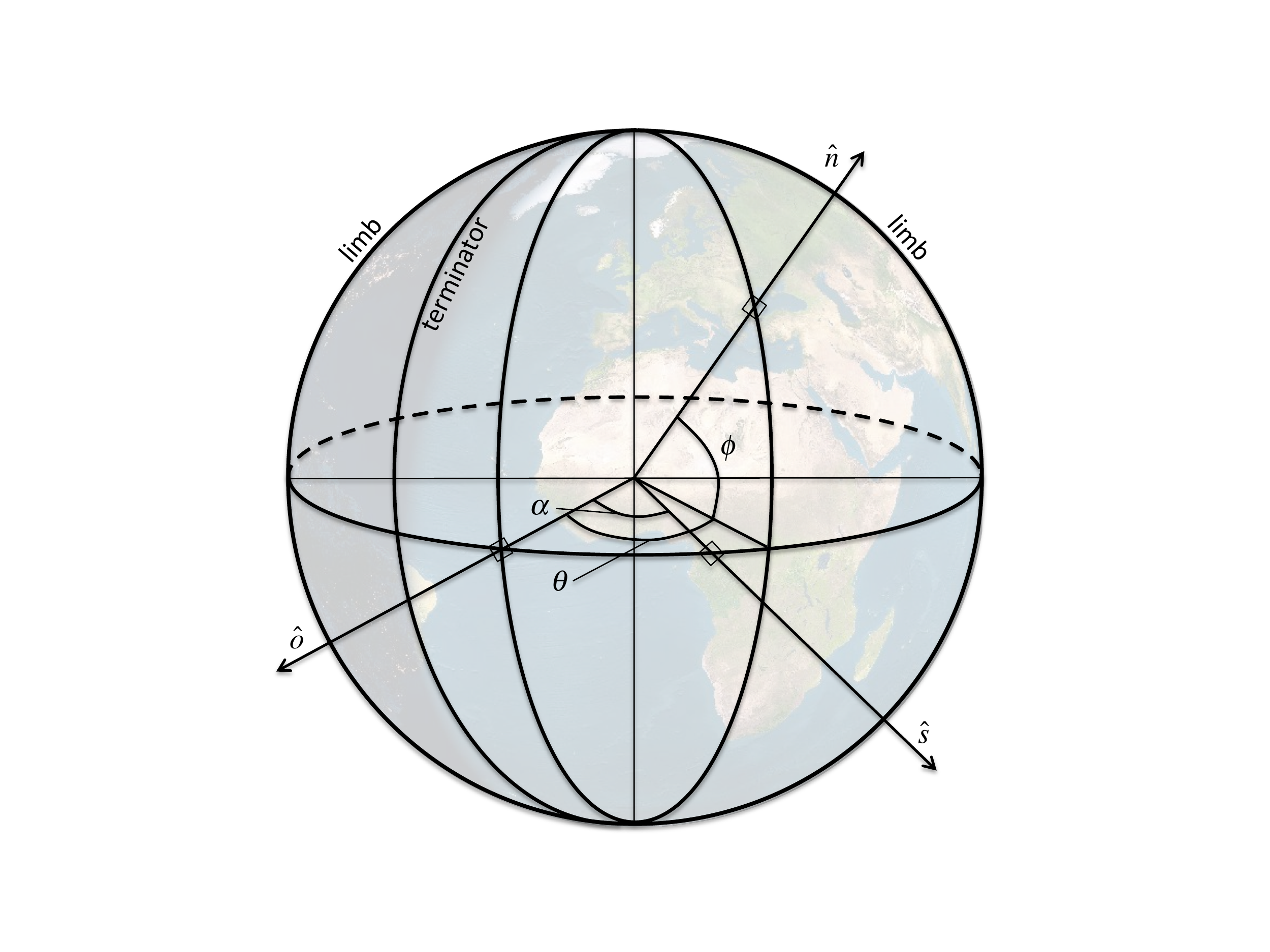}
 \caption{\small Geometry for modeling disk-integrated Earth observations.  The surface normal vector, and the vectors in the direction of the observer and Sun are ${\bf \hat{n}}$, ${\bf \hat{o}}$, and ${\bf \hat{s}}$, respectively.  The angle $\alpha$ is the phase angle, while $\phi$ and $\theta$ are the coordinates of latitude and longitude, respectively.  Earth view generated by the Earth and Moon Viewer, first implemented by J. Walker (http://www.fourmilab.ch/cgi-bin/Earth).}
 \label{fig:geometry}
\end{figure*}

In general, three-dimensional models of the Pale Blue Dot compute the spatially-resolved radiance over the planetary disk, and then integrate this radiance over solid angle to produce a disk-integrated quantity.  More formally, three-dimensional models of Earth aim to compute the integral of the projected area weighted intensity in the direction of an observer, which is written as,
\begin{equation}
F_{\lambda}\left({\bf \hat{o}},{\bf \hat{s}}\right) = \frac{R_{\rm{E}}^{2}}{d^{2}}\int_{2\pi} I_{\lambda}\!\left({\bf \hat{n}},{\bf \hat{o}},{\bf \hat{s}}\right) \left({\bf \hat{n}} \cdot {\bf \hat{o}}\right){\rm d}\omega \ ,
\label{eqn:fluxint}
\end{equation}
where $F_{\lambda}$ is the disk-integrated specific flux density received from a world of radius $R_{\rm E}$ at a distance $d$ from the observer, $I_{\lambda}\!\left({\bf \hat{n}},{\bf \hat{o}},{\bf \hat{s}}\right)$ is the location-dependent specific intensity in the direction of the observer, ${\rm d}\omega$ is an infinitesimally small unit of solid angle on the globe, ${\bf \hat{n}}$ is a surface normal unit vector for the portion of the surface corresponding to ${\rm d}\omega$, and ${\bf \hat{o}}$ and ${\bf \hat{s}}$ are unit vectors in the direction of the observer and the Sun, respectively (see Figure~\ref{fig:geometry}).  The integral in Equation~\ref{eqn:fluxint} is over the entire observable hemisphere ($2\pi$ steradians) and the dot product at the end of the expression ensures that an element of area $R_{\rm{E}}^{2}{\rm d}\omega$ near the limb is weighted less than an element of equal size near the sub-observer point.  Note that, for reflected light, $I_{\lambda}$ will be zero at locations on the night side of the world (i.e., where ${\bf \hat{n}} \cdot {\bf \hat{s}} < 0$), but is non-zero at all locations when considering thermal emission.

The most straightforward (quasi) three-dimensional models use empirical bi-directional reflectance distribution functions \citep[e.g., the previously-mentioned scene models from ][]{manalosmithetal98} to specify the reflectivity of a patch on the disk as a function of viewing geometry.  These three-dimensional tools can either be spectrally-resolved \citep{fordseager&turner01} or broadband \citep{mccullough2006,palleetal03,palleetal08,williams&gaidos2008,oakley&cash09}.  Atmospheric effects (e.g., gas absorption and scattering) are typically omitted, although \citet{fujiietal10} produced a three-dimensional reflectance model that blended wavelength-dependent bi-directional reflectance distribution functions from a variety of sources and also included an additive atmospheric Rayleigh scattering term.  Time-dependent distributions of clouds and surface types are derived from Earth science datasets, such as the International Satellite Cloud  Climatology Project \citep[ISCCP;][]{schiffer&rossow1983}.

\begin{figure*}
 \epsscale{1.5}
 \plotone{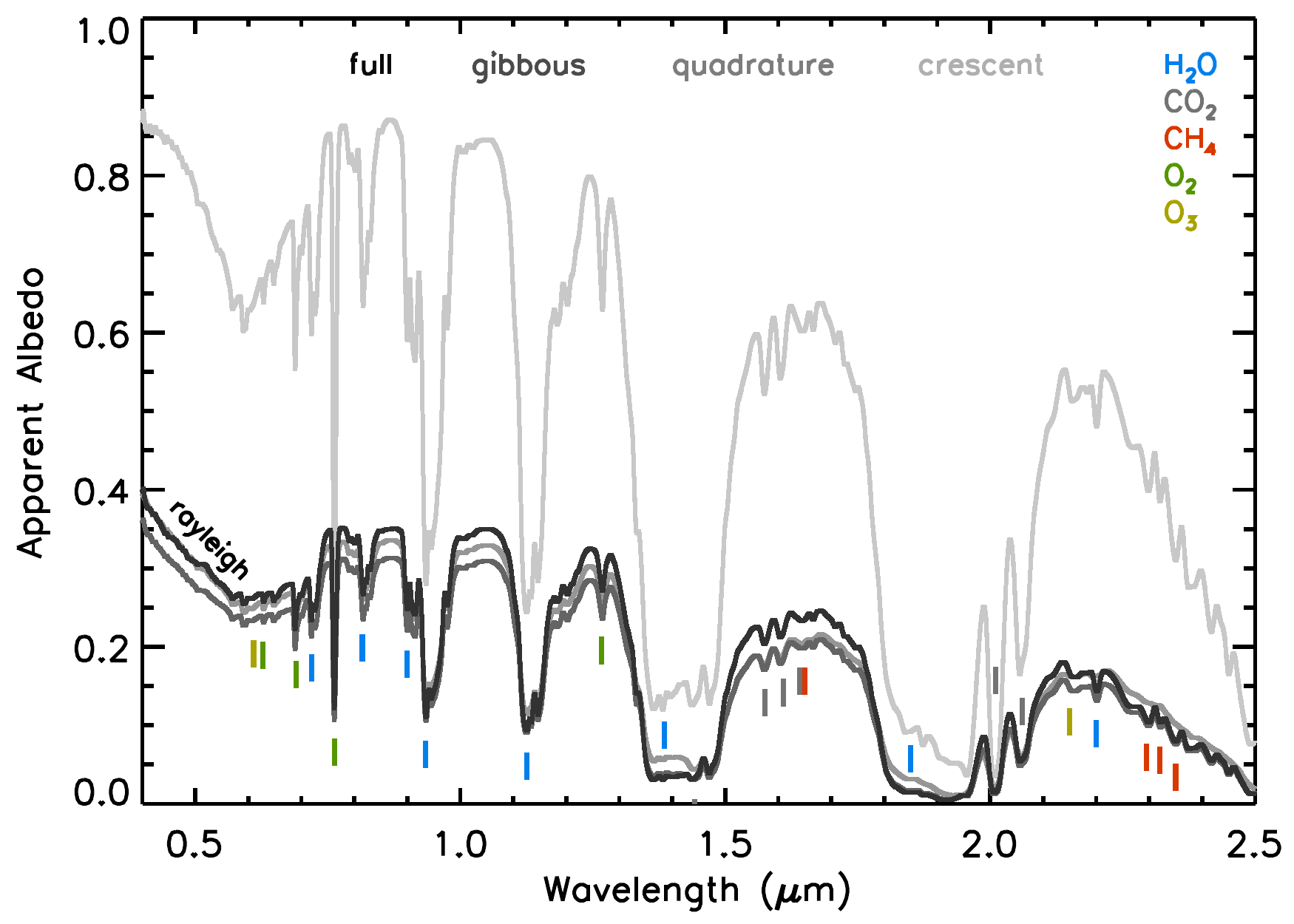}
 \caption{\small Simulations of Earth's phase- and wavelength-dependent apparent albedo \citep[from][]{robinsonetal2010}.  Models are averaged over a full rotation at each phase, and the angles of the given phases are 0$^{\circ}$ (full), 45$^{\circ}$ (gibbous), 90$^{\circ}$ (quadrature), and 135$^{\circ}$ (crescent).  Large apparent albedos at crescent phase are primarily due to ocean glint and cloud forward scattering.  The similarity in apparent albedo scales for the quadrature, gibbous, and full spectra indicate that Earth largely scatters like a Lambert sphere across these phase angles.  A slight enhancement in apparent albedo at full phase is due to cloud back scattering.}
 \label{fig:phases}
\end{figure*}

The most complex three-dimensional tools for simulating observations of the distant Earth solve the full plane-parallel, multiple-scattering radiative transfer equation to determine the emergent radiance over the planetary disk.  By including realistic atmospheric radiative effects, these fully multiple-scattering tools can more self-consistently capture gas and cloud absorption and scattering \citep{tinettietal06a,fujiietal11,robinsonetal2011,fengetal2018} as well as polarization effects \citep{stam2008}.  Like the previously-discussed reflectance models, cloud and surface type coverages are typically derived from Earth science datasets, while cloud optical thicknesses must also be adopted from similar datasets to include in the multiple-scattering calculation.  Such sophisticated three-dimensional models can serve as virtual ``laboratories'' for studying the Pale Blue Dot across a wide range of timescales, wavelengths, and viewing geometries (see Figure~\ref{fig:phases}), insofar as they are validated against observations (that, admittedly, do not span all possible combinations of planetary phase, wavelength coverage, and spectral resolution; see Section~\ref{sec:earthobs}).  In any case, our observational analysis and models of the Pale Blue Dot must also be extended to confront the realization that Earth's major characteristics and remotely observable properties have changed considerably throughout the course of planetary evolution.

%
\bigskip
\section{DECIPHERING EXO-EARTH OBSERVATIONS}
\label{sec:infocontent}
\bigskip
%

The observations and models discussed in previous sections provide insights into remote sensing approaches to understanding distant habitable worlds.  At wavelengths spanning the ultraviolet through the infrared, and for both broadband photometry and spectroscopy across a range of resolutions, data (or simulated data) for the distant Earth --- at any of its evolutionary stages --- contain a great deal of information about the planetary environment.  The sections below discuss the information content of observations of reflected light and of thermal emission.  Additional information can be found in reviews by \citet{meadows2008}, \citet{kalteneggeretal2010}, \citet{kalteneggeretal2012}, and \citet{robinson2018}.

\bigskip
\noindent
\subsection{Visible Photometry}
\label{subsec:photom}
\bigskip

Single-instance broadband photometry of a distant Earth-like world provides limited information about the planetary environment.  As Earthshine observations have shown \citep{qiuetal2003,palleetal03}, photometric data could constrain planetary albedo --- which is central to an understanding of planetary energy balance --- as long as planetary size and phase are known.  (If the planetary radius is unknown, the planetary reflectivity and size are degenerate, although see Section~\ref{subsec:ir} for a discussion of using thermal infrared observations to constrain the planetary radius.)  Additionally, broad absorption features can be detected using photometric observations \citep[e.g., as was the case for the 950~nm H$_2$O band in {\it EPOXI} observations;][]{livengoodetal2011}, although constraining atmospheric abundances from low-resolution observations is extremely challenging \citep{lupuetal2016}.  Planetary color derived from broadband observations has been suggested as a means of identifying exo-Earth candidates \citep{traub2003}, and, regarding Figure~\ref{fig:earthevol}, visible photometry could differentiate a hazy Archean Earth from a non-hazy Earth at all other evolutionary stages, even at low signal-to-noise.  Distinguishing our planet from certain non-Earth-like planets might be more problematic for Earth at any geological phase other than the Phanerozoic \citep{krissansentottonetal2016b} and could be confused by planetary phase effects as well as our lack of knowledge on realistic colors of temperate or cool exoplanets.

\begin{figure*}
 \epsscale{1.5}
 \plotone{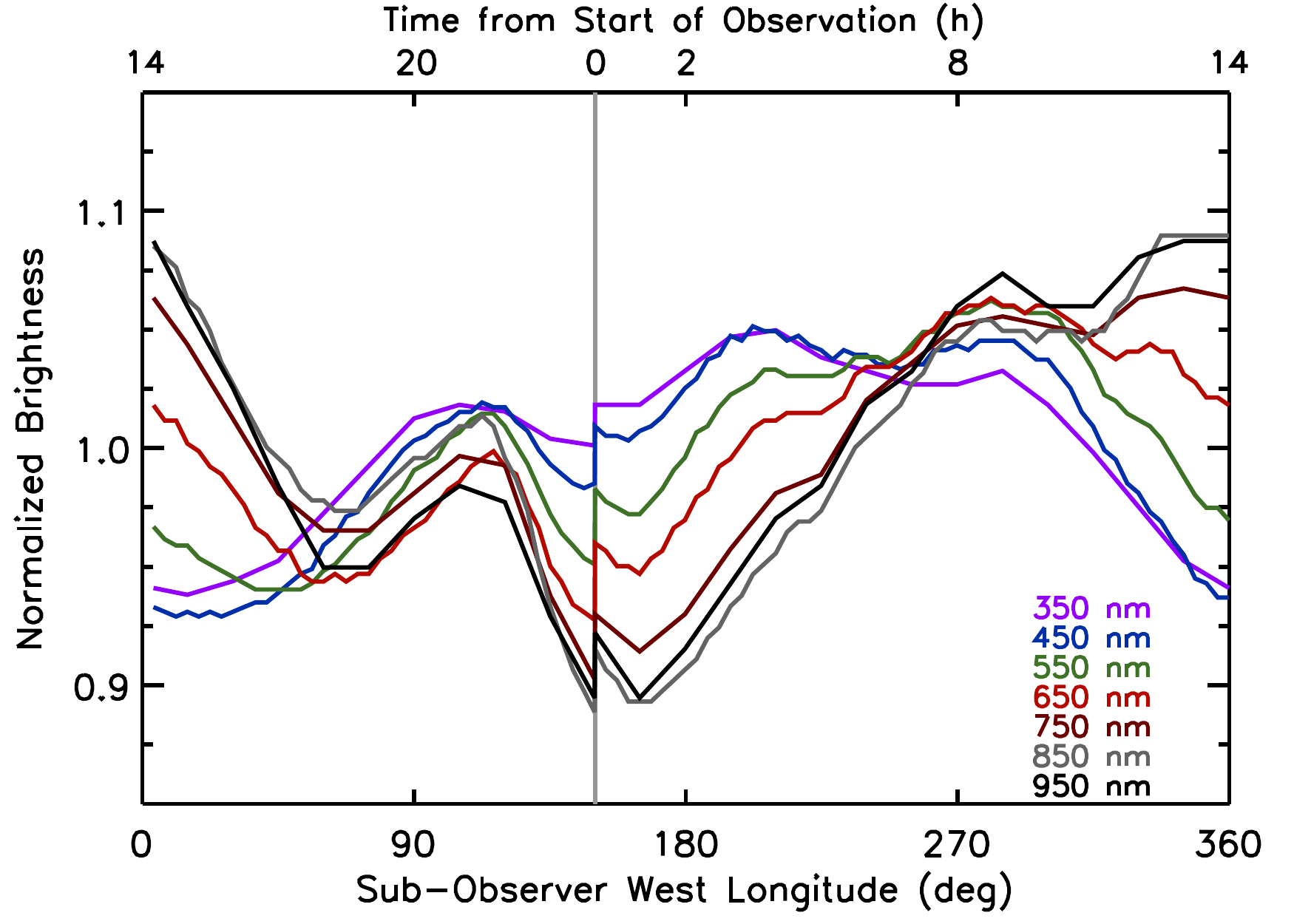}
 \caption{\small Full-rotation lightcurves for Earth from the March 2008 {\it EPOXI} dataset in all shortwave filters.  Filter bandpasses are 100~nm wide, and filter center wavelengths are indicated.  Overall brightness is driven largely by clouds at shorter wavelengths and by both clouds and continents at longer wavelengths \citep{robinsonetal2011}.  Variability is primarily due to Earth's rotation, however differences in brightness after a full rotation are due to longer-term evolution of cloud patterns.}
 \label{fig:lightcurves}
\end{figure*}

Disk-integrated photometric observations that resolve the rotation of an exo-Earth yield much more powerful diagnostics than single-instance photometry (Figure~\ref{fig:lightcurves}).  When acquired over multiple days (rotations), the rotation rate of the planet can be determined from diurnal variability \citep[which is typically 10--20\%;][]{fordseager&turner01,livengoodetal2011}, even in the presence of evolving weather patterns \citep{palleetal08,oakley&cash09}.  Once the rotation rate of an exo-Earth is known, the correspondence between time and sub-observer longitude enables longitudinally-resolved mapping \citep{cowanetal09,fujiietal10,kawahara&fujii10,fujiietal11,fujii&kawahara2012,cowan&strait2013,lustigyaegeretal2018}, although degeneracies can occur in mapping approaches, especially with regard to spectral unmixing \citep{fujiietal2017}.  Additionally, studying photometric variability inside absorption bands of well-mixed gases (e.g., O$_2$) as compared to variability inside bands of other species (e.g., H$_2$O) can reveal condensation processes in planetary atmospheres \citep{fujiietal2013}.  Depending on the optical thickness of a potential haze in the atmosphere of the Archean Earth \citep{arneyetal2016}, it might be necessary to push photometric observations to red or near-infrared wavelengths to have surface and near-surface sensitivity in lightcurves.

Photometric exo-Earth observations resolved at both rotational and orbital timescales could reveal additional information about the planetary surface.  Due to the obliquity of the planetary rotational axis \citep[that could be constrained from lightcurves; ][]{schwartzetal2016,kawahara2016}, and depending on orbital inclination, maps resolved in latitude and longitude could be produce from high-quality data \citep{fujii&kawahara2012,cowanetal2013}.  Even in the absence of rotationally-resolved photometry, surface oceans --- whose presence directly confirms the habitility of an exoplanet --- could be detected via the effect of specular reflectance on a planetary phase curve \citep{mccullough2006,williams&gaidos2008}, especially at red and near-infrared wavelengths where observations at large phase angles are less strongly impacted by Rayleigh scattering and, thus, have better surface sensitivity \citep{robinsonetal2010,zuggeretal11a} \citep[observations at these wavelengths would also be less influenced by any hazes on the Archean Earth;][]{arneyetal2016}.  Additionally, scattering at the Brewster angle will maximize the polarization signature from an exo-ocean and could be detected in the polarization phase curve of an exo-Earth \citep{mccullough2006,stam2008,williams&gaidos2008,zuggeretal10}.  Finally, polarization and reflectance phase curves for Earth-like exoplanets can also reveal cloud properties through scattering effects \citep{bailey2007,karalidietal2011,karalidietal2012,karalidi&stam2012}, although little work has been done to understand how photochemical hazes (e.g., like those that may have been present on the Archean Earth) would impact polarimetric observations.

\bigskip
\noindent
\subsection{Visible Spectroscopy}
\label{subsec:spect}
\bigskip

Spectroscopic observations in reflected light provide powerful information about the atmospheric and surface environment of the Pale Blue Dot at any stage in its evolution.  In addition to the insights offered from photometry (as spectra can always be degraded to lower resolution), observations at even moderate spectral resolution enable the detection of trace atmospheric gases.  For example, \citet{drossartetal1993} used near-infrared {\it Galileo} data to constrain the abundances of CO$_2$, H$_2$O, CO, O$_3$, CH$_4$, and N$_2$O in the atmosphere of Earth.  While the \citet{drossartetal1993} study used spatially-resolved observations, the same gaseous absorption features appear in the disk-integrated {\it EPOXI} dataset \citep{livengoodetal2011}.

Beyond trace gas detection, spectroscopic reflected-light observations can also constrain atmospheric pressure --- a key determinant of habitability --- through Rayleigh scattering effects, broadening of gas absorption lines and bands, and through collision-induced absorption and dimer features.  Pressure-induced absorption features due to O$_2$ and N$_2$ occur throughout the near-infrared \citep[and into the mid-infrared;][]{misraetal2014a,schwietermanetal2015}.  Of course, interpretation of Rayleigh scattering features and pressure-broadened absorption bands is not straightforward.  The former depends on surface gravity and atmospheric mean molecular weight and can be masked by surface or haze absorption at blue wavelengths (see Figure~\ref{fig:earthevol}), while the latter is impacted by the composition of the background atmosphere \citep[e.g.,][]{hedges&madhusudhan2016}.  However, despite difficulties associated with detection, constraining O$_2$ and N$_2$ levels in the atmosphere of an Earth-like exoplanet would be key to deciphering the disequilibrium signature of N$_2$-O$_2$-H$_2$O \citep{krissansentottonetal2016a}.

\citet{fengetal2018} investigated retrievals of planetary and atmospheric properties for the modern Pale Blue Dot from visible-wavelength (0.4--1.0~$\upmu$m) spectroscopy at a variety of spectral resolutions and signal-to-noise ratios.  Here, firm constraints on key gas mixing ratios (for H$_2$O, O$_3$, and O$_2$), total atmospheric pressure, and planetary radius could be achieved with simulated Earth observations at a V-band signal-to-noise ratio of 20 and spectral resolution of 140.  Thus, observations of this quality for modern exo-Earth twin could be sufficient to indicate that the planet is either super-Earth or Earth-sized and that O$_2$ is a major atmospheric constituent, which is strong evidence that the planet may be inhabited.

Low resolution, low signal-to-noise ultraviolet observations of an Earth-like planet could rapidly distinguish the ozone-free Archean Earth from Earth at different evolutionary stages, and the ultraviolet Hartley-Huggins band may have exhibited dramatic seasonal variations during the Proterozoic \citep{olsonetal2018a}.  For the Archean, spectral models have demonstrated strong features due to methane and haze \citep{arneyetal2016}, but retrieval investigations have yet to show how observations of different quality and wavelength coverage could be used to infer methane and haze concentrations for an Archean Earth-like exoplanet.  Here, and as opposed to a haze-free Earth, planetary radius may be difficult to constrain from reflected light observations as the size determination relies strongly on measuring a Rayleigh scattering feature. (Since the Rayleigh scattering properties of a gas are well-defined, the planet-to-star flux ratio in a Rayleigh scattering feature is dependent primarily on the planetary size and orbital distance.)

\bigskip
\noindent
\subsection{Thermal Infrared Observations}
\label{subsec:ir}
\bigskip

Owing to the great technical challenges posed by techniques for observing Earth-like planets around Sun-like stars at long wavelengths, relatively little attention has been focused on understanding disk-integrated observations of our planet at thermal infrared wavelengths.  Nevertheless, infrared spectra of the distant Earth --- even at relatively low spectral resolution --- provide a great deal of information about the atmospheric and surface environment.  Most fundamentally, and unlike reflected-light data, infrared observations can directly constrain the radius of an exo-Earth.  The flux received from a true blackbody depends on its temperature, size, and distance.  If the distance to a target star is known, and with the temperature constrained via Wien's displacement law, the size of a planet can then be determined from low-resolution thermal infrared observations.  Of course, Earth does not emit like a true blackbody, which would introduce some uncertainty into a fitted planetary radius.

Beyond planetary size, infrared gas absorption features, by definition, reveal the key greenhouse gases of a planetary atmosphere.  For Earth, observations (Figure~\ref{fig:earthIR}) plainly reveal signatures of CO$_2$, H$_2$O, O$_3$, CH$_4$, and N$_2$O \citep{christensenandpearl97,heartyetal09}.  Regarding our evolving planet, the 9.7~$\upmu$m ozone band becomes apparent in Earth's emission spectrum after the rise of oxygen and the carbon dioxide bands at 9.4, 10.4, and 15~$\upmu$m strongly track decreasing atmospheric CO$_2$ levels with time \citep{meadows2008,rugheimer&kaltenegger2018}.  Critically, these would be observable at modest resolving powers and characteristic spectral signal-to-noise ratios of 5 \citep{rugheimer&kaltenegger2018}.  Additionally, pressure-induced absorption features can be used to indicate bulk atmospheric composition and pressure, and one such feature from N$_2$ has been detected in observations of the distant Earth near 4~$\upmu$m \citep{schwietermanetal2015}.

Critically, as molecular absorption bands are pressure broadened, and because high-opacity regions of a molecular band probe lower atmospheric pressures than do low-opacity regions, infrared observations can be used to probe the thermal structure of the atmosphere and surface of an exo-Earth.  This idea would apply for Earth at any stage in its evolution, even for a hazy Archean Earth as such hydrocarbon aerosols are typically transparent at infrared wavelengths \citep{arneyetal2016}.   Finally, thermal infrared lightcurves could also reveal variability due to weather (and associated condensational processes), rotation, and seasons \citep{heartyetal09,selsisetal2011,robinson2011,gomezlealetal2012,cowanetal2012b}.

 Using spatially-resolved {\it Galileo}/NIMS Earth observations, along with adopted {\it a priori} knowledge of total atmospheric pressure and the CO$_2$ mixing ratio, \citet{drossartetal1993} derived the thermal structures of cloud-free regions on Earth from the 4.3~$\upmu$m CO$_2$ band.  More recently, \citet{vonparisetal2013} used retrieval techniques on simulated infrared observations of a distant, modern Earth to show that low-resolution ($\lambda/\Delta\lambda = 20$) observations at signal-to-noise ratios of 10--20 could constrain thermal structure and atmospheric composition reasonably well.  However, like the \citet{drossartetal1993} retrievals, the results from \citet{vonparisetal2013} emphasize a cloud-free atmosphere.  Thus, it remains unclear how realistic patchy clouds would influence our ability to understand the atmosphere and surface of an exo-Earth from thermal infrared observations, and how this might impacts attempts to characterize exo-Earths in  different thermal states (e.g., a snowball state versus a clement or hothouse state).

%
\bigskip
\section{SUMMARY}
\label{sec:summary}
\bigskip
%

Exoplanetary science is rapidly progressing towards its long-term goal of discovering and characterizing Earth-like planets around our nearest stellar neighbors.  We now know that exoplanets, including potentially habitable Earth-sized worlds, are quite common, and small worlds orbiting within the Habitable Zone of nearby cool stars have already been discovered.  Advances in observational technologies, especially with regards to exoplanet direct imaging techniques, will enable the detection of Pale Blue Dots around other stars, potentially in the not-too-distant future.

Flybys of Earth by the {\it Galileo} spacecraft in the early 1990s enabled the remote detection of habitability and life on our planet using planetary science remote sensing techniques.  A combination of spatially-resolved visible and near-infrared spectral observations argued conclusively for the presence of liquid water on Earth's surface.  These same datasets indicated an atmosphere that was in a state of strong chemical disequilibrium --- a sign of life --- and observations at radio wavelengths contained features that indicated the presence of intelligent organisms.

More recently, a variety of observational approaches have yielded datasets that, effectively, allow us to view Earth as a distant exoplanet.  While observations from spacecraft at or beyond the Moon's orbit are ideal for understanding habitability and life signatures from the Pale Blue Dot, such data are rarely acquired.  Critically, satellite and Earthshine observations complement, and fill in certain gaps between, spacecraft data for the distant Earth.

Beyond observational datasets, models have proved effective tools for simulating and characterizing Earth as an exoplanet.  These tools span a wide range of complexities, including one-dimensional (vertical) spectral simulators, simple reflectance tools that capture the broadband reflectivity of Earth at visible wavelengths, and complex three-dimensional models that can simulate observations of the distant Earth at arbitrary viewing geometry across wavelengths that span the ultraviolet through the thermal infrared.  Especially in the absence of frequent spacecraft observations, models of the Pale Blue Dot can serve as testing grounds for proposed approaches to detecting and characterizing Earth-like exoplanets.

Geological and bio-geochemical studies of the long-term evolution of Earth reveal a world that, while being continuously habitable and inhabited, has progressed through a variety of surface and atmospheric states.  Abundances of key atmospheric constituents, including biosignature gases, have varied by many orders of magnitude.  As these gases imprint information about their concentrations on spectra of Earth, applying the aforementioned spectral simulation tools to the Pale Blue Dot at different geologic epochs reveals planetary spectra (and associated biosignatures) quite distinct from modern Earth.  Especially for the Archean Earth, the term ``Pale Blue Dot'' may not even apply.

Combining an understanding of remote sensing techniques relevant to exoplanets with knowledge of the conditions on the current and ancient Earth yields insights into approaches for detecting and studying Earth-like worlds around other stars.  Broadband observations have the potential to reveal habitable environments on ocean-bearing exoplanets, and time-resolved photometry can be used to extract spatial information from spatially-unresolved data.  More powerfully, spectroscopic observations at moderate resolutions can uncover key details about the surface and atmospheric state on a potentially Earth-like planet, including fundamental details relevant to life detection.  Only by uncovering the key signatures that indicate the habitability and inhabitance of Earth --- at any point in its evolution --- can we properly design the observational tools needed to discover and fully characterize other Earths.

%
\textbf{ Acknowledgments.} This research has made use of the Planetary Data System (PDS) and USGS Integrated Software for Imagers and Spectrometers (ISIS).  TR gratefully acknowledges support from NASA through the Sagan Fellowship Program executed by the NASA Exoplanet Science Institute and through the Exoplanets Research Program (award \#80NSSC18K0349).  Both TR and CR would like to acknowledge support from the NASA Astrobiology Institute, both through a grant to the  Virtual Planetary Laboratory (under Cooperative Agreement No. NNA13AA93A) and to the University of California, Riverside ``Alternative Earths'' team.  The results reported herein benefited from collaborations and/or information exchange within NASA's Nexus for Exoplanet System Science (NExSS) research coordination network sponsored by NASA's Science Mission Directorate.  We thank M.~Turnbull, P.~Christensen, T.~Hearty, E.~Pall{\'e}, E.~Schwieterman, and G.~Arney for openly sharing data used in this chapter.  Both authors also thank V.~Meadows, G.~Arney, N.~Cowan and an anonymous reviewer for detailed and constructive comments on versions of this manuscript.
%

%
\bigskip
\bibliographystyle{apa-good}

%

%
\end{document}